\documentclass[a4paper,11pt]{article}
\usepackage{jheppub} 
\usepackage{lineno}
\usepackage{bbold}
\usepackage{xcolor,multirow,makecell}
\usepackage[normalem]{ulem}
\usepackage{subfig}
\usepackage{xspace}
\usepackage[countmax]{subfloat}

\usepackage{booktabs}
\usepackage[normalem]{ulem}

\newcommand{\nc}{\newcommand}
\nc{\non}{\nonumber}
\nc{\hc}{\hbox {H.c.}} 
\nc{\noi}{\noindent}
\nc{\barx}{\bar{x}}
\nc{\pbarn}{\;\hbox {pb}}
\nc{\fbarn}{\;\hbox {fb}}
\nc{\gev}{\rm GeV}
\nc{\lsp}{\;\;\;\;\;}
\nc{\Lsp}{\;\;\;\;\;\;\;\;\;\;}  
\nc{\LLsp}{\lspace \lspace}
\nc{\lra}{\longrightarrow}
\nc{\beq}{\begin{equation}}  \nc{\eeq}{\end{equation}}
\nc{\bea}{\begin{eqnarray}}  \nc{\eea}{\end{eqnarray}}
\nc{\baa}{\begin{array}}     \nc{\eaa}{\end{array}}
\nc{\bit}{\begin{itemize}}   \nc{\eit}{\end{itemize}}
\nc{\ben}{\begin{enumerate}} \nc{\een}{\end{enumerate}}
\nc{\bce}{\begin{center}}    \nc{\ece}{\end{center}}
\nc{\bpm}{\begin{pmatrix}}   \nc{\epm}{\end{pmatrix}}
\nc{\bvt}{\begin{verbatim}}  \nc{\evt}{\end{verbatim}}

\begin{document}
\title{ALP and $Z^\prime $ boson at the Electron-Ion collider  }

\author[a]{Amit Adhikary,}
\author[a]{Dilip Kumar Ghosh,} 
\author[b,c]{Sk Jeesun} 
\author[a]{Sourov Roy}

\affiliation[a]{School of Physical Sciences, Indian Association for the Cultivation of Science, \\
2A $\&$ 2B, Raja S.C. Mullick Road, Kolkata, West Bengal-700032, India} 

\affiliation[b]{State Key Laboratory of Dark Matter Physics, Tsung-Dao Lee Institute \& School of Physics and Astronomy, Shanghai Jiao Tong University, Shanghai 200240, China}
\affiliation[c]{Key Laboratory for Particle Astrophysics and Cosmology (MOE) \& Shanghai Key Laboratory for Particle Physics and Cosmology, Shanghai Jiao Tong University, Shanghai 200240, China}

\emailAdd{amit93.adhikary@gmail.com}
\emailAdd{tpdkg@iacs.res.in}
\emailAdd{skjeesun48@gmail.com}
\emailAdd{tpsr@iacs.res.in}

\abstract{We study the sensitivity of the upcoming electron–ion (EIC) collider to 
purely electrophilic new physics in the GeV mass range. Within an 
effective field theory framework, we consider two different 
scenarios: an axion-like particle (ALP) and a new heavy neutral vector 
gauge boson $Z^\prime $, each couples to electrons only. We analyze 
electron–proton collisions at $\sqrt{s}= 141$ GeV with an integrated 
luminosity of $100~{\rm fb}^{-1}$, focusing primarily on the tri-electron 
final state. Additionally, loop-induced ALP–photon couplings driven 
photon final states are also explored. Incorporating realistic detector 
effects and systematic uncertainties, we obtain projected exclusion 
limits on the relevant cross-sections and couplings. We find that the results from
EIC can significantly extend the sensitivity to electrophilic axion-like 
particles and $Z^\prime $ bosons in regions of parameter space that remain 
weakly constrained by existing experiments.}

\maketitle

\section{Introduction}
Despite the triumph of standard model (SM) after the discovery of 125 GeV Higgs boson at the large hadron collider (LHC) \cite{ATLAS:2012yve,CMS:2012qbp}, it still fails to address some of the fundamental aspects of particle physics like dark matter (DM) \cite{Zwicky:1933gu,Rubin:1970zza,Clowe:2006eq,Planck:2018vyg}, non zero neutrino mass \cite{DayaBay:2012fng,RENO:2012mkc,MINOS:2013xrl}, baryon asymmetry \cite{Planck:2018vyg,ParticleDataGroup:2024cfk,Sakharov:1967dj}, strong CP problem \cite{ParticleDataGroup:2024cfk,Peccei:1977hh} etc.
To address these shortcomings of SM at its present setup, beyond the standard model (BSM) particles are often introduced.
These BSM particles may leave their trace in colliders \cite{ParticleDataGroup:2024cfk,Harris:2014hga,Alekhin:2015byh,Essig:2013vha}, cosmology \cite{Planck:2018vyg,DESI:2024mwx} and astrophysical environments \cite{Raffelt:1996wa} depending on their fundamental nature e.g. mass, interaction strength and interaction type \cite{ParticleDataGroup:2024cfk}.
Colliders like LHC, large electron positron collider (LEP) etc have served as one of the most effective way to look for these BSM particles from the past several decades since we have the very precise information of the expected SM background compared to the astrophysical or cosmological probes, and thus have put stringent constraints on several BSM scenarios \cite{ParticleDataGroup:2024cfk,Harris:2014hga,Alekhin:2015byh,Essig:2013vha}.

The Electron-Ion Collider (EIC) will be operational in the existing 
Relativistic Heavy Ion Collider (RHIC) infrastructure at Brookhaven National Laboratory \cite{Accardi:2012qut,AbdulKhalek:2021gbh}. 
This facility will enable high-luminosity deep inelastic scattering by 
colliding a polarized electron beam, capable of reaching energies of 
$E_e = 21 \text{ GeV}$ against hadron beams including polarized protons 
at $E_p \leq 250 \text{ GeV}$ and fully stripped heavy ions at
$E_A \leq 100 \text{ GeV/u}$. This configuration yields the center-of-mass 
energies that span $30–145 \text{ GeV}$ for polarized $ep$ interactions 
and $20–90 \text{ GeV}$ for $eA$ collisions. The EIC is designed to 
achieve peak luminosities between $10^{33}$ and $10^{34} 
\text{ cm}^{-2}\text{s}^{-1}$, providing the high-statistics data 
required for the study of various scenarios beyond the standard model.
In our collider analysis, we set the center-of-mass energy for the unpolarized $e^-p$ collisions, with 
$E_e = 18 $ GeV and $E_p = 275 $ GeV \cite{Balkin:2023gya,Batell:2022ogj}, respectively, providing 
$\sqrt{s}=141$ GeV with an integrated luminosity of ${\cal L}=100$ ${\rm ~fb}^{-1}$. 
EIC was originally proposed to probe the QCD frontier, but it can also be invaluable to probe various aspects of BSM physics scenarios. Several studies have already explored this possibility in the future EIC and have shown promising detection prospects for previously unexplored BSM physics parameter space \cite{Jiang:2025frv, Deng:2025hio, Davoudiasl:2025rpn, Bellafronte:2025ubi, Balkin:2025rtc, Wen:2024cfu, Wang:2024zns, Davoudiasl:2024vje, Balkin:2023gya, Davoudiasl:2023pkq, Batell:2022ogj, Yan:2022npz, Li:2021uww, Yan:2021htf, Cirigliano:2021img, Davoudiasl:2021mjy, Liu:2021lan}.
Although the LHC is useful for probing the energy frontier, the EIC provides a cleaner experimental environment with less contamination from SM backgrounds and measurements with unprecedented precision.

The axion is one of such well-motivated BSM particles that was proposed to solve the strong CP problem \cite{Peccei:1977hh, Weinberg:1977ma, Wilczek:1977pj,Kim:2008hd}. The axion is the pseudo-Nabu-Goldstone boson (pNGB) of  spontaneously broken global chiral 
$U(1)$ symmetry dubbed as the Peccei-Quinn (PQ) symmetry and has a mass inversely proportional to the breaking scale of the $U(1)$ symmetry ($f_a$). 
To realize such an axion, there exist several well known
ultraviolet (UV) complete BSM models, e.g., Peccei-Quinn-Weinberg-Wilczek (PQWW)~
\cite{Peccei:1977hh,Weinberg:1977ma,Wilczek:1977pj}, Kim-Shifman-Vainshtein-Zakharov (KSVZ) \cite{Kim:1979if,Shifman:1979if} and Dine-Fischler-Srednicki-Zhitnitsky (DFSZ) \cite{Zhitnitsky:1980tq,Dine:1981rt} model.
On the other hand, Axion-like particle (ALP) is coined to broadly signify the pNGBs not specifically related to the strong CP problem and can arise in multiple BSM theories with $U(1)$ symmetries broken at some scale higher than the electroweak (EW) scale \cite{Biekotter:2025fll,DiLuzio:2020wdo,Bharucha:2022lty}.
ALPs can couple to SM fermions and can lead to a rich phenomenology.
Axion and ALPs have been extensively studied in the context of DM \cite{Dine:1982ah,Preskill:1982cy, 
Abbott:1982af, DiLuzio:2020wdo, Co:2019jts,Arias:2012az,Jaeckel:2014qea}, portal to DM \cite{Bharucha:2022lty,Ghosh:2023tyz}, cosmology \cite{Marsh:2015xka,Caloni:2022uya,Cadamuro:2010cz} and astrophyiscs \cite{Turner:1987by,Caputo:2024oqc,Fiorillo:2025gnd,Candon:2025sdm}.
Collider and fixed target experiments provide the most direct probe of such ALP interactions, and numerous experimental bounds have been placed on the effective ALP-SM couplings in the GeV and sub-GeV mass ranges \cite{Bauer:2017ris}.
Examples of such existing works include LEP~\cite{OPAL:2002vhf,Jaeckel:2015jla,Yue:2021iiu,Tian:2022rsi,BESIII:2022rzz}, LHC~\cite{Adhikary:2024mzi, CMS:2012cve,ATLAS:2014jdv,Mimasu:2014nea,Brivio:2017ije, Ebadi:2019gij,Bonilla:2022pxu,Bharucha:2022lty,Mitridate:2023tbj,Dutta:2023abe}, muon collider \cite{Bao:2022onq}, beam dump \cite{CHARM:1985anb, Bjorken:1988as,Blumlein:1990ay,Dobrich:2017gcm,NA64:2020qwq,Afik:2023mhj,Ema:2023tjg}, BaBar \cite{BaBar:2014zli}, and Belle experiment \cite{Inguglia:2016acz}.

Similarly, $U(1)_X$ gauge extended BSM models are also motivated from several particle physics problems, including the DM portal \cite{Okada:2016tci,Mohapatra:2020bze,Okada:2021nwo}, neutrino mass \cite{Ma:2015raa,Okada:2021nwo}, baryon asymmetry \cite{Qi:2024pqe}, flavor anomalies \cite{Bonilla:2017lsq} and other phenomenological aspects as well
\cite{Bauer:2018onh,Coloma:2020gfv,Coloma:2022umy,AtzoriCorona:2022moj,Majumdar:2021vdw,Chakraborty:2021apc,Demirci:2023tui}. 
The gauge boson $Z^\prime$ realised in such models has also been searched for the past several decades in numerous present and future experiments \cite{Bauer:2018onh}. 
For example, in the TeV or sub-TeV masses LHC \cite{CMS:2016cfx,ATLAS:2019erb,Das:2016zue,Accomando:2017qcs} and LEP \cite{Essig:2009nc};  in the GeV regime LHCb \cite{LHCb:2017trq}, in the sub-GeV regime Belle\cite{Inguglia:2016acz}, BaBar \cite{BaBar:2014zli}, and fixed target experiments \cite{Bauer:2018onh,Bross:1989mp} have placed stringent constraint on such $Z'$ through their coupling with quarks or leptons. In the MeV - sub-MeV mass region neutrino scattering experiments \cite{Coloma:2022umy,COHERENT:2021xmm}, cosmological probes \cite{Escudero:2019gzq,Ghosh:2024cxi}; in the keV regime  astrophysical constraints like stellar cooling \cite{Kolb:1987qy,Croon:2020lrf,Akita:2023iwq,Li:2023vpv} constrain $Z'$-electron interaction.

These BSM particles are mostly studied through their effective interactions in the effective field theory (EFT) framework. 
For both ALP and $Z'$, LHC places the strongest constraint on their interactions with quarks or photons in the $\mathcal{O}(100)$ GeV mass range due to the high center-of-mass energy. 
However, the effective interactions of such ALPs and $Z'$ with electrons remain less explored in the mass range between $\sim$10 GeV and 100 GeV.
The EIC collider can provide a unique opportunity to probe such electrophilic GeV-scale BSM particles.
EIC can be more effective in probing such particles due to the low QCD background and the high integrated luminosity ($100{\rm ~fb}^{-1}$) \cite{AbdulKhalek:2021gbh}.
Previously, several works have considered EIC to constrain ALP-photon couplings \cite{Davoudiasl:2021mjy, Liu:2021lan, Balkin:2023gya}, heavy neutral leptons \cite{Batell:2022ogj}. 
However, the ALP-electron and $Z'$-electron couplings have not yet been explored at the EIC, which serves as the primary goal of this work.

In this work, we probe the electro-philic ALP and $Z'$ in the upcoming $e-p$ collider. We consider an EFT framework where an ALP is linearly coupled to the electron only. 
We considered the ALP-induced $e^-p\to e^-e^+e^-j$ process as signal events. The same signature can be mimicked by the SM photon and the $Z$ boson, and we consider them to be the SM background. Using proper cuts and statistical analysis, we place the projected limit of the future $e-p$ collider.
We follow the same methodology to probe an electrophilic $Z'$ gauge boson coupled to the electron through the fermion current.
Using the state-of-the-art analysis, we demonstrate the potential reach of the future $e-p$ collider in probing such electrophilic BSM sectors.

This paper is organized as follows. In Section~\ref{sec:theory}, we present the model of interest, featuring an electrophilic ALP in Section~\ref{sec:theory_alp}
and an electrophilic $Z'$ in Section~\ref{sec:theory_zpr}. An in-depth collider analysis for the ALP scenario is discussed in Section~\ref{sec:collider}, for three possible final states. Section~\ref{sec:analysis_zpr} presents the collider analysis for the $Z'$ scenario. In Section~\ref{sec:result}, we compare our results with the existing bounds. Finally, we conclude and outline future directions in Section~\ref{sec:conclusion}.

\section{Theoretical Framework}
\label{sec:theory}

As mentioned in the Introduction, we consider two example BSM scenarios where the SM particle content is augmented by (1) an electrophilic ALP and (2) an electrophilic vector boson.
As the names suggest, for each of these BSM scenarios, we consider coupling with the electron only in a model-independent way. 

\subsection{Electrophilic ALP }
\label{sec:theory_alp}
In this section, we extend the SM by an ALP, which arises as the pNGB, associated with the spontaneous breaking of global $U(1)$ symmetry at an energy scale $\Lambda$, assumed to be 
well above the energy scale $(E)$ associated with the processes considered in this analysis. For our considered energy range, $E \ll \Lambda$, the interaction of ALP with SM fields can be described using the effective field theory framework. In this analysis, we restrict ourselves to the ALP interaction with the electron only, and the corresponding effective interaction Lagrangian is given by : 
\begin{equation}
    \mathcal{L}_{\rm ALP}\supset \frac{1}{2}(\partial_\mu a)(\partial^\mu a)-\frac{1}{2} m_a^2 a^2-i g_{aee} a \bar{e} \gamma^5 e,
\label{eq:Lalp}
\end{equation}
where, $g_{aee}$ and $m_a$ are the coupling and mass of the ALP, respectively. Such kind of couplings can be realized in several UV complete models \cite{Dine:1981rt,Bharucha:2022lty} where effective operators of dimension-5 can be realized between ALP and SM fermions featuring a derivative coupling with the
 axial current of the fermion.
One can use the Dirac equation and field redefinition to obtain the pseudo Yukawa-like interaction in Eq.~\eqref{eq:Lalp}.
In principle, one can assume ALP coupled to all SM fermions, although simultaneous contributions from multiple couplings in the physical processes may jeopardize our understanding of the BSM paradigm. 
For a better understanding of ALP-electron interactions, we neglect all other ALP-fermion couplings (simply by setting other Wilson coefficients to zero \cite{DiLuzio:2020wdo,Bharucha:2022lty}) and adhere to Eq.~\eqref{eq:Lalp} throughout this work.

The presence of such an ALP will have striking consequences in the upcoming $e-p$ collider.
The scattering of  $e$ and $p$ can produce a distinct tri-lepton signature (see Fig.~\ref{fig:FD_eee} ) i.e.
\begin{eqnarray}
    e^-+p\to e^-+e^+ +e^-+j.
\end{eqnarray}
In addition to the contribution of the SM from the $\gamma$ and $Z$ boson, ALP also contributes to this process.
Thus, any statistically significant event that exceeds the expected SM background will result in a projected limit on the parameter space, as will be demonstrated later in this paper.

It is worth highlighting that, despite the absence of any tree-level ALP-photon coupling, it may be generated through an electron loop, leading to the following
loop induced ALP di-photon coupling \cite{Eberhart:2025lyu,Bauer:2017ris}:
\begin{eqnarray}
    \mathcal{L}_{\rm a \gamma}&\supset&  \frac{1}{4} g_{a\gamma\gamma}^{\rm eff} a F^{\mu \nu} \Tilde{F}_{\mu \nu},
\label{eq:Lga}
\end{eqnarray}    
where,
the loop induced effective coupling of ALP-photons, $g_{a\gamma\gamma}^{\rm eff }$, henceforth referred to as $g_{a\gamma\gamma}$,
is given by~\cite{Eberhart:2025lyu,Bauer:2017ris},
\begin{equation}
\label{eq:gaggl}
g_{a\gamma\gamma}=\frac{\alpha}{\pi } \frac{g_{aee}}{m_e} ~({\rm B_1}(\tau) - 1).
\end{equation}
with,
\bea
{\rm B_1}(\tau)=1-\tau f^2(\tau),~{\rm with}~\tau=\frac{4 m_e^2}{m_a^2},\nonumber\\
{\rm and}~f(\tau)= \begin{cases}
              \sin^{-1}\left(\frac{1}{\sqrt{\tau}} \right) {\rm for}~ \tau\geq 1 ,\\
              \frac{\pi}{2}+\frac{i}{2} \ln\dfrac{1+\sqrt{1-\tau}}{1-\sqrt{1-\tau}} {\rm for}~ \tau<1.
             \end{cases}
\eea
We have also incorporated this interaction into our analysis.
However, for our region of interest ($m_a\gtrsim 5.5$ GeV), we found that $g_{a\gamma\gamma}$ is always suppressed by $\sim \mathcal{O}(m_e/m_a^2)$ and hence does not contribute significantly to the event rates.
The partial decay width for the ALP to an electron pair and a photon pair can be written as

\bea
\label{eq:G_alp}
\Gamma_{a\to e^+e^-} &=& \frac{g_{aee}^2 m_a }{8 \pi }\sqrt{1-\frac{4 m_e^2}{m_a^2}},~\nonumber\\
    \Gamma_{a\to \gamma \gamma} &=& \frac{m_a^3 \left(  g_{a\gamma\gamma}\right)^2}{64 \pi}
\eea

\subsection{Electrophilic $Z'$ boson}
\label{sec:theory_zpr}
Similar to the previous subsection, here we consider a $Z'$ coupled to an electron,
\begin{equation}
    \mathcal{L}_{Z'}\supset \frac{1}{4} Z'_{\mu \nu}Z'^{\mu \nu}-\frac{1}{2} m_{\rm Z'}^2 {Z'}^2- g_{\rm Z'}  \bar{e} \gamma^\mu e Z'_\mu
    \label{eq:Zpr}
\end{equation}
where, $g_{Z'}$ and $m_Z'$ are the coupling and mass of the BSM $Z'$, respectively. $Z'_{\mu \nu} =\partial_\mu Z'_\nu -\partial_\nu Z'_\mu$ is the $Z'$ field strength tensor.
$Z'$ arise from the breaking of a new local $U(1)_X$ abelian gauge symmetry under which the SM particles are charged.
 SM fermion charges can be chosen arbitrarily, though their charge combinations are fixed from anomaly cancellation conditions \cite{Coloma:2022umy}.
Thus, a pure electrophilic $Z'$ is not possible in a gauge-invariant theory. However, in anomaly-free 
constructions, such as the $L_e-L_{\mu/\tau}$ framework, $Z'$ can couple with electrons without quark couplings.
Hence, we work with Eq.~\eqref{eq:Zpr} and obtain the conservative limits on such BSM gauge bosons.

Similar to ALPs, $Z'$ can also induce tri-electron events, leading to excess events over the expected background. At this point, it is worth noting that since leptons are part of doublets, $Z'$ will also couple to $\nu_e$ as well. In our analysis, we have considered this coupling as well, although neutrinos do not lead to any visible signature and are hence not useful for us.
The partial decay widths of the $Z'$ are given by
\bea
\label{eq:G_Zp}
\Gamma_{Z'\to e^+e^-} &=& \frac{g_{\rm Z'}^2 m_{\rm Z'} }{12 \pi }(1+\frac{2 m_e^2}{m_{\rm Z'}^2})\sqrt{1-\frac{4 m_e^2}{m_{\rm Z'}^2}}\\
\Gamma_{Z'\to \nu \bar{\nu}} &=& \frac{g_{\rm Z'}^2 m_{\rm Z'} }{24 \pi }
\eea

The parameter space of interest in this work mainly focuses on GeV-scale mass for the ALP or $Z'$ boson, with coupling strength roughly in the range of $\mathcal{O}(10^{-3}-1)$. 
Using Eqs.~\ref{eq:G_alp} and \ref{eq:G_Zp}, we find that the ALP and the \(Z'\) boson decay promptly into two on-shell electrons in the aforementioned parameter region with a typical decay width \(\Gamma \sim 10^{-7}\,{\rm GeV}\) which corresponds to
$
c\tau=\frac{\hbar c}{\Gamma}
\simeq 2.0\times 10^{-9}\,{\rm m}
\simeq 2~{\rm nm}.$
This proper decay length is much smaller than detector-scale displaced-vertex resolutions.
Thus, we consider the reconstruction of a prompt $e^+e^-$ resonance throughout the paper unless mentioned explicitly. After having a detailed discussion of the theoretical framework, we now move to the collider analysis of the aforementioned two BSM scenarios.

\section{Collider Analysis of leptophilic ALP}
\label{sec:collider}
In this section, we perform a collider study to search for a pseudoscalar with a mass above $\sim 5.5$ GeV in the upcoming electron-ion collider, where the electron and each nucleon in the ion carry energy, $E_e=18$ GeV and $E_p=275$ GeV, respectively~\cite{Balkin:2023gya,Batell:2022ogj}. As discussed in Section~\ref{sec:theory}, we are particularly interested in electrophilic ALPs. Thus, we first focus on the three-electron final state, $e^- p \to e^-e^+e^-j$, where j stands for light-quark jets. However, the ALP can couple to a photon pair with an electron loop. Although the effect of adding such an effective coupling is mild, we explore the three-electron channel in the presence of this coupling as well. In addition, we study final states containing photons, such as single-photon channel, $e^- p \to e^-\gamma j$ and di-photon channel, $e^- p \to e^-\gamma\gamma j$. We implement the ALP model discussed in Section~\ref{sec:theory_alp}, specifically Eq.~\ref{eq:Lalp}, in \texttt{FeynRules}~\cite{Alloul:2013bka} and generate the \texttt{UFO} model file to simulate events for the signal process. The parton-level signal and background events are generated with {\tt MadGraph5\_aMC@NLO}~\cite{Alwall:2014hca} in the leading order. {\tt NNPDF2.3LO} PDF set~\cite{NNPDF:2014otw} is used for the proton beam. To account for the detector effects in the analysis, the tracking and calorimeter energy resolutions are taken from~\cite{Adkins:2022jfp}. They are psudorapidity, $\eta$ dependent functions as shown in Table~\ref{tab:det_eff}.
We consider a flat identification efficiency of $\epsilon_\gamma\sim 70\%$ for a single photon~\cite{Balkin:2023gya} in the case of the final states that contain a photon. Note that identification efficiency is a function of transverse momentum $p_T$ and $\eta$, and must be handled carefully for more precise analysis.

\begin{table}[tb!]
\center
\begin{tabular}{c|c}
\bottomrule
Psudorapidity ($\eta$) range &  resolution $[\%]$ \\ \toprule 

\multicolumn{2}{c}{Tracking ($\Delta p/p$)}\\\bottomrule
$[-1.0,+1.0]$                    & $0.03 \times p/{\gev}\oplus 0.5$  \\
$[-2.5,-1.0]$ $\&$ $[+1.0,+2.5]$ & $0.03 \times p/{\gev}\oplus 1.0$  \\
$[-3.0,-2.5]$ $\&$ $[+2.5,+3.0]$ & $0.06 \times p/{\gev}\oplus 2.2$   \\
$[-3.5,-3.0]$ $\&$ $[+3.0,+3.5]$ & $0.11 \times p/{\gev}\oplus 4.0$   \\\toprule

\multicolumn{2}{c}{Electromagnetic calorimeter ($\Delta E/E$)}\\\bottomrule
$[+1.3,+3.5]$ & $7.1/\sqrt{E/{\gev }}$   \\
$[-1.7,+1.3]$ & $1.6/\sqrt{E/{\gev }}\oplus 0.7$  \\
$[-3.5,-1.7]$ & $1.8/\sqrt{E/{\gev }}\oplus 0.8$   \\\toprule
\end{tabular}
\caption{The projected detector resolutions applied to the collider analysis in this work~\cite{Adkins:2022jfp}. In case of an electron, we apply tracking resolutions on the three momentum and calorimeter resolutions on the energy. For a photon, calorimeter resolutions are used to smear the 4-momentum.}
\label{tab:det_eff}
\end{table}

The following signal significance formula, $\mathcal{S}$ is used in this work~\cite{Cowan:2010js, Cowan:2012}:
\begin{equation}
\label{S}
{\cal S} = \sqrt{2 \left[({\rm S+B}) \log (1+\frac{\rm S}{\rm B})-{\rm S} \right]} \,.
\end{equation}
The signal yield, S, and the total background yield, B, are calculated from $\sigma\cdot\epsilon\cdot\mathcal{L}$ where $\sigma$ is the cross-section of the process, $\epsilon$ is the efficiency after collider analysis, and $\mathcal{L}$ refers to the integrated luminosity. In the presence of systematic uncertainty, $\sigma_{sys\_un}$, the signal significance formula takes the following form:
\begin{equation}
\label{S_sys}
{\cal S}_\text{sys} = \sqrt{2 \left[({\rm S+B}) \log \left(\frac{({\rm S+B})({\rm B}+\sigma_{\rm B}^{2})}{{\rm B}^{2}+({\rm S+B})\sigma_{\rm B}^{2}}\right)-\frac{{\rm B}^{2}}{\sigma_{\rm B}^{2}}\log \left(1+\frac{\sigma_{\rm B}^{2} {\rm S}}{{\rm B}({\rm B}+\sigma_{\rm B}^{2})} \right) \right]} \,,
\end{equation}
where $\sigma_{\rm B}=\sigma_{sys\_un}\times {\rm B}$. 

In the following subsections, we discuss the generation-level cuts for signal and background processes, kinematic observables constructed from the final state objects, optimization procedure to reduce backgrounds, and finally the exclusion limits on the cross-section of the signal process and interaction coupling, for the above mentioned channels. 
We begin the discussion by studying the prospects of the three electron channel in the context of an electron-ion collider.

\subsection{The $e^-e^+e^-$ channel}
\label{sec:eee}

\begin{figure}[!tb]
\begin{center}
\subfloat[]{\label{fig:11}
\includegraphics[width=.15\textwidth]{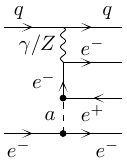}
}\qquad
\subfloat[]{\label{fig:12}
\includegraphics[height=.15\textwidth]{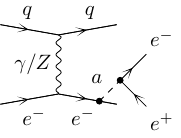}
}\qquad
\subfloat[]{\label{fig:13}
\includegraphics[width=.15\textwidth]{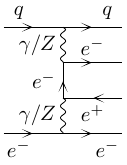}
}\qquad
\subfloat[]{\label{fig:14}
\includegraphics[height=.15\textwidth]{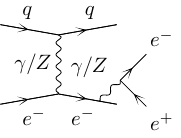}
}
\end{center}
\caption{{ The leading order (LO) Feynman diagrams for signal process ($(a) \& (b)$) and background process ($(c)\& (d)$) in $e^-p\to e^+e^-e^+j$ final state.}}
\label{fig:FD_eee}
\end{figure}

\begin{figure}[tb!]
\centering
\includegraphics[scale=0.45]{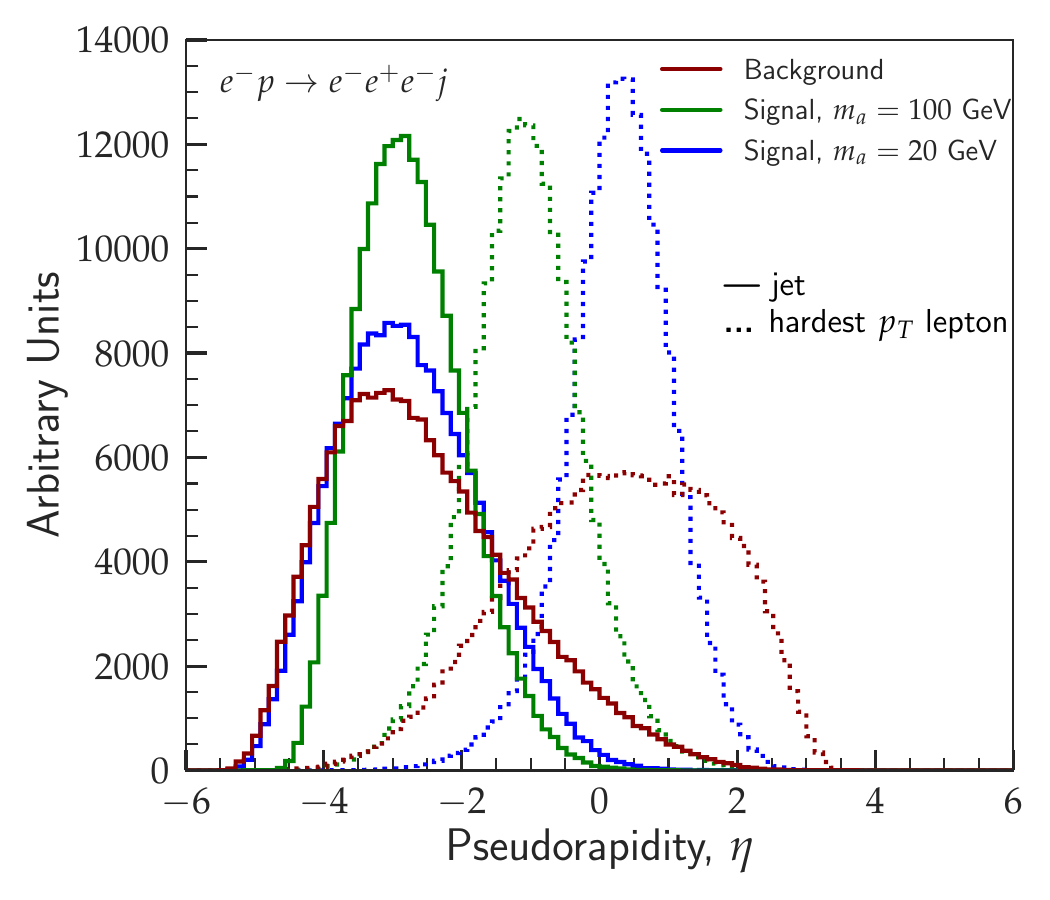}
\caption{The pseudorapidity distributions are shown for the signal process with chosen ALP masses of $m_a=20$ GeV and $m_a=100$ GeV, along with the background process. The solid and dotted lines correspond to jet and hardest $p_T$ electron, respectively.}
\label{fig:eta}
\end{figure}

This section focuses on the $e^- p \to e^-e^+e^-j$ final state. The Feynman diagrams of the signal process are shown in Fig.~\ref{fig:FD_eee}. In the chosen mass range between 5.5 GeV and 100 GeV, the ALP is dominantly produced on-shell as shown in Fig.~\ref{fig:FD_eee} (b) and has a dominant contribution to the total production cross-section. 
The background in this channel comes from the SM production of a final state with three electrons. We generate this background inclusively, containing all the contributing Feynman diagrams. The representative Feynman diagrams are shown in Fig.~\ref{fig:FD_eee} (c) and (d). The signal and background events are generated in {\tt MadGraph5} with specific requirements: $p_{T,e}>2$ GeV, $|\eta|<3.5$, $\Delta R_{ee}>0.3$~\footnote{The angular distance in the $\eta-\phi$ plane between two final state particles is $\Delta R =\sqrt{(\Delta\eta)^2 + (\Delta\phi)^2}$, where $\phi$ refers to the azimuthal angle and pseudorapidity ($\eta$) can be written in terms of polar angle ($\theta$) as $\eta=-{\rm ln~tan}(\theta/2)$. }~\footnote{ The matrix element for signal production involves pole structures in electron and proton momentum transfers, $t_e = (p_e-k_{e})^2$ and
$t_p = (p_p-k_{p})^2$, respectively. Here, the incoming and outgoing momentum are referred with ``$p$" and ``$k$", respectively. We checked explicitly and applied these generation level cuts to ensure stability of the signal generation in {\tt MadGraph5} under collinear/soft enhancements.}. 
As shown with dotted lines in Fig.~\ref{fig:eta}, the electrons are mainly produced in the central region. The hadronic activity due to the accompanying jet, in both the signal and background processes, is in the forward region, as illustrated with solid colored lines. Thus we choose the baseline selection without any requirement on the jet activity.
We would like to mention that this does not affect the analysis strategy proposed in the following, as the invariant mass distribution of the two hardest $p_T$ electrons, $m_{ee}$, remains the same with or without a jet pseudorapidity requirement. However, far-forward instrumentation and forward proton/ion reconstruction may be feasible at the EIC~\cite{Pitt:2024utg}, which might help to achieve better sensitivity for some specific scenarios. We present a discussion on vetoing jets in the forward region and their effect on the signal sensitivity at the end of this section.

We select events with exactly three electrons in the final state. The electrons are sorted according to the transverse momentum. We apply the tracking resolutions to the three-momentum and calorimeter energy resolutions to the energy of the electrons. This will induce the smearing effect of the EIC detector on the final state particles. At EIC, the hadronization objects from the recoiling ion/proton will be closer to the beam direction, and their four-momentum cannot be measured. We select the two hardest $p_T$ electrons in the final state and construct their invariant mass $\mathrm{m}_{ee}$. $\mathrm{m}_{ee}$ is used in the analysis to discriminate the ALP signal from background events. The choice of selecting the two hardest $p_T$ electrons comes from the fact that they predominantly originate from the decay of ALP in the chosen range of $m_a$, as shown in Fig.~\ref{fig:FD_eee} (b). Thus, the kinematic distribution of $\mathrm{m}_{ee}$ appears as a peak at each ALP benchmark mass. However, as the ALP mass increases, the contribution from the Feynman diagram in Fig.~\ref{fig:FD_eee} (a) becomes dominant and the variance of the $\mathrm{m}_{ee}$ distribution increases.

\begin{figure}[tb!]
\centering
\includegraphics[scale=0.45]{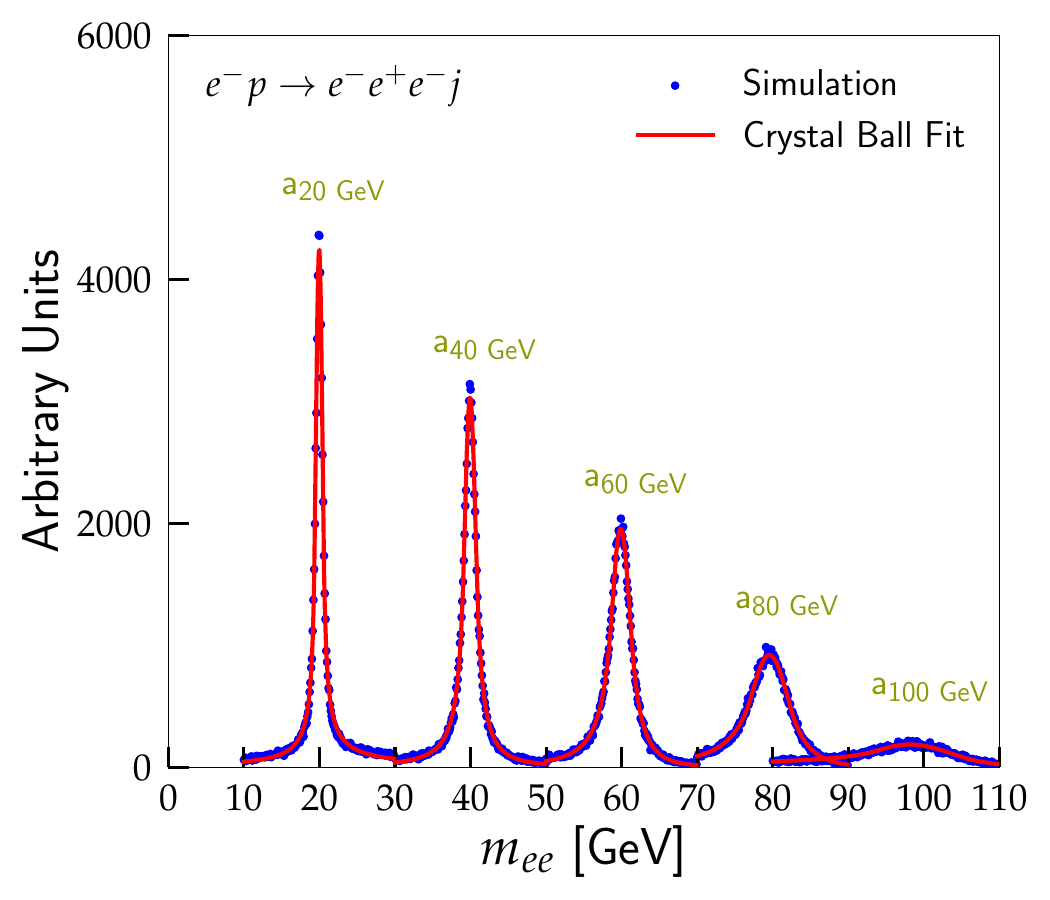}
\caption{The $\mathrm{m}_{ee}$ distribution for the ALP signal and Crystal Ball fits, in blue and red color, respectively. Fits are shown for signal events with $m_a=$ 20, 40, 60, 80, and 100 GeV. As example, $a_{\rm 20~GeV}$ refers to an ALP with mass $m_a=$ 20 GeV.}
\label{fig:fit_mee}
\end{figure}

We estimate the signal efficiency and background yields by fitting the $\mathrm{m}_{ee}$ distribution of the signal process with a double-sided Crystal Ball function~\cite{Oreglia:1980cs,ATLAS-CONF-2014-031}. The double-sided Crystal Ball function is a combination of a Gaussian function and a power law function that is used to fit the core of the distribution, and both the low and high $\mathrm{m}_{ee}$ regions, respectively.
The Crystal Ball function is defined as:
 \begin{equation}
 \begin{split}
   & f(\mathrm{m}_{ee}|N,\mu,\sigma,\alpha_{\mathrm{low}},\alpha_{\mathrm{high}},n_{\mathrm{low}},n_{\mathrm{high}}) = \\ \\
   & N\times 
    \begin{cases}
    \mathrm{e}^{-0.5(\frac{\mathrm{m}_{ee}-\mu}{\sigma})^2}, & \mbox{if $-\alpha_{\mathrm{low}} \leq \frac{\mathrm{m}_{ee}-\mu}{\sigma} \leq \alpha_{\mathrm{high}}$~,}\\
    \mathrm{e}^{-0.5\alpha_{\mathrm{low}}^{2} } \left[\frac{\alpha_{\mathrm{low}}}{n_{\mathrm{low}}} \left(\frac{n_{\mathrm{low}}}{\alpha_{\mathrm{low}}} - \alpha_{\mathrm{low}} -\frac{\mathrm{m}_{ee}-\mu}{\sigma} \right)\right]^{-n_{\mathrm{low}}},  & \mbox {if $\frac{\mathrm{m}_{ee}-\mu}{\sigma} < -\alpha_{\mathrm{low}}$~,}\\
    \mathrm{e}^{-0.5\alpha_{\mathrm{high}}^{2} } \left[\frac{\alpha_{\mathrm{high}}}{n_{\mathrm{high}}} \left(\frac{n_{\mathrm{high}}}{\alpha_{\mathrm{high}}} - \alpha_{\mathrm{high}}  + \frac{\mathrm{m}_{ee}-\mu}{\sigma} \right)\right]^{-n_{\mathrm{high}}},  & \mbox {if $\frac{\mathrm{m}_{ee}-\mu}{\sigma} > \alpha_{\mathrm{high}}$~.}\\
    \end{cases}
    \label{eq:CBF}
\end{split}
 \end{equation}
The function consists of a normalization factor $N$. $\mu$ and $\sigma$ are the mean value and width of the central Gaussian part of the Crystal Ball distribution, respectively. The two power law functions take over the Gaussian function at the $\alpha_{\mathrm{low}}$ and $\alpha_{\mathrm{high}}$ points, while $n_{\mathrm{low}}$ and $n_{\mathrm{high}}$ correspond to the exponents of the functions, respectively. The simulated distributions $\mathrm{m}_{ee}$ for signal events are shown in Fig.~\ref{fig:fit_mee} with blue color. The corresponding Crystal Ball fits are also shown in red for ALP masses of 20 GeV, 40 GeV, 60 GeV, 80 GeV, and 100 GeV. We select the mass window in the interval $[\mu - \sigma, \mu + \sigma]$ to search for each hypothesis of the ALP mass. The signal efficiency and background yield for electron-proton collisions at the center-of-mass energy of $\sqrt{s}=$ 141 GeV and 100 ${\rm fb}^{-1}$ of integrated luminosity are shown in Table~\ref{tab:yield_eee}.

\begin{table}[tb!]
\center
\begin{tabular}{c|c|c|c|c|c|c}
\bottomrule
\multirow{3}{*}{ $m_a$ [GeV] }  & \multicolumn{3}{c|}{ Fitted parameters [GeV]}  &  \multirow{3}{*}{ $\epsilon$ } & \multicolumn{2}{c}{Yield at $\mathcal{L}=100$ fb$^{-1}$} \\\cline{2-4}\cline{6-7}

& \multirow{2}{*}{$\mu$ } & \multirow{2}{*}{$\sigma$} & \multirow{2}{*}{$\chi^2$/d.o.f} && \multirow{2}{*}{\makecell{S ($g_{\mathrm{aee}}=0.1$})} &  \multirow{2}{*}{B }\\ 
&& &&&\\ \hline

 $5.5$  & $5.68\pm 0.02$ & $1.10\pm 0.03$ & 2.11  & $0.24$ & 43680  & 5188 \\
 $10$  & 10.00$\pm 0.00$ & $0.52\pm 0.00$   & 2.77  & $0.27$ & 15660  & $3467$ \\
 $20$  & $19.96\pm 0.00$ & $0.46\pm 0.00$ & 3.61  & $0.34$ & 3614 & $1300$ \\
 $30$  & $29.96\pm 0.00$ & $0.59\pm 0.00$ & 2.78  & $0.38$ & 1110 & $699$ \\
 $40$  & $39.93\pm 0.00$ & $0.77\pm 0.00$ & 2.86  & $0.40$ & 392 & $404$ \\
 $50$  & $49.92\pm 0.00$ & $1.00\pm 0.00$ & 1.86  & $0.43$ & 143 & $233$ \\
 $60$  & $59.87\pm 0.01$ & $1.28\pm 0.01$ & 1.86  & $0.43$ & 52 & $121$ \\
 $70$  & $69.75\pm 0.01$ & $1.80\pm 0.01$ & 1.19  & $0.44$ & 18 & $55$ \\
 $80$  & $79.54\pm 0.02$ & $2.50\pm 0.03$ & 1.07  & $0.40$ & 5 & $22$ \\
 $90$  & $89.23\pm 0.04$ & $3.60\pm 0.06$ & 1.03  & $0.31$ & 1 & $9$ \\
 $100$ & $98.17\pm 0.06$ & $5.47\pm 0.08$ & 1.12  & $0.18$ & 0.1 & $5$  \\
 \toprule
\end{tabular}
\caption{The mean values, $\mu$ and widths, $\sigma$ obtained after fitting the ALP reconstructed $\mathrm{m}_{ee}$ mass variable with the Crystal Ball function. To assess the goodness of fit, the $\chi^2$/d.o.f values are also provided. The notation $\pm 0.00$ indicates that the observed variation occurs only at the third decimal place or beyond. After considering the $[\mu - \sigma, \mu + \sigma]$ interval for each ALP mass, the signal efficiency, $\epsilon$, the signal yields, S for $g_{\mathrm{aee}}=0.1$ and background yields, B, are shown at integrated luminosity of $100$ fb$^{-1}$. }
\label{tab:yield_eee}
\end{table}

\begin{figure}[tb!]
\centering
\subfloat[]{\label{fig:3a}
\includegraphics[width=.45\textwidth]{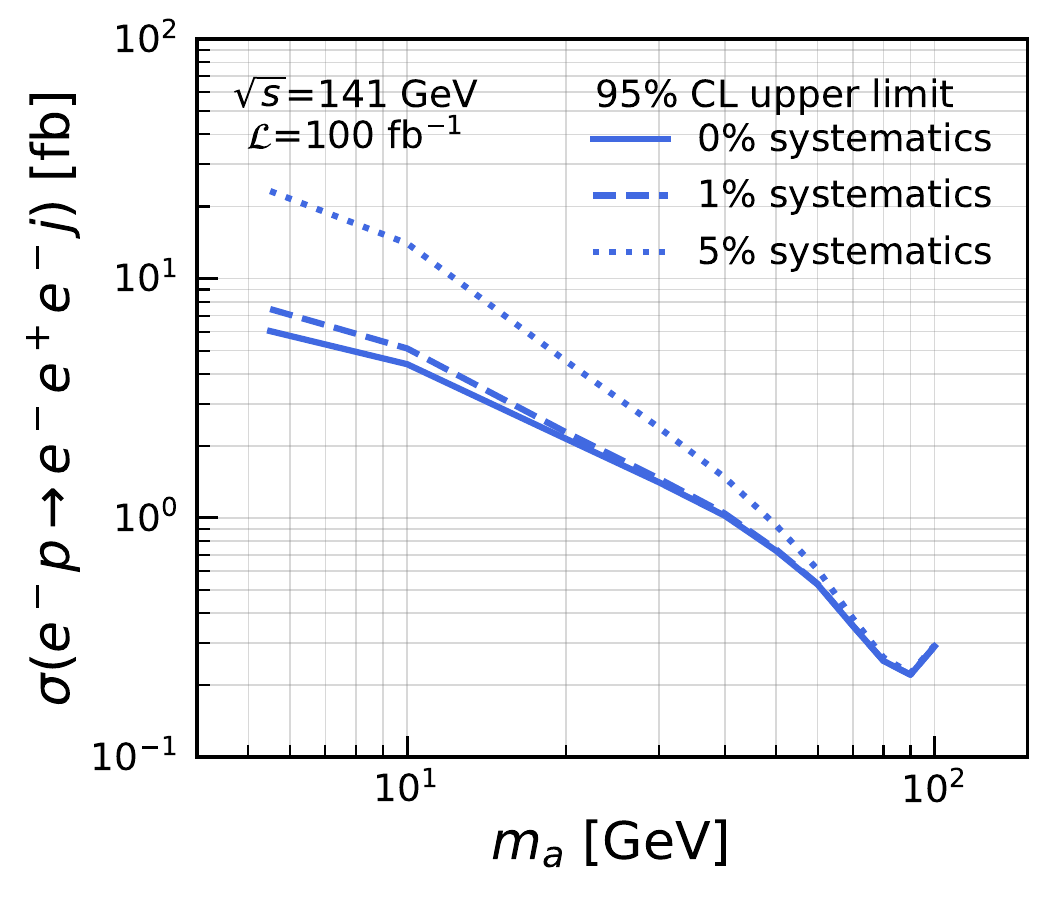}
}\qquad
\subfloat[]{\label{fig:3b}
\includegraphics[width=.45\textwidth]{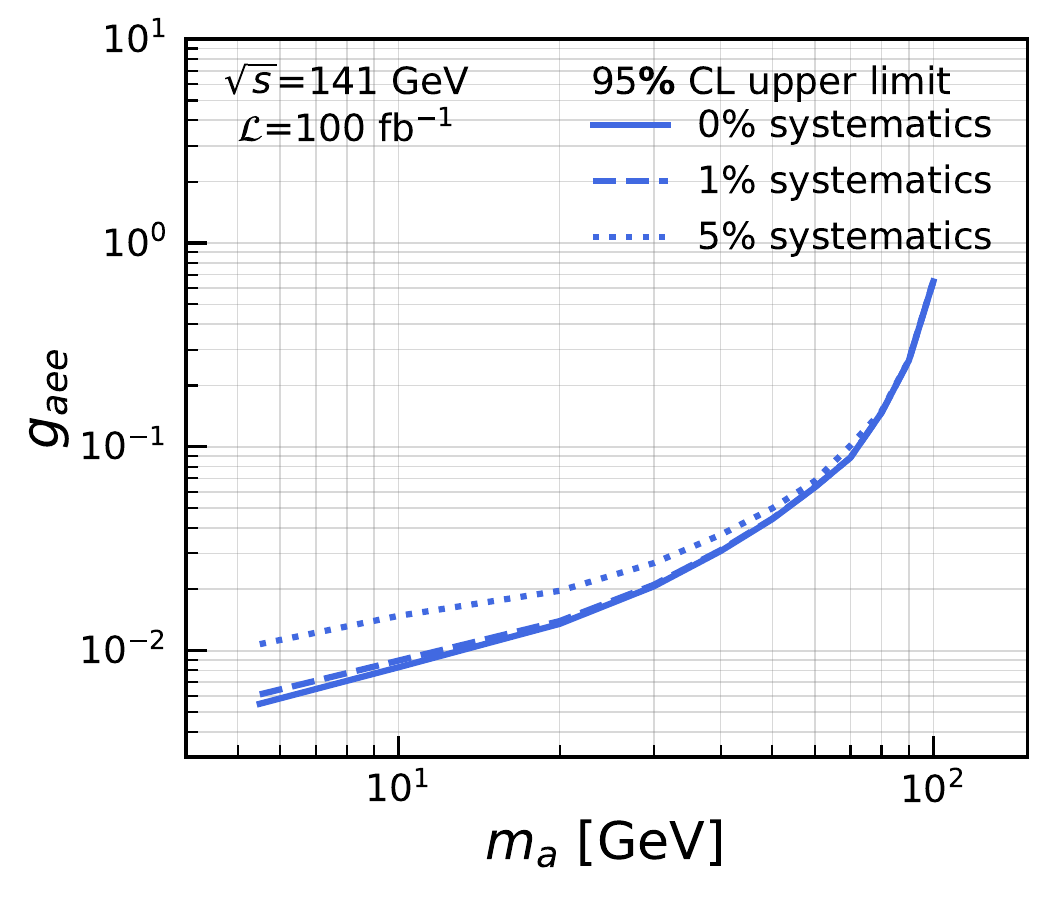}
}
\caption{Upper limit at $95\%$ CL on (a) signal production cross-section, $\sigma(e^- p\to e^-e^+e^-j)$ and (b) ALP-electron coupling, $g_{aee}$ as a function of ALP mass, $m_a$. The solid blue line corresponds to adding null systematic uncertainty, while dashed and dotted lines include $1\%$ and $5\%$ systematic uncertainties, respectively.}
\label{fig:ul_gaee_3e}
\end{figure}

The upper limit on the cross-section is calculated using the following formula: ${\cal S}> {N}_{\rm CL}$, where ${N}_{\rm CL}=2$ is taken for the $95\%$ confidence interval. As defined earlier in Eq.~\ref{S}, the signal significance ${\cal S}$ depends on the signal yield (${\rm S}$) and background yield (${\rm B}$). Here, the signal yield can be written as $\epsilon\times\sigma(e^- p\to e^- e^+ e^- j)_{\text{UL}}\times{\cal L}$. Thus, we obtain the upper limit on the cross-section of signal production, $\sigma(e^- p\to e^- e^+ e^- j)_{\text{UL}}$, by numerically solving the Eq.~\ref{S} without the systematics and Eq.~\ref{S_sys} with systematics. Henceforth, the upper limits on cross-sections are denoted without a suffix, as $\sigma(e^- p\to e^- e^+ e^- j)$.
The expected integrated luminosity for electron-ion collisions is taken to be $\mathcal{L}=100~{\rm fb}^{-1}$. We perform the analysis for each $m_a$ and calculate $\epsilon$ and B. They are tabulated in Table~\ref{tab:yield_eee}, along with the signal yield corresponding to $g_{aee}=0.1$.
In Fig.~\ref{fig:ul_gaee_3e} (a), we show the derived exclusion limits at $95\%$ CL in the plane of $\sigma(e^- p\to e^- e^+ e^- j)$ and $m_a$. The solid blue line corresponds to the addition of no systematic uncertainties, while the $1\%$ and $5\%$ systematic uncertainties are considered for the dashed and dotted lines, respectively. The exclusion limits vary in the range of 6.03 fb and 0.29 fb for an ALP mass of 5.5 GeV and 100 GeV, respectively. The upper limits become weaker after adding systematic uncertainty of $1\%$ ($5\%$) with cross-section values of 7.45 fb (23.19 fb) for $m_a=5.5$ GeV. However, the effect of adding systematics decreases with higher ALP masses and becomes negligible around an ALP mass of 100 GeV. This happens because the background yield becomes smaller (see Table~\ref{tab:yield_eee}) and adding systematics to the background has negligible effect in the signal significance formula. 

Next, the upper limit on $\sigma(e^- p\to e^- e^+ e^- j)$ is translated into an upper limit on the ALP-electron coupling $g_{aee}$, as shown in Fig.~\ref{fig:ul_gaee_3e} (b), at $\sqrt{s}=141$ GeV with $\mathcal{L}=100~{\rm fb}^{-1}$. 
Couplings that lie above the blue solid line are excluded at a $95\%$ CL when assuming zero systematic uncertainty. If a systematic uncertainty of $1\%$ (or $5\%$) is assumed, the couplings above the blue dashed (or dotted) line are similarly ruled out. The upper limit is stronger for an ALP mass of 5.5 GeV with $g_{aee}\lesssim 0.005$, and varies between $g_{aee}\lesssim (0.008-0.64)$ in the range $m_a\sim (10-100)$ GeV. Upon adding systematic uncertainty of $1\%$ and $5\%$, the sensitivity to coupling decreases to $g_{aee}\lesssim (0.009-0.09)$ and $(0.015-0.1)$, respectively, for $m_a\sim (10-70)$ GeV.
Further, we have scrutinised the effects of showering and hadronisation on the simulated samples. The signal sensitivity mildly changes with at most $\sim 10\%$ weaker exclusion limit on $g_{aee}$ without any systematics. This happens primarily because the signal $m_{ee}$ distribution is smeared after showering and the signal efficiency drops upon choosing a $1\sigma$ mass window around the fitted mean value.

\begin{table}[tb!]
\center
\begin{tabular}{c|c|c|c|c|c|c}
\bottomrule
\multirow{3}{*}{ $m_a$ [GeV] }  & \multicolumn{3}{c|}{ Fitted parameters [GeV]}  &  \multirow{3}{*}{ $\epsilon$ } & \multicolumn{2}{c}{Yield at $\mathcal{L}=100$ fb$^{-1}$} \\\cline{2-4}\cline{6-7}

& \multirow{2}{*}{$\mu$ } & \multirow{2}{*}{$\sigma$} & \multirow{2}{*}{$\chi^2$/d.o.f} && \multirow{2}{*}{\makecell{S ($g_{\mathrm{aee}}=0.1$})} &  \multirow{2}{*}{B }\\ 
&& &&&\\ \hline

 $5.5$  & $5.55\pm 0.02$ & $0.61\pm 0.06$ & 2.91  & $0.16$ & 28408  & 2155 \\
 $10$  & 9.99$\pm 0.00$ & $0.50\pm 0.00$ & 3.43  & $0.26$ & 14942  & $2491$ \\
 $20$  & $19.96\pm 0.00$ & $0.45\pm 0.00$ & 4.45  & $0.32$ & 3459 & $904$ \\
 $30$  & $29.95\pm 0.00$ & $0.57\pm 0.00$ & 3.32  & $0.36$ & 1075 & $459$ \\
 $40$  & $39.94\pm 0.00$ & $0.77\pm 0.00$ & 2.85  & $0.40$ & 391 & $272$ \\
 $50$  & $49.92\pm 0.00$ & $0.99\pm 0.00$ & 2.84  & $0.42$ & 141 & $159$ \\
 $60$  & $59.86\pm 0.00$ & $1.33\pm 0.01$ & 1.90  & $0.44$ & 52 & $88$ \\
 $70$  & $69.75\pm 0.01$ & $1.78\pm 0.01$ & 1.70  & $0.43$ & 17 & $40$ \\
 $80$  & $79.55\pm 0.01$ & $2.53\pm 0.02$ & 1.29  & $0.40$ & 5 & $16$ \\
 $90$  & $89.22\pm 0.03$ & $3.60\pm 0.04$ & 1.05  & $0.31$ & 1 & $7.3$ \\
 $100$ & $98.47\pm 0.04$ & $5.26\pm 0.05$ & 0.99  & $0.19$ & 0.1 & $4.1$  \\
 \toprule
\end{tabular}
\caption{The mean values, $\mu$ and widths, $\sigma$ obtained after fitting the ALP reconstructed $\mathrm{m}_{ee}$ mass variable with the Crystal Ball function. To assess the goodness of fit, the $\chi^2$/d.o.f values are also provided. The notation $\pm 0.00$ indicates that the observed variation occurs only at the third decimal place or beyond. After considering the $[\mu - \sigma, \mu + \sigma]$ interval for each ALP mass, the signal efficiency, $\epsilon$, the signal yields, S for $g_{\mathrm{aee}}=0.1$ and background yields, B, are shown at integrated luminosity of $100$ fb$^{-1}$, in the $e^- p\to e^- e^+ e^- j$ channel with jet requirements. }
\label{tab:yield_eee_jet}
\end{table}

\begin{figure}[tb!]
\centering
\subfloat[]{\label{fig:3a}
\includegraphics[width=.45\textwidth]{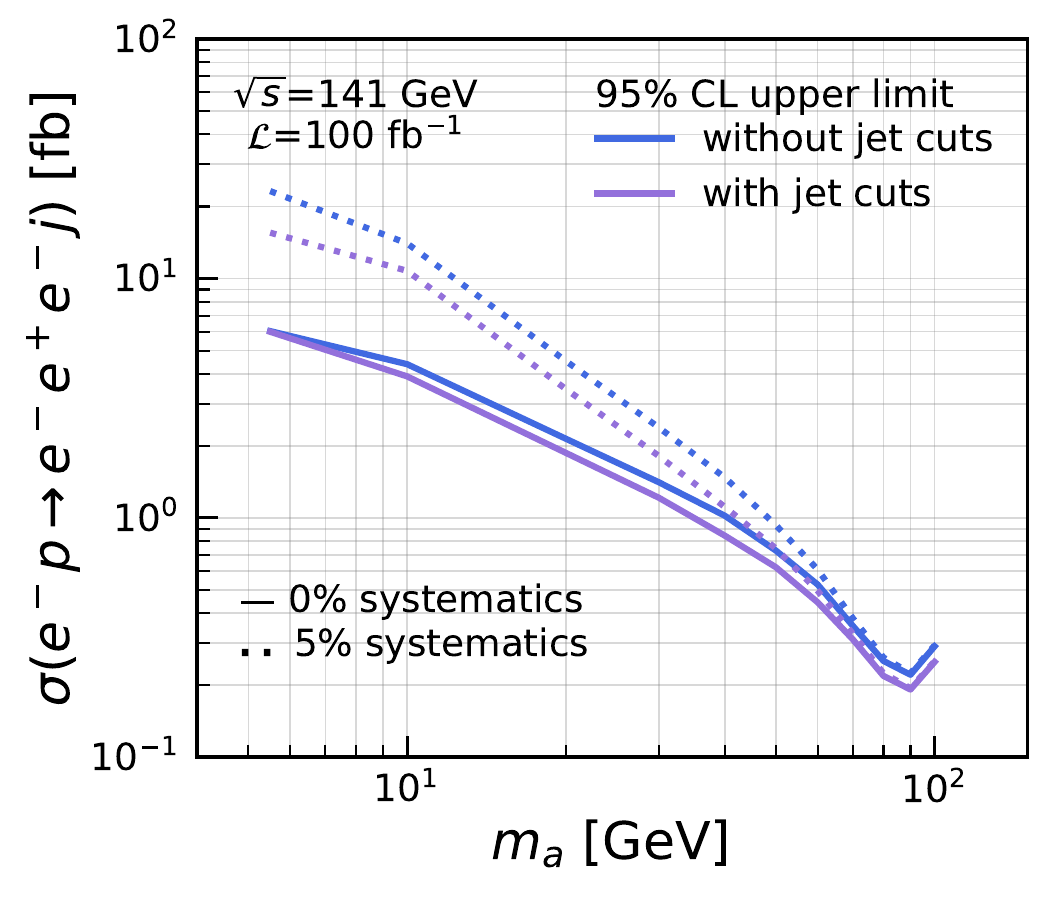}
}\qquad
\subfloat[]{\label{fig:3b}
\includegraphics[width=.45\textwidth]{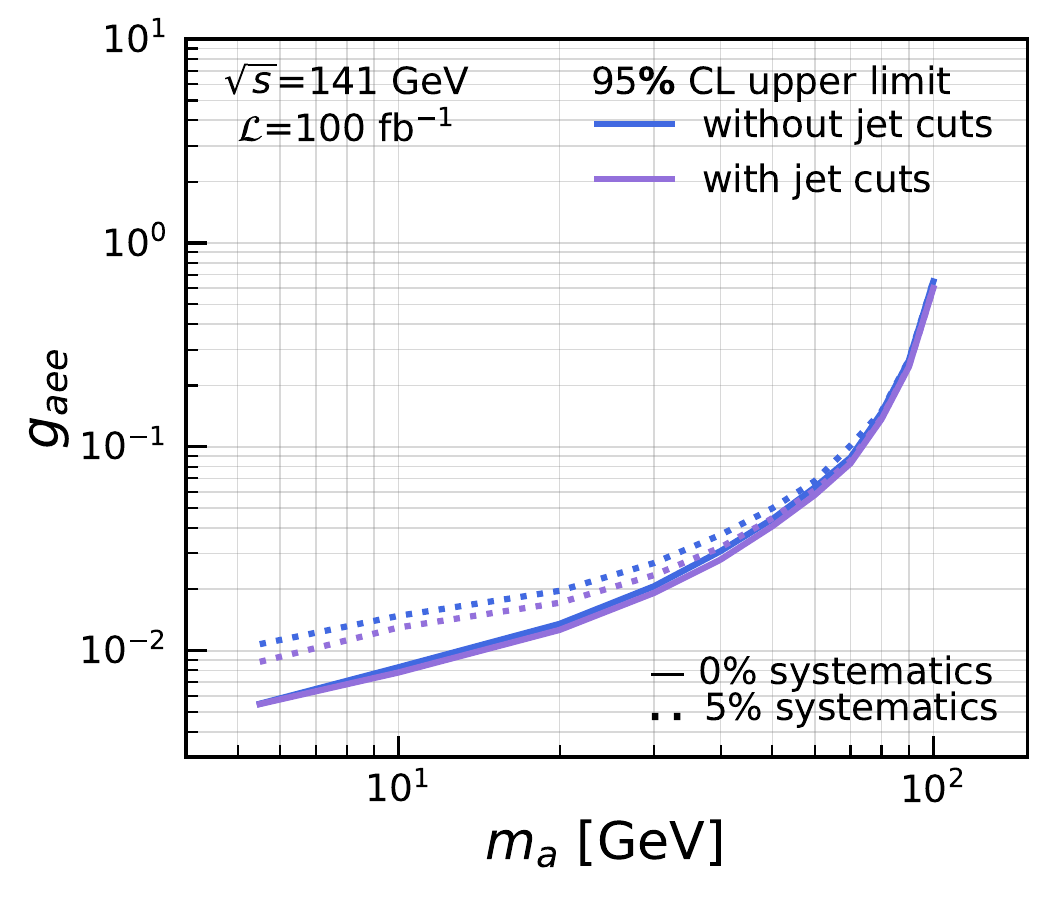}
}
\caption{The $95\%$ CL upper limits on the (a) signal production cross-section, $\sigma(e^- p\to e^-e^+e^-j)$ and (b) ALP-electron coupling, $g_{aee}$ as a function of ALP mass, $m_a$. Solid lines correspond to adding null systematic uncertainty, while dotted lines include $5\%$ systematic uncertainty. The slightly stronger limit after jet-veto requirements in the forward region is shown with purple-colored line and compared with previous results without any jet requirements in blue.}
\label{fig:ul_gaee_3e_jet}
\end{figure}

Finally, we discuss the consequences of vetoing the forward jets in this channel. As detailed in Ref.~\cite{Pitt:2024utg}, one of the primary objectives at EIC detector setup is to utilise  “far-forward” (FF) and “far-backward” (FB) detectors. It will enable the reconstruction of the forward scattered electron, and tagging and reconstruction of hadronic activity close to the beam direction. To explore this facet, we generate the signal and background events with the following generation level requirements: $p_{T,e}>2$ GeV, $|\eta_e|<3.5$, $\Delta R_{ee}>0.3$, $p_{T,j}>2$ GeV, $|\eta_j|<3.5$, $\Delta R_{ej}>0.4$. The event selection criteria, final state reconstruction and the fitting procedure of the $m_{ee}$ distribution is identical to the analysis performed above. In Table~\ref{tab:yield_eee_jet}, the signal efficiency, signal yield for $g_{aee}=0.1$ and background yield are tabulated after repeating the fitting analysis for each ALP benchmark masses. The mean values and width from the Crystal Ball fit are also shown. As expected, the background yield decreases after vetoing the forward jets in the events with $|\eta_j|<3.5$, and  reduces by roughly $30\%$ as compared to previous analysis without any cuts on the accompanying jet. We further evaluate the exclusion limits at 95$\%$ CL and illustrate them in Fig.~\ref{fig:ul_gaee_3e_jet} on the (a) signal production cross-section, $\sigma(e^- p\to e^- e^+ e^- j)$ and (b) $g_{aee}$ coupling, as a function of $m_a$. The solid and dashed purple-colored lines corresponds to exclusion limits without and with $5\%$ systematic uncertainty, respectively. The upper limits in blue are taken from the previous analysis without any jet requirements. In comparison, the exclusion limits without any systematics improves by roughly $10\%$ for the signal production cross-section and by $\sim 6\%$ for the ALP-electron coupling, after vetoing the forward jets. Thereby the forward and backward detectors may improve the possible extraction of potential new physics at the EIC.

\hspace{1cm}
\subsubsection*{The $e^-e^+e^-$ channel with loop-induced $g_{a\gamma\gamma}$}

\begin{figure}[!tb]
\begin{center}
\subfloat[]{\label{fig:11}
\includegraphics[height=.22\textwidth]{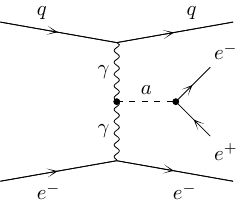}
}\qquad
\subfloat[]{\label{fig:12}
\includegraphics[height=.22\textwidth]{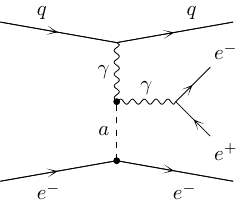}
}\qquad
\subfloat[]{\label{fig:13}
\includegraphics[height=.12\textheight]{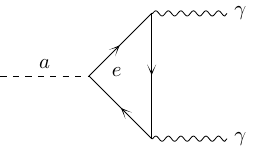}
}
\end{center}
\caption{ Feynman diagrams for the signal process in the $e^-p\to e^-e^+e^-j$ final state are shown in (a) and (b), which include the ALP-photon coupling $g_{a\gamma\gamma}$, while (c) illustrates the decay of the ALP into a photon pair via an electron loop.
}
\label{fig:FD_ega}
\end{figure}

The goal of the present section is to understand the effects of including the ALP-photon coupling through the electron loop, in the $e^-e^+e^-$ channel. Here, we repeat the analysis performed in the previous Section~\ref{sec:eee} by including the $g_{a\gamma\gamma}$ coupling in the signal production. In addition to the Feynman diagrams in Fig.~\ref{fig:FD_eee}, the signal process also includes the Feynman diagrams presented in Fig.~\ref{fig:FD_ega}. To generate the signal process in \texttt{Madgraph5}, we implement Eq.~\ref{eq:Lalp} and Eq.~\ref{eq:Lga} in \texttt{FeynRules} and obtain the corresponding \texttt{UFO} model file. The same model file is used in the following sections for final states containing a photon.

\begin{figure}[tb!]
\centering
\includegraphics[scale=0.45]{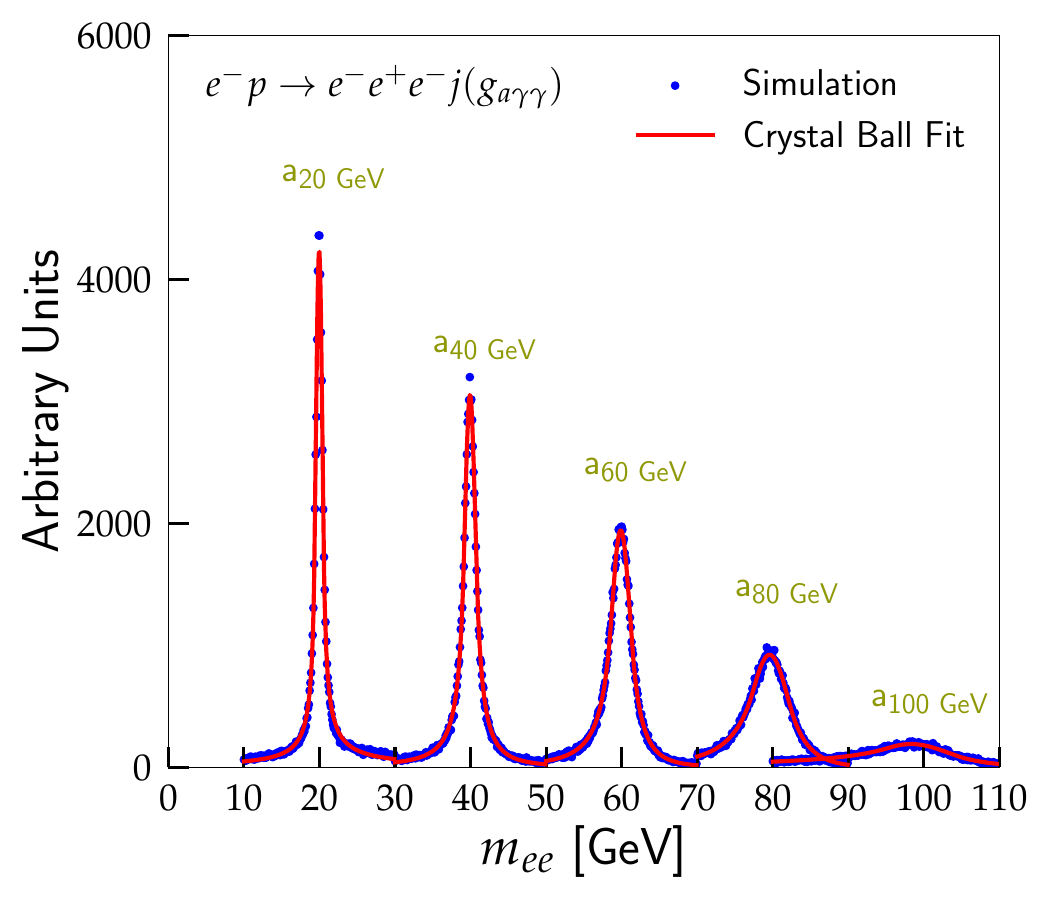}
\caption{The $\mathrm{m}_{ee}$ distribution for the ALP signal and Crystal Ball fits, in blue and red color, respectively. Fits are shown for signal events with $m_a=$ 20, 40, 60, 80, and 100 GeV. As example, $a_{\rm 20~GeV}$ refers to an ALP with mass $m_a=$ 20 GeV.}
\label{fig:fit_mee_ga}
\end{figure}

The events are selected if they contain three electrons in the final state. The 4-momentum of electrons is smeared with the projected detector resolutions of the EIC detector. After that, we reconstruct the invariant mass of the two hardest $p_T$ electrons ${\rm m}_{ee}$. To estimate the signal and background events, the ${\rm m}_{ee}$ distribution for each ALP mass is fitted with a double-sided Crystal Ball function. In Fig.~\ref{fig:fit_mee_ga}, the simulated signal samples are shown in blue, while the corresponding Crystal Ball fits are shown with red lines for a few ALP masses of $m_a=$ 20, 40, 60, 80, and 100 GeV. After repeating the analysis for each ALP mass, we obtain the fitted mass window $\mu\pm\sigma$ of the Gaussian core distribution. They are tabulated in Table~\ref{tab:yield_3e_ga}, along with signal efficiency, background yield, and signal yield for a chosen $g_{aee}=0.1$ at the integrated luminosity of $\mathcal{L}=100~{\rm fb}^{-1}$. We observe that the change in signal efficiency and background yield is almost negligible compared to the previous analysis of three electron channels without the $g_{a\gamma\gamma}$ coupling. This is expected from the fact that the ${\rm m}_{ee}$ distribution for each ALP mass has a negligible change after adding the loop-generated ALP-photon interaction in the electrophilic ALP model. 

\begin{table}[tb!]
\center
\begin{tabular}{c|c|c|c|c|c|c}
\bottomrule
\multirow{3}{*}{ $m_a$ [GeV] }  & \multicolumn{3}{c|}{ Fitted parameters [GeV]}  &  \multirow{3}{*}{ $\epsilon$ } & \multicolumn{2}{c}{Yield at $\mathcal{L}=100$ fb$^{-1}$} \\\cline{2-4}\cline{6-7}

& \multirow{2}{*}{$\mu$ } & \multirow{2}{*}{$\sigma$} & \multirow{2}{*}{$\chi^2$/d.o.f} && \multirow{2}{*}{\makecell{S ($g_{\mathrm{aee}}=0.1$})} &  \multirow{2}{*}{B }\\ 
&& &&&\\ \hline

 $5.5$ & $5.69\pm 0.02$  & $1.11\pm 0.03$ & 1.64  & $0.24$ & 43200  & 5250 \\
 $10$  & $9.99\pm 0.00$  & $0.52\pm 0.00$ & 2.36  & $0.27$ & 15660  & $3469$ \\
 $20$  & $19.96\pm 0.00$ & $0.46\pm 0.00$ & 3.82  & $0.34$ & 3740 & $1300$ \\
 $30$  & $29.95\pm 0.00$ & $0.59\pm 0.00$ & 2.23  & $0.37$ & 1110 & $700$ \\
 $40$  & $39.93\pm 0.00$ & $0.75\pm 0.00$ & 2.91  & $0.40$ & 400 & $393$ \\
 $50$  & $49.92\pm 0.00$ & $0.99\pm 0.00$ & 1.69  & $0.42$ & 147 & $230$ \\
 $60$  & $59.88\pm 0.01$ & $1.30\pm 0.01$ & 1.72  & $0.44$ & 52 & $123$ \\
 $70$  & $69.74\pm 0.01$ & $1.83\pm 0.01$ & 1.32  & $0.44$ & 18 & $56$ \\
 $80$  & $79.57\pm 0.02$ & $2.48\pm 0.03$ & 1.02  & $0.39$ & 5 & $22$ \\
 $90$  & $89.17\pm 0.04$ & $3.54\pm 0.06$ & 0.98  & $0.30$ & 1.1 & $9$ \\
 $100$ & $98.31\pm 0.05$ & $5.07\pm 0.09$ & 1.11  & $0.17$ & 0.1 & 5  \\
 \toprule
\end{tabular}
\caption{The mean values, $\mu$ and widths, $\sigma$ obtained after fitting the ALP reconstructed $\mathrm{m}_{ee}$ mass variable with the Crystal Ball function, along with the signal efficiency, $\epsilon$. To assess the goodness of fit, the $\chi^2$/d.o.f values are also provided. The notation $\pm 0.00$ indicates that the observed variation occurs only at the third decimal place or beyond. The signal yields, S for $g_{\mathrm{aee}}=0.1$ and background yields, B, are also shown at integrated luminosity of $100$ fb$^{-1}$, for the simulated MC samples.}
\label{tab:yield_3e_ga}
\end{table}

\begin{figure}[tb!]
\centering
\subfloat[]{\label{fig:6a}
\includegraphics[width=.45\textwidth]{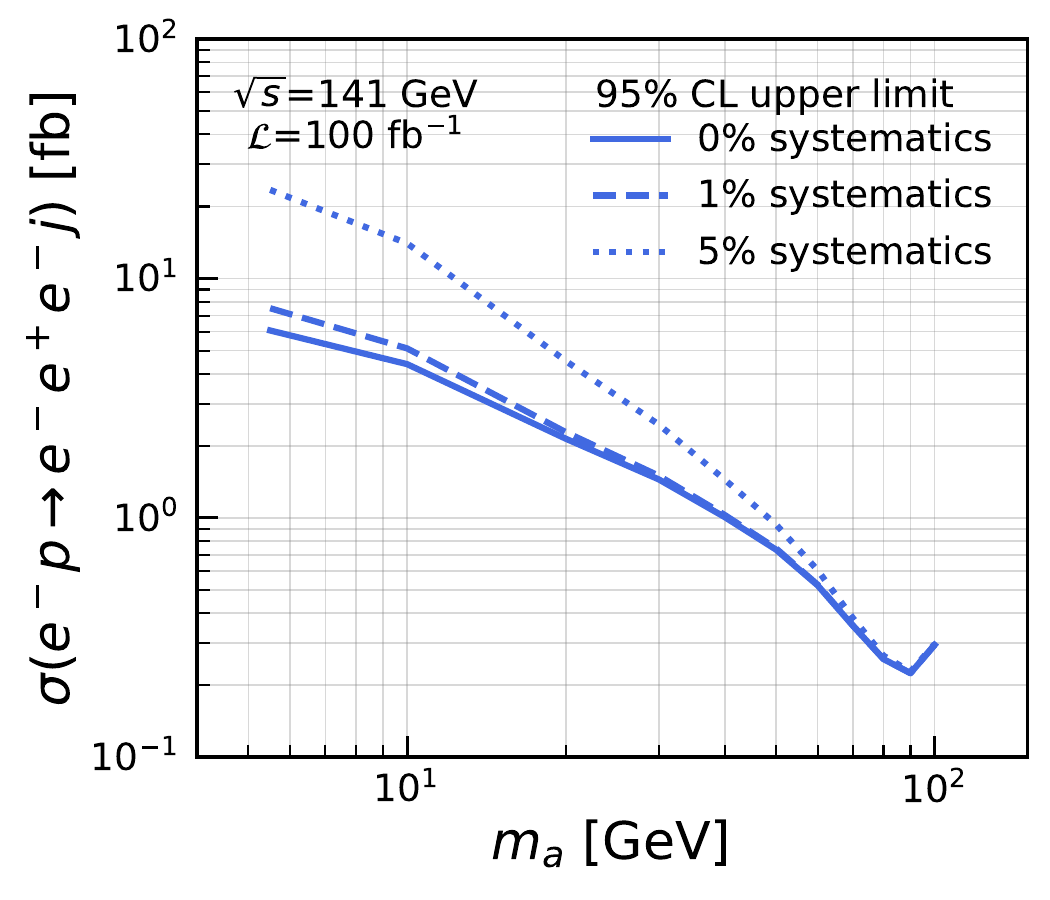}
}\qquad
\subfloat[]{\label{fig:6b}
\includegraphics[width=.45\textwidth]{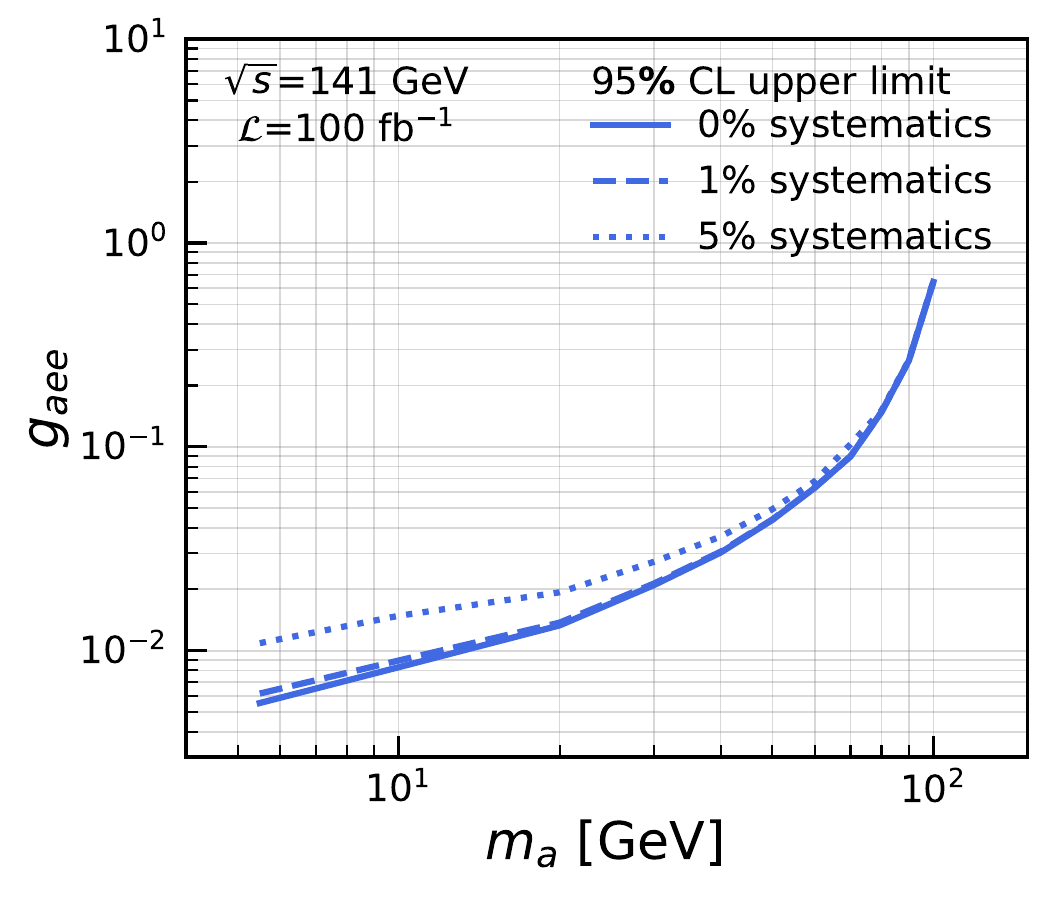}
}
\caption{Upper limit at $95\%$ CL on (a) the signal production cross-section, $\sigma(e^- p\to e^-e^+e^-j)$ including ALP-$\gamma$ interaction and (b) the ALP-electron coupling, $g_{aee}$ as a function of ALP mass, $m_a$. The solid blue line corresponds to adding null systematic uncertainty, while dashed and dotted lines include $1\%$ and $5\%$ systematic uncertainties, respectively.}
\label{fig:ul_gaee_3e_ga}
\end{figure}

We show the results in Fig.~\ref{fig:ul_gaee_3e_ga} (a) as exclusion limits at 95$\%$ CL on the signal production cross-section, $\sigma(e^- p\to e^- e^+ e^- j)$, as a function of $m_a$. The limit is weaker at low ALP mass and gets stronger as the ALP mass increases, varying between 6.07 fb and 0.29 fb for $m_a=5.5$ and 100 GeV, respectively. Adding $1\%$ ($5\%$) systematic uncertainty, the upper limits weaken to 7.51 fb (23.46 fb) and 0.36 fb (0.38 fb) for an ALP mass of 5.5 GeV and 70 GeV, respectively. The translated upper limit on the ALP-electron coupling as a function of the ALP mass is illustrated in Fig.~\ref{fig:ul_gaee_3e_ga} (b). The upper limit varies in the range $g_{aee}\leq$ (0.006 - 0.64) for $m_a\sim (5.5-100)$ GeV. As is evident, these results are similar to the ones obtained in the $e^-e^+e^-j$ channel in the absence of the $g_{a\gamma\gamma}$ coupling.

\subsection{The $e^-\gamma\gamma$ channel}

\begin{figure}[!tb]
\begin{center}
\subfloat[]{\label{fig:7a}
\includegraphics[height=.22\textwidth]{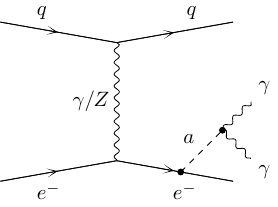}
}\qquad
\subfloat[]{\label{fig:7b}
\includegraphics[height=.22\textwidth]{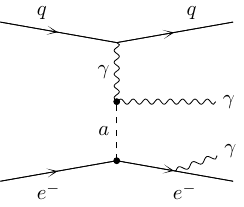}
}\qquad
\subfloat[]{\label{fig:7c}
\includegraphics[height=.22\textwidth]{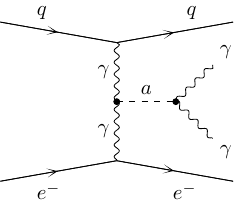}
}
\end{center}
\caption{ The Feynman diagrams for the signal process leading to the $e^-p \to e^- \gamma\gamma j$ final state are as follows: diagrams (a) and (b) involve both the ALP–electron and ALP–photon couplings, while diagram (c) depends only on the ALP–photon coupling. In diagram (b), the photon may originate from the electron leg or from the quark leg; here we show one diagram for illustrative purposes. In contrast, the electrophilic ALP emission in diagram (a) can only occur from the electron leg.}
\label{fig:FD_2ph}
\end{figure}

Having introduced the loop-generated ALP-photon interaction, the ALP can be searched in di-photon final state as well. In this section, we discuss the reach of the $e^- p \to e^-\gamma\gamma j$ final state for electron-proton collisions at the center-of-mass energy of $\sqrt{s}=$ 141 GeV with 100 ${\rm fb}^{-1}$ of integrated luminosity. The representative Feynman diagrams of the signal process are illustrated in Fig.~\ref{fig:FD_2ph} that include resonant (Fig.~\ref{fig:FD_2ph} (a) and (c)) and non-resonant (Fig.~\ref{fig:FD_2ph} (b)) ALP production. We simulate the signal process, $e^- p \to e^-\gamma\gamma j$, in \texttt{Madgraph5}. As is obvious, the production rate drops significantly compared to the $e^-e^+e^-$ final state (see Section~\ref{sec:eee}) with tree-level ALP-electron interaction. The background in this channel arises from the electroweak processes, where the final state photons come from initial/final state radiation. We include all the Feynman diagrams corresponding to the hard process and generate the background inclusively. Both signal and background samples are simulated with specific requirements at the parton level: $p_{T,e/\gamma/j}>2$ GeV, $|\eta_{e/\gamma}|<3.5$, $|\eta_{j}|<5.0$ and $\Delta R_{e,\gamma,j}>0.4$, where $\Delta R_{a,b,c}$ represents the $\Delta R$ separation between all combinations of $a,b$ and $c$.

Events are selected if they contain exactly two photons and one electron in the final state. To apply the detector effects at the EIC, we smear the photon 4-momentum with the calorimeter resolutions as summarised in Table~\ref{tab:det_eff}. Next, we reconstruct the invariant mass of the two photons, $m_{\gamma\gamma}$. The $m_{\gamma\gamma}$ distribution for each signal process has a narrow peak around the chosen ALP masses as $\Gamma_{a\to\gamma\gamma}/m_a\ll 1$. We use this variable to reduce the background events in this channel. This is done by fitting the $\mathrm{m}_{\gamma\gamma}$ distribution for each ALP signal with a double-sided Crystal Ball function. As before, the function is given by:
 \begin{equation}
 \begin{split}
   & f(\mathrm{m}_{\gamma\gamma}|N,\mu,\sigma,\alpha_{\mathrm{low}},\alpha_{\mathrm{high}},n_{\mathrm{low}},n_{\mathrm{high}}) = \\ \\
   & N\times 
    \begin{cases}
    \mathrm{e}^{-0.5(\frac{\mathrm{m}_{\gamma\gamma}-\mu}{\sigma})^2}, & \mbox{if $-\alpha_{\mathrm{low}} \leq \frac{\mathrm{m}_{\gamma\gamma}-\mu}{\sigma} \leq \alpha_{\mathrm{high}}$~,}\\
    \mathrm{e}^{-0.5\alpha_{\mathrm{low}}^{2} } \left[\frac{\alpha_{\mathrm{low}}}{n_{\mathrm{low}}} \left(\frac{n_{\mathrm{low}}}{\alpha_{\mathrm{low}}} - \alpha_{\mathrm{low}} -\frac{\mathrm{m}_{\gamma\gamma}-\mu}{\sigma} \right)\right]^{-n_{\mathrm{low}}},  & \mbox {if $\frac{\mathrm{m}_{\gamma\gamma}-\mu}{\sigma} < -\alpha_{\mathrm{low}}$~,}\\
    \mathrm{e}^{-0.5\alpha_{\mathrm{high}}^{2} } \left[\frac{\alpha_{\mathrm{high}}}{n_{\mathrm{high}}} \left(\frac{n_{\mathrm{high}}}{\alpha_{\mathrm{high}}} - \alpha_{\mathrm{high}}  + \frac{\mathrm{m}_{\gamma\gamma}-\mu}{\sigma} \right)\right]^{-n_{\mathrm{high}}},  & \mbox {if $\frac{\mathrm{m}_{\gamma\gamma}-\mu}{\sigma} > \alpha_{\mathrm{high}}$~,}\\
    \end{cases}
    \label{eq:CBF_gaga}
\end{split}
 \end{equation}

\begin{figure}[tb!]
\centering
\includegraphics[scale=0.45]{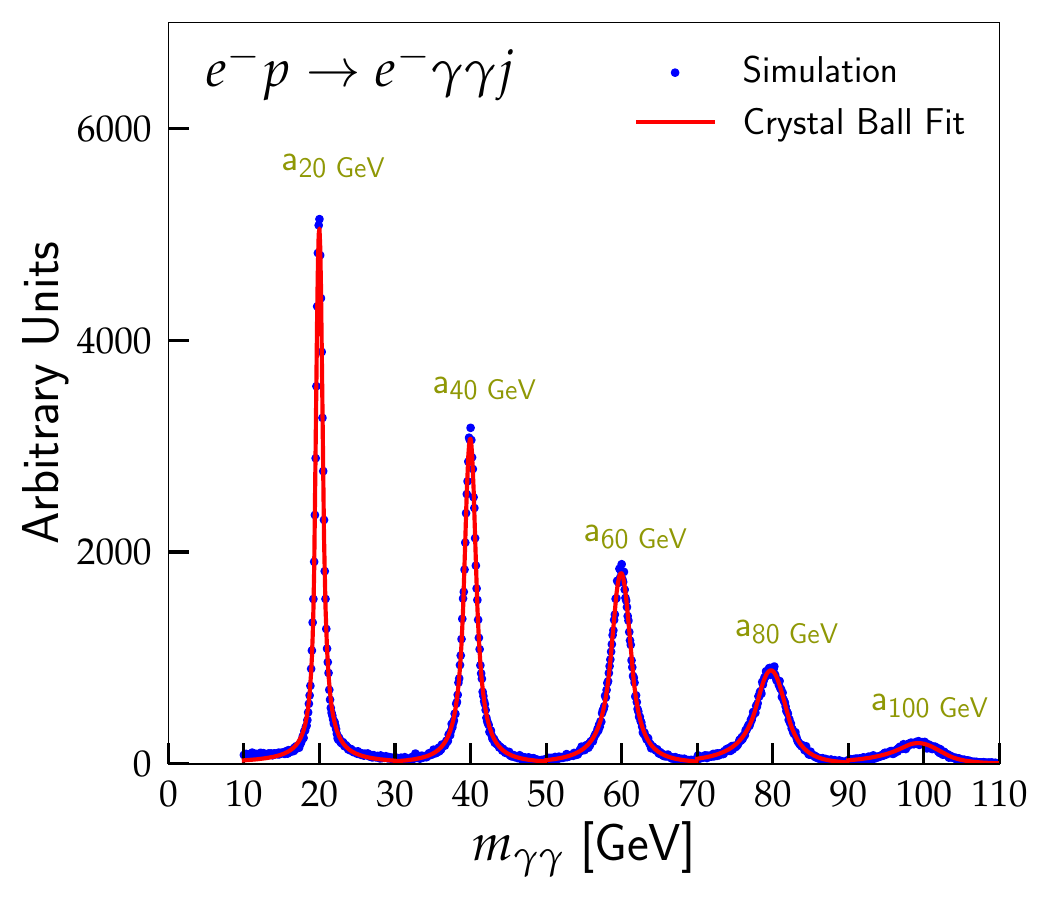}
\caption{The $\mathrm{m}_{\gamma\gamma}$ distribution for the ALP signal and Crystal Ball fits, in blue and red color, respectively. Fits are shown for signal events with $m_a=$ 20, 40, 60, 80, and 100 GeV. For example, $a_{\rm 20~GeV}$ refers to an ALP with mass $m_a=$ 20 GeV.}
\label{fig:fit_2ga}
\end{figure}

In Fig.~\ref{fig:fit_2ga}, we show the $\mathrm{m}_{\gamma\gamma}$ distribution for signal processes in blue and the corresponding Crystal Ball fits with a solid red line for ALP masses from 20 GeV to 100 GeV, in steps of 20 GeV. A mass window is chosen to search for each ALP mass that varies in the range $[\mu - \sigma, \mu + \sigma]$ and is tabulated in Table~\ref{tab:yield_2ph}. In addition, we provide the estimated signal efficiency and background yield at $\sqrt{s}$ = 141 GeV and ${\cal L}=$100 ${\rm fb}^{-1}$, along with the signal yield for each ALP mass corresponding to $g_{aee}=1.0$.

\begin{table}[tb!]
\center
\begin{tabular}{c|c|c|c|c|c|c}
\bottomrule
\multirow{3}{*}{ $m_a$ [GeV] }  & \multicolumn{3}{c|}{ Fitted parameters [GeV]}  &  \multirow{3}{*}{ $\epsilon$ } & \multicolumn{2}{c}{Yield at $\mathcal{L}=100$ fb$^{-1}$} \\\cline{2-4}\cline{6-7}

& \multirow{2}{*}{$\mu$ } & \multirow{2}{*}{$\sigma$} & \multirow{2}{*}{$\chi^2$/d.o.f} && \multirow{2}{*}{\makecell{S$\times 10^{-5}$ ($g_{\mathrm{aee}}=1.0$})} &  \multirow{2}{*}{B }\\ 
&& &&&\\ \hline
 $5.5$ & $5.61\pm 0.01$  & $0.75\pm 0.02$  & 2.10 & $0.12$ & 528 & 13369 \\
 $10$  & $10.01\pm 0.00$  & $0.47\pm 0.00$ & 2.16  & $0.19$ & 120 & 5583 \\
 $20$  & $19.97\pm 0.00$ & $0.47\pm 0.00$  & 2.48 & $0.20$ & 8 & 1598 \\
 $30$  & $29.97\pm 0.00$ & $0.59\pm 0.00$  & 2.23 & $0.20$ & 1 & 815 \\
 $40$  & $39.96\pm 0.00$ & $0.76\pm 0.00$  & 2.72 & $0.20$ & 0.2 & 437 \\
 $50$  & $49.94\pm 0.00$ & $0.97\pm 0.00$  & 1.89 & $0.19$ & 0.06 & 201 \\
 $60$  & $59.92\pm 0.00$ & $1.27\pm 0.01$  & 1.21 & $0.19$ & 0.02 & 87 \\
 $70$  & $69.85\pm 0.01$ & $1.56\pm 0.01$  & 1.30 & $0.18$ & 4$\times 10^{-3}$ & 34 \\
 $80$  & $79.75\pm 0.01$ & $1.96\pm 0.02$  & 1.10 & $0.15$ & 1$\times 10^{-3}$ & 11 \\
 $90$  & $89.53\pm 0.02$ & $2.44\pm 0.03$  & 0.95 & $0.10$ & 2$\times 10^{-4}$ & 3 \\
 $100$ & $99.23\pm 0.04$ & $2.94\pm 0.07$  & 1.21 & $0.05$ & 4$\times 10^{-5}$ & $0.8$\\
 \toprule
\end{tabular}
\caption{The fitted $\mu$ and $\sigma$ after fitting the $\mathrm{m}_{\gamma\gamma}$ distribution with the Crystal Ball function. To assess the goodness of fit, the $\chi^2$/d.o.f values are also provided. The notation $\pm 0.00$ indicates that the observed variation occurs only at the third decimal place or beyond. The signal efficiency, $\epsilon$, is multiplied by the identification efficiency of $70\%$ for each photon. The signal yields, S for $g_{\mathrm{aee}}=1.0$ and background yields, B, are also shown in the last two columns for $\mathcal{L}=100~{\rm fb}^{-1}$.}
\label{tab:yield_2ph}
\end{table}

\begin{figure}[tb!]
\centering
\includegraphics[width=.45\textwidth]{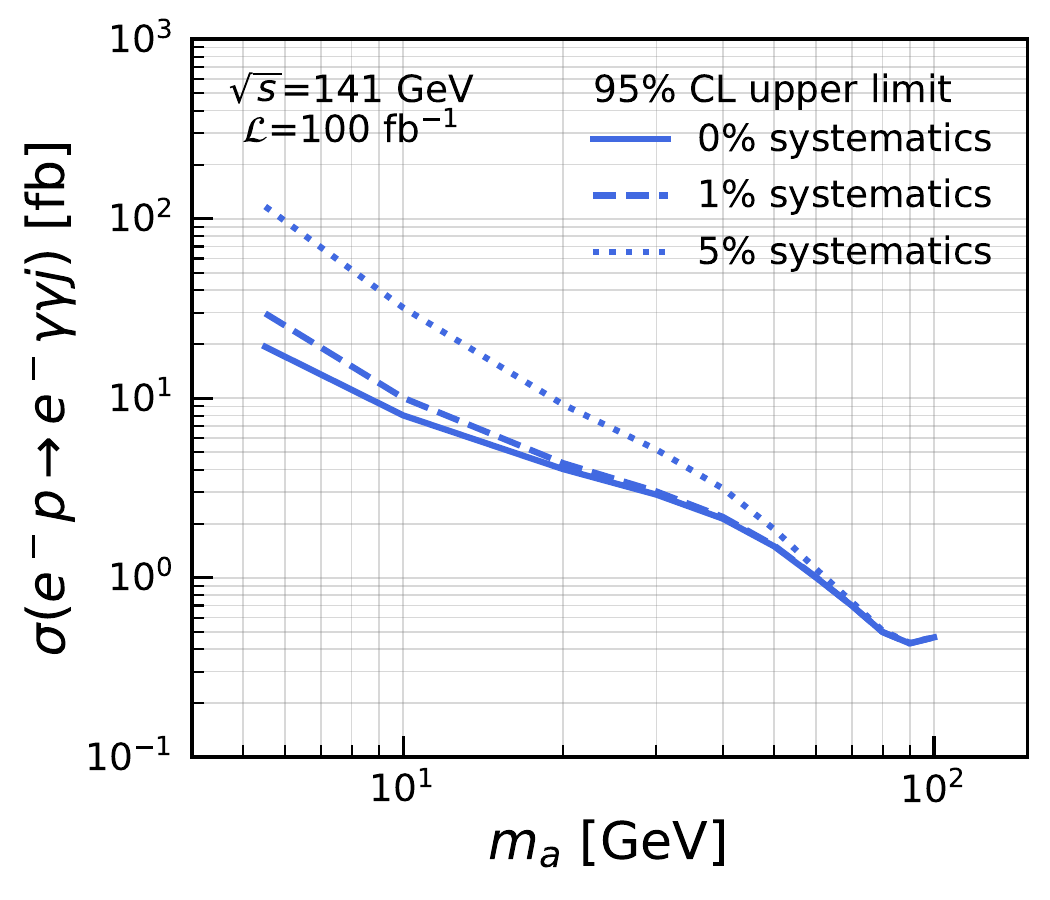}
\caption{Upper limit at $95\%$ CL on the signal production cross-section, $\sigma(e^- p\to e^-\gamma\gamma j)$ as a function of ALP mass, $m_a$. The solid blue line corresponds to adding null systematic uncertainty, while dashed and dotted lines include $1\%$ and $5\%$ systematic uncertainties, respectively.}
\label{fig:ul_gaee_e2ph}
\end{figure}

In Fig.~\ref{fig:ul_gaee_e2ph}, we present the 95$\%$ CL exclusion limits on the production cross-section of the $e^- p\to e^- \gamma\gamma j$ process. The upper limits are between 19.33 fb and 0.47 fb for an ALP mass of 5.5 GeV and 100 GeV, respectively. Adding systematic uncertainties $1\%$ ($5\%$), the limits weaken in the mass range of 5.5 GeV and 70 GeV, with $\sigma(e^- p\to e^- \gamma\gamma j)<$ 29.66 (116.86) fb and 0.70 (0.73) fb, respectively. We translate the cross-section exclusions into the $g_{aee}$ coupling plane as a function of the ALP mass.
These upper limits are extremely weaker than the $e^- p\to e^- e^+ e^- j$ channel, and we do not show them in this paper.

\subsection{The $e\gamma$ channel}

\begin{figure}[tb!]
\centering 
\subfloat[]{\label{fig:10a}
\includegraphics[height=.22\textwidth]{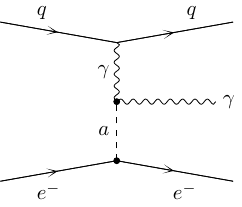}
}\qquad
\subfloat[]{\label{fig:10b}
\includegraphics[height=.22\textwidth]{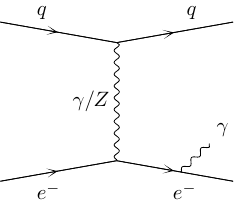}
}\qquad
\subfloat[]{\label{fig:10c}
\includegraphics[height=.22\textwidth]{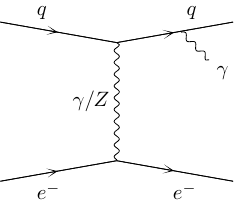}
}
\caption{(a) The Feynman diagram of the hard process for signal production is shown, along with (b) and (c) initial/final state photon radiation for the background production, in the $e^-p\to e\gamma j$ channel.}
\label{fig:FD_e1ph}
\end{figure}

This section focuses on the single photon production, \textit{viz.} $e^- p\to e^-\gamma j$ channel. In Fig.~\ref{fig:FD_e1ph} (a), the Feynman diagram of the signal process is shown. We generate this process in \texttt{Madgraph} using the \texttt{UFO} model file discussed in Section~\ref{sec:eee}, including the $g_{a\gamma\gamma}$ coupling. The background processes in this channel include single photon emission from the electron or quark leg, as shown in Fig.~\ref{fig:FD_e1ph} (b) and (c). We simulate the signal and background events with the following requirements. The transverse momentum and pseudorapidity of the final state particles must be $p_{T,e/\gamma/j}>2$ GeV, and $|\eta_{e/\gamma}|<3.5$, $|\eta_{j}|<5.0$, respectively. The angular separation in the $\eta-\phi$ plane between any two final state particles must satisfy $\Delta R_{e,\gamma,j}>0.4$. These requirements are tabulated in Table~\ref{tab:cuts_e1ph}.

\begin{figure}[tb!]
\centering
\includegraphics[width=.45\textwidth]{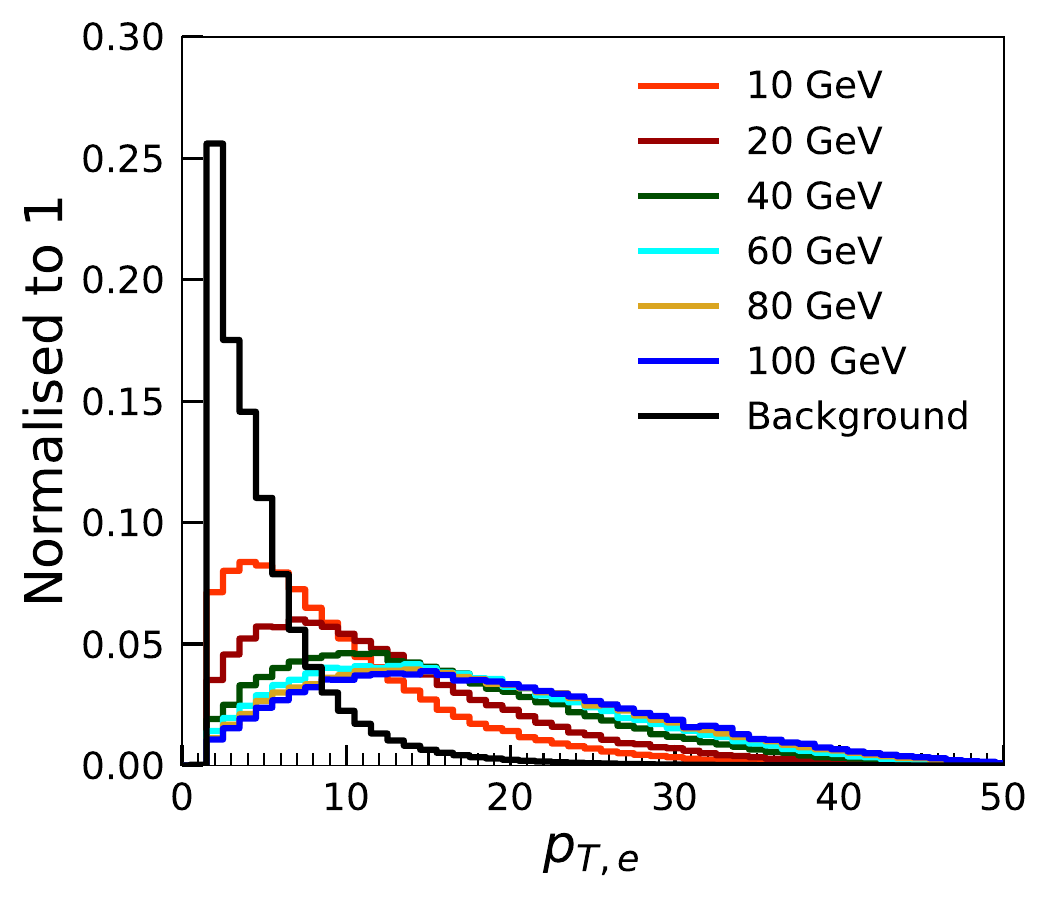}~\includegraphics[width=.45\textwidth]{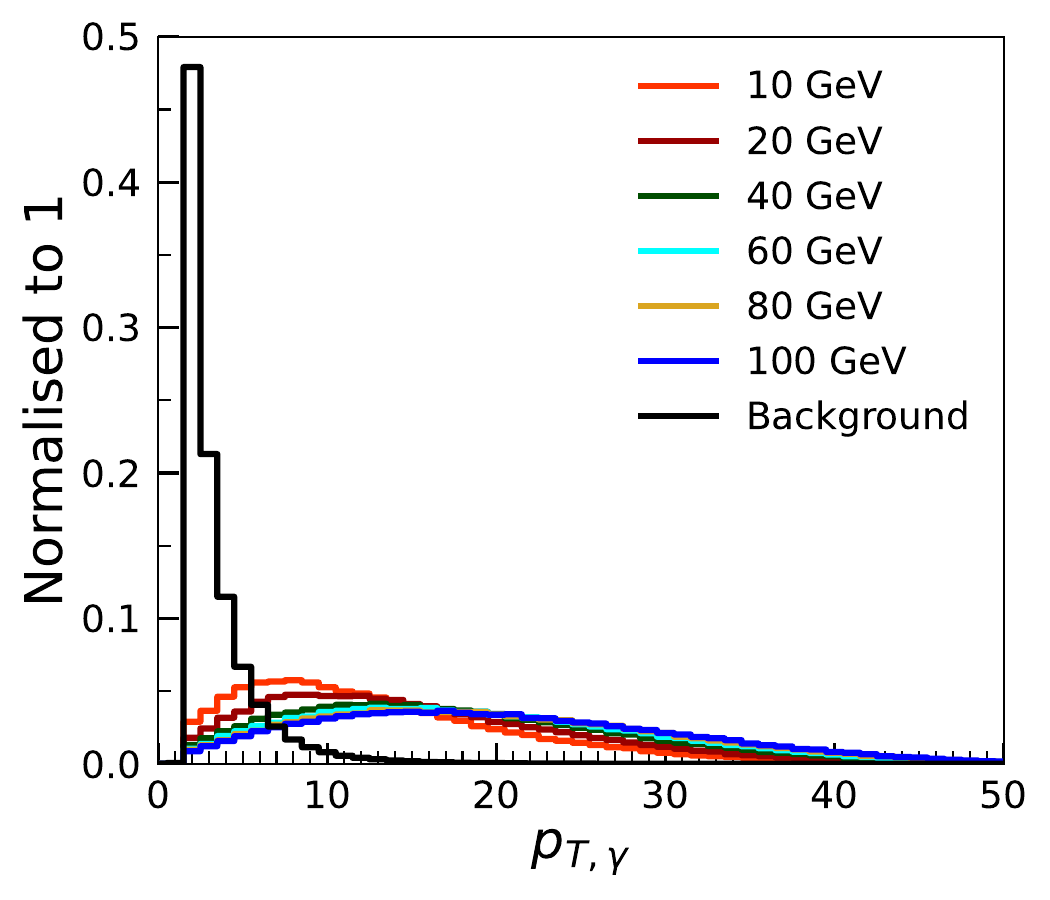}\\
\includegraphics[width=.45\textwidth]{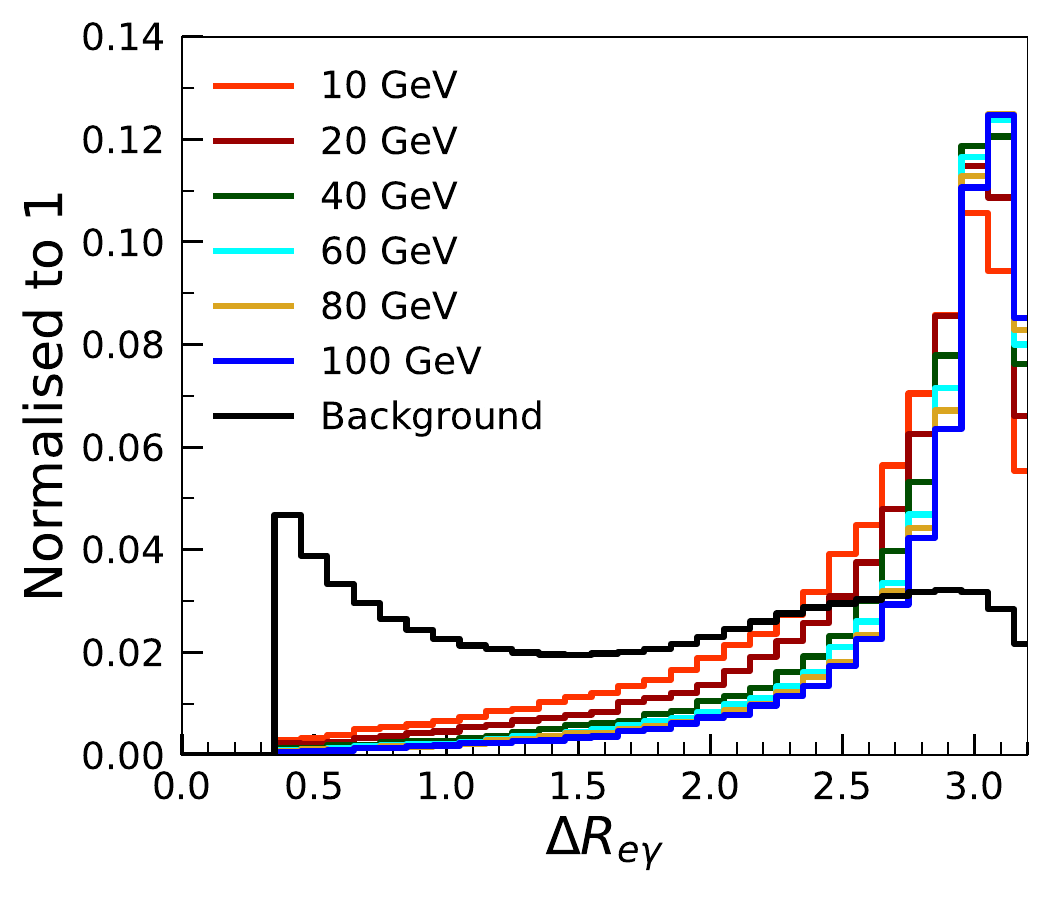}
\caption{Normalised distributions of transverse momentum of the electron, $p_{T,e}$ and photon $p_{T,e},~p_{T,\gamma}$, and $\Delta R$ separation between electron and photon in the final state $\Delta R_{e\gamma}$. The background is shown with a black line, while colored lines correspond to ALP benchmark masses.}
\label{fig:var_1ph}
\end{figure}

The events are selected if they contain exactly one electron and one photon in the final state. Next, we perform a cut-based analysis to improve the signal sensitivity and reduce background events. For this purpose, we choose three kinematic observables constructed from the final state electron and photon: the transverse momentum of the electron $p_{T,e}$ and photon $p_{T,\gamma}$, $\Delta R$ separation between the electron and photon, $\Delta R_{e\gamma}$. The schematic kinematic distributions of these observables are shown in Fig.~\ref{fig:var_1ph}. The colored distributions correspond to ALP signals with mass $m_a=10,20,40,60,80$ and 100 GeV,
while the background distribution is shown in black. We observe that the transverse momentum of the electron and photon is softer in the case of the background process and falls in the higher $p_{T}$ region for the ALP signal. In the case of the $\Delta R$ distribution, the electron and photon are almost back-to-back in the $\eta-\phi$ plane for the ALP signal. Note that the source of the final state photon in background events is from initial state or final state radiation. Thus, we observe a peak at low $\Delta R_{e\gamma}$ when the photon is radiated off the electron, while excess events at higher $\Delta R_{e\gamma}$ correspond to cases where the photon is emitted from the quark. The strategy here is to optimise the cuts on these kinematic observables and increase the signal significance (see Eq.~\ref{S}). We perform the optimisation analysis for an ALP mass of 5.5 GeV, then from 10 GeV to 100 GeV in steps of 10 GeV. The optimised cuts for each ALP mass are summarised in Table~\ref{tab:cuts_e1ph}. Further, we tabulate the signal efficiency, signal yield for $g_{aee}=1.0$, and background yield at ${\cal L}=100~{\rm fb}^{-1}$ in Table~\ref{tab:yield_ga}. We include a photon identification efficiency of $70\%$ in the signal efficiency.

\begin{table}[tb!]
\centering
\begin{tabular}{c|c}
\toprule
\multicolumn{2}{c}{Cuts applied} \tabularnewline \bottomrule
Generation cuts & Optimised cuts\\ 
& for $m_{a} = [5.5,10,20,30,40,50,60,70,80,90,100]$ GeV \\ \hline
$p_{T,e/\gamma/j}>2$ GeV,  &
$p_{T,e}\geq [2,3,6,10,15,17,19,22,22,22,26]$ GeV \\ 

$|\eta_{e/\gamma}|<3.5$, $|\eta_{j}|<5.0$ &
$p_{T,\gamma}\geq [15,17,20,21,24,25,27,28,28,29,29]$ GeV\\

$\Delta R_{e,\gamma,j}>0.4$ & $\Delta R_{e\gamma}\geq [1.0,1.2,1.4,2.0,1.8,2.3,2.5,2.6,2.6,2.6,2.7]$   \\\toprule
\end{tabular}
\caption{The generation level cuts for the signal and background generation, along with optimised cuts imposed on the cut-based analysis.}
\label{tab:cuts_e1ph}
\end{table}

\begin{table}[tb!]
\centering
\begin{tabular}{c|c|c|c} 
\toprule
\multicolumn{4}{c}{After the cut-based analysis} \tabularnewline \bottomrule

\multirow{3}{*}{ $m_a$ [GeV] } & \multirow{3}{*}{ $\epsilon$ } & \multicolumn{2}{c}{Yield at $\mathcal{L}=100$ fb$^{-1}$} \\ \cline{3-4}

& & \multirow{2}{*}{\makecell{S$\times 10^{-4}$ ($g_{\mathrm{aee}}=1.0$})} &  \multirow{2}{*}{B }\\
&& &\\ \hline

5.5 & 0.20 & 4000 & 208192 \\
10 & 0.19  & 266 & 118843 \\
20 & 0.18  & 8 & 52754 \\
30 & 0.19  & 1 & 34940 \\
40 & 0.15  & 0.15 & 16685 \\
50 & 0.15  & 0.04 & 12494 \\
60 & 0.13  & 0.01 & 7810 \\
70 & 0.12  & 4$\times 10^{-3}$ & 5629 \\
80 & 0.13  & 2$\times 10^{-3}$ & 5629 \\
90 & 0.12  & 7$\times 10^{-4}$ & 4650 \\
100 & 0.12 & 3$\times 10^{-4}$ & 3825 \\
\toprule
\end{tabular}
\caption{The signal efficiency $\epsilon$, signal yield S, and background yield B at $\mathcal{L}=100~{\rm fb}^{-1}$ are shown after the cut-based analysis. Here, $\epsilon$ includes a flat $70\%$ single photon identification efficiency.}
\label{tab:yield_ga}
\end{table}

\begin{figure}[tb!]
\centering
\includegraphics[width=.45\textwidth]{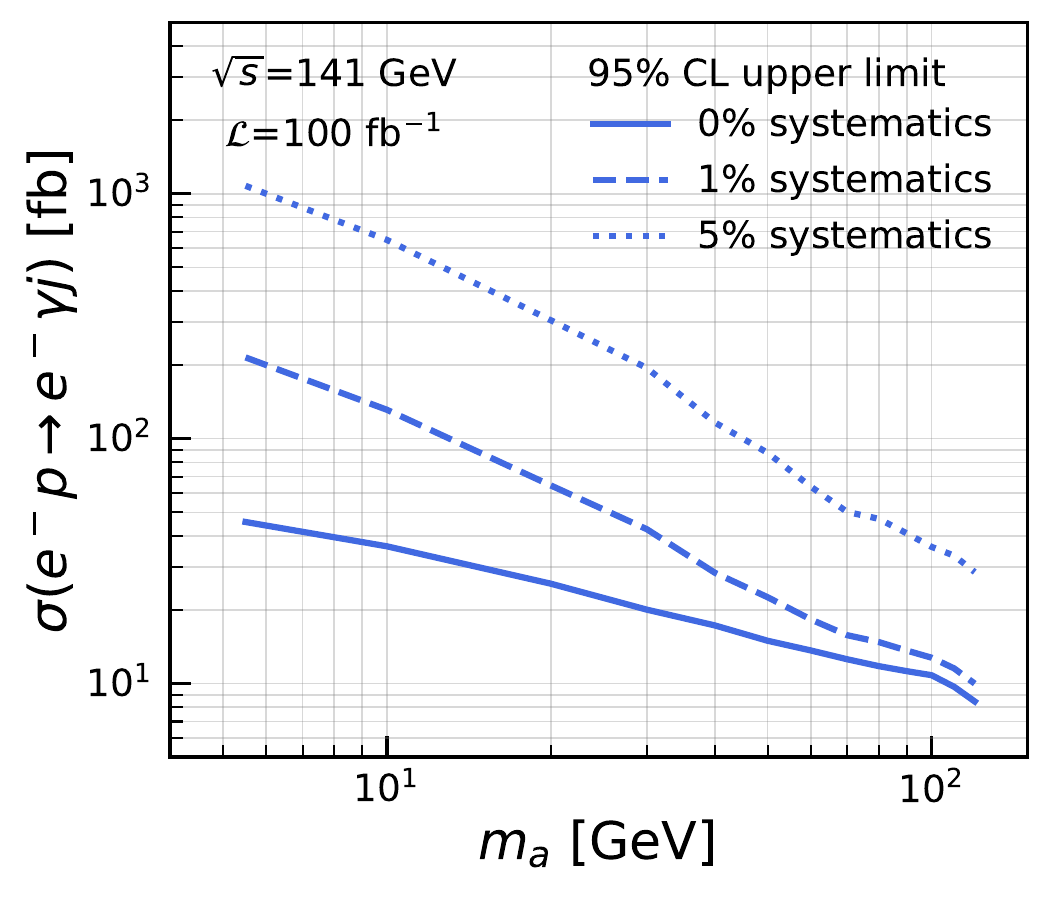}
\caption{Upper limit at $95\%$ CL on the signal production cross-section, $\sigma(e^- p\to e^-\gamma j)$ as a function of the ALP mass, $m_a$. The solid blue line corresponds to adding null systematic uncertainty, while dashed and dotted lines include $1\%$ and $5\%$ systematic uncertainties, respectively.}
\label{fig:ul_gaee_e1ph}
\end{figure}

After the cut-based analysis, we derive the upper limits at $95\%$ CL on the production cross-section of the $e^- p\to e^- \gamma j$ process. The limits are shown as a function of the ALP mass in Fig.~\ref{fig:ul_gaee_e1ph}. For $m_a$ between 5.5 GeV and 100 GeV, we obtain $\sigma(e^- p\to e^- \gamma j)\lesssim 45.66$ fb and $\lesssim 10.81$ fb, respectively, with null systematic uncertainty. Adding systematic uncertainty, the limits become weaker because the signal-to-background ratio, S/B, is poor in this channel, and they are shown with dashed (dotted) lines for $1\%$ ($5\%$) systematics in Fig.~\ref{fig:ul_gaee_e1ph}. Furthermore, we calculate the translated upper limits at $95\%$ CL on the $g_{aee}$ coupling as a function of mass.
Similar to the previous $e^- p\to e^- \gamma\gamma j$ channel, these exclusion limits are feeble than the $e^- p\to e^-  e^+ e^- j$ channel.

\section{Collider Analysis of leptophilic $Z'$}
\label{sec:analysis_zpr}

\begin{figure}[!tb]
\begin{center}
\subfloat[]{\label{fig:13a}
\includegraphics[width=.16\textwidth]{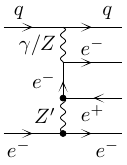}
}~~~~~~~~~~~~\qquad
\subfloat[]{\label{fig:13b}
\includegraphics[height=.18\textwidth]{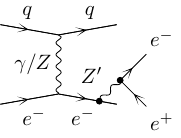}
}
\end{center}
\caption{ The Feynman diagrams for the signal process in the $e^-p \to e^+e^-e^+j$ channel are as follows: in diagram $(a)$, a $Z'$ boson appears as a propagator, whereas in diagram $(b)$, the $Z'$ is produced on-shell for the selected $Z'$ mass range.}
\label{fig:FD_eee_zpr}
\end{figure}

We now focus on the leptophilic $Z'$ scenario where $Z'$ has a coupling to electrons alone (see Section~\ref{sec:theory_zpr}). The final state of interest here is the three electron channel, $e^- p \to e^-e^+e^-j$. The signal process involves $Z'$ production, either off-shell or on-shell, and the corresponding Feynman diagrams are shown in Fig.~\ref{fig:FD_eee_zpr}. The $Z'$ mass is chosen to vary between 5.5 GeV and 100 GeV. The background to this channel comes from three electron production from SM processes, and we generate the background events inclusively by considering all the contributing Feynman diagrams at LO. We generate signal and background events in {\tt MadGraph5} with the following requirements: $p_{T,e}>2$ GeV, $|\eta_e|<3.5$, $\Delta R_{ee}>0.3$.

\begin{figure}[tbh!]
\centering
\includegraphics[scale=0.45]{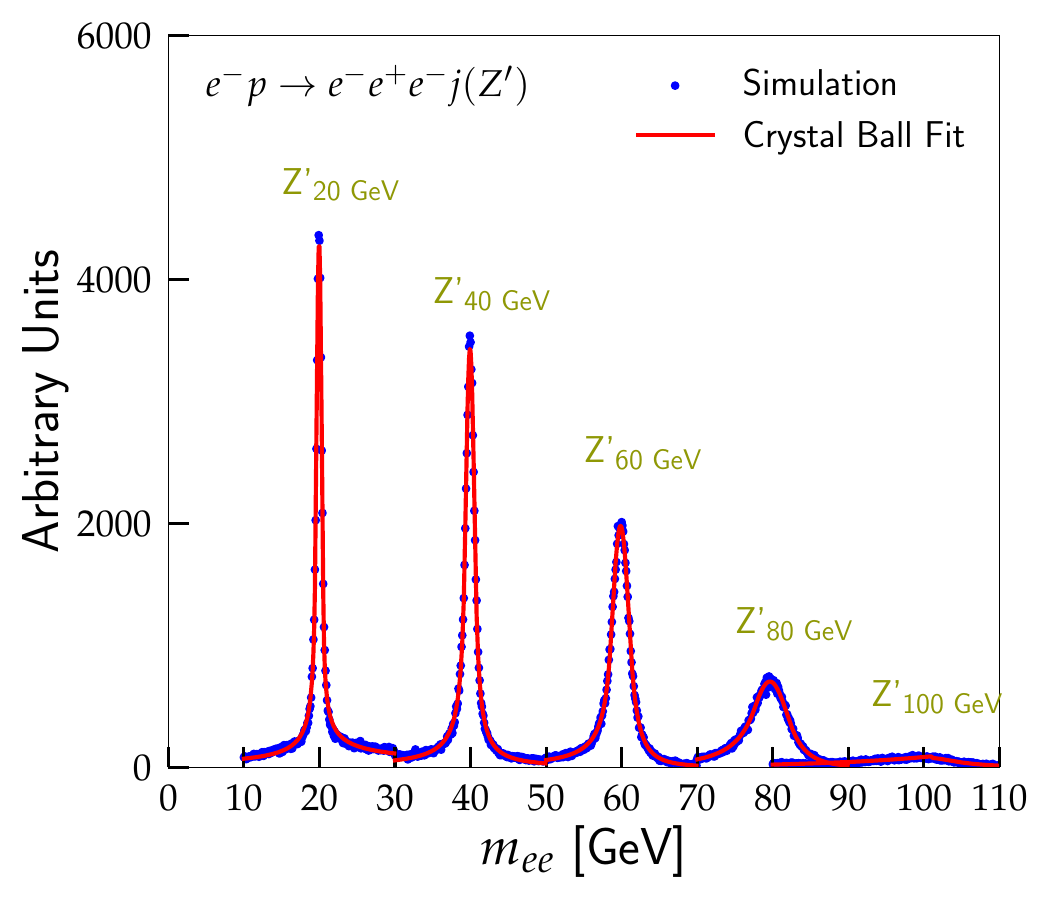}
\caption{The $\mathrm{m}_{ee}$ distribution for the $Z'$ signal and Crystal Ball fits, in blue and red color, respectively. Fits are shown for signal events with $m_{\rm Z'}=$ 20, 40, 60, 80, and 100 GeV. For example, $Z'_{\rm 20~GeV}$ refers to $Z'$ with mass $m_{Z'}=$ 20 GeV.}
\label{fig:fit_zpr}
\end{figure}

\begin{table}[tb!]
\center
\begin{tabular}{c|c|c|c|c|c|c}
\bottomrule
\multirow{3}{*}{ $m_{\rm Z'}$ [GeV] }  & \multicolumn{3}{c|}{ Fitted parameters [GeV]}  &  \multirow{3}{*}{ $\epsilon$ } & \multicolumn{2}{c}{Yield at $\mathcal{L}=100$ fb$^{-1}$} \\\cline{2-4}\cline{6-7}

& \multirow{2}{*}{$\mu$ } & \multirow{2}{*}{$\sigma$} & \multirow{2}{*}{$\chi^2$/d.o.f} && \multirow{2}{*}{\makecell{S ($g_{\mathrm{Z'}}=0.1$})} &  \multirow{2}{*}{B }\\ 
&& &&&\\ \hline

 $5.5$  & $5.66\pm 0.04$ & $1.05\pm 0.04$  & 4.58 & $0.22$ & 162800  & 3573 \\
 $10$  & $9.98\pm 0.00$ & $0.42\pm 0.00$  & 2.71 & $0.20$ & 42000  & $2030$ \\
 $20$  & $19.95\pm 0.00$ & $0.37\pm 0.00$  & 3.85 & $0.28$ & 9240 & $759$ \\
 $30$  & $29.94\pm 0.00$ & $0.47\pm 0.00$  & 3.39 & $0.33$ & 2970 & $404$ \\
 $40$  & $39.93\pm 0.00$ & $0.60\pm 0.00$  & 2.19 & $0.36$ & 1080 & $228$ \\
 $50$  & $49.90\pm 0.00$ & $0.80\pm 0.00$  & 2.56 & $0.38$ & 342 & $136$ \\
 $60$  & $59.85\pm 0.00$ & $1.14\pm 0.01$  & 1.57 & $0.39$ & 117 & $78$ \\
 $70$  & $69.80\pm 0.01$ & $1.51\pm 0.01$  & 1.68 & $0.34$ & 34 & $33$ \\
 $80$  & $79.62\pm 0.02$ & $2.35\pm 0.03$  & 1.03 & $0.29$ & 9 & $15$ \\
 $90$  & $89.32\pm 0.04$ & $3.46\pm 0.07$  & 1.09 & $0.18$ & 2 & $7$ \\
 $100$ & $99.89\pm 0.17$ & $4.20\pm 0.35$  & 1.11 & $0.06$ & 0.1 & $2$  \\
 \toprule
\end{tabular}
\caption{The mean values, $\mu$ and widths, $\sigma$ obtained after fitting the $Z'$ reconstructed $\mathrm{m}_{ee}$ mass variable with the Crystal Ball function, along with the signal efficiency, $\epsilon$. To assess the goodness of fit, the $\chi^2$/d.o.f values are also provided. The notation $\pm 0.00$ indicates that the observed variation occurs only at the third decimal place or beyond. The signal yields, S, for $g_{\mathrm{Z'}}=0.1$ and the background yields, B, are also shown at an integrated luminosity of $100$ fb$^{-1}$.}
\label{tab:yield_zpr}
\end{table}

We select events that contain exactly three electrons in the final state with $p_{T}>2$ GeV, $|\eta|<3.5$. The 4-momentum of the electrons is smeared with the $\eta$ dependent functions, as summarized in Table~\ref{tab:det_eff}. We adopt a similar strategy for the cut-based analysis, as previously discussed for the ALP signal, and reconstruct the invariant mass of the two hardest $p_T$ electrons, denoted as $m_{ee}$. The $m_{ee}$ distribution shows a peak for the signal events at a chosen $Z'$ mass. However, the variance of the peak increases for a higher $Z'$ mass. Next, we fit the signal distribution $m_{ee}$ with a double-sided Crystal Ball function (see Eq.~\ref{eq:CBF}). In Fig.~\ref{fig:fit_zpr}, the simulated signal samples for $m_{\rm Z'}=$ 20, 40, 60, 80, 100 GeV, and the fitted Crystal Ball function are shown with blue dots and a solid red line, respectively. The mean $\mu$ and variance $\sigma$ of the fitted distribution are summarized in Table~\ref{tab:yield_zpr} for each $Z'$ mass. The $\mu\pm 1\sigma$ region is chosen to calculate the final signal and background events after the analysis. The corresponding signal efficiency, signal yield for a chosen $g_{\rm Z'}=0.1$ and background yields, at the integrated luminosity of $\mathcal{L}=100~{\rm fb}^{-1}$ are tabulated in Table~\ref{tab:yield_zpr}.

\begin{figure}[tb!]
\centering
\subfloat[]{\label{fig:15a}
\includegraphics[width=.45\textwidth]{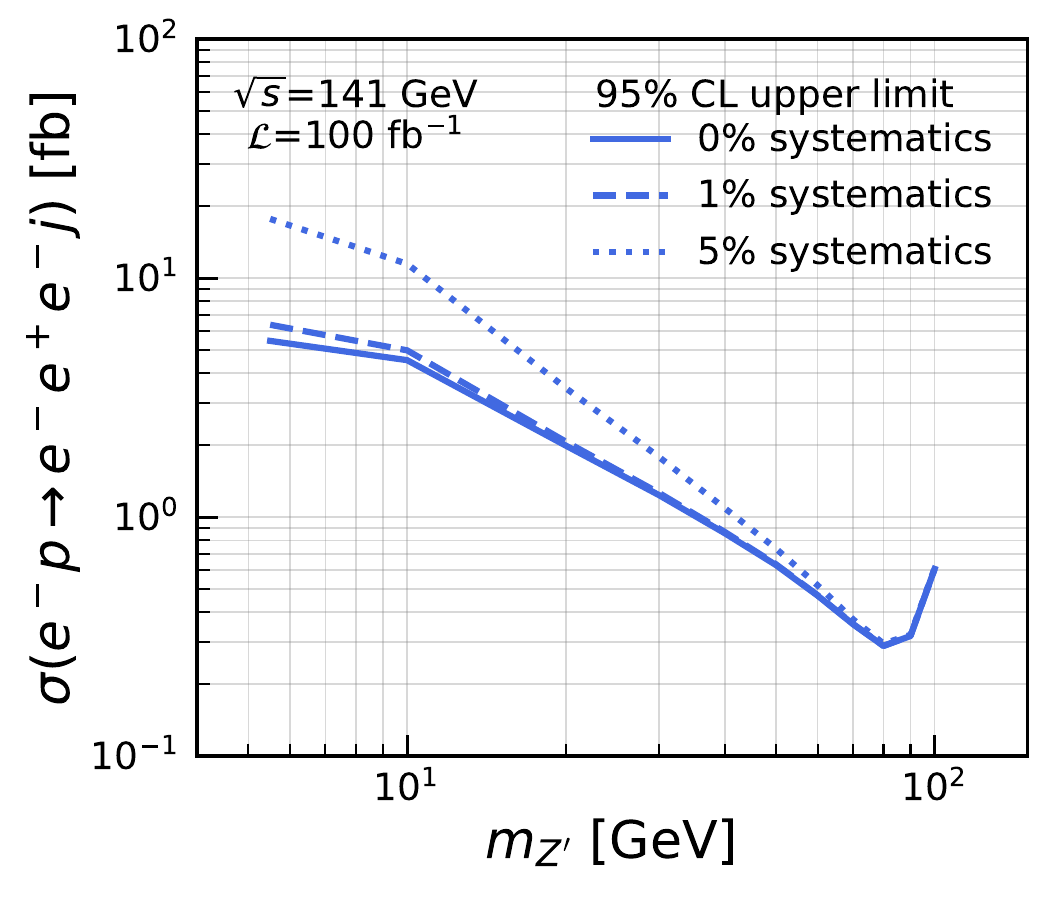}
}\qquad
\subfloat[]{\label{fig:15b}
\includegraphics[width=.45\textwidth]{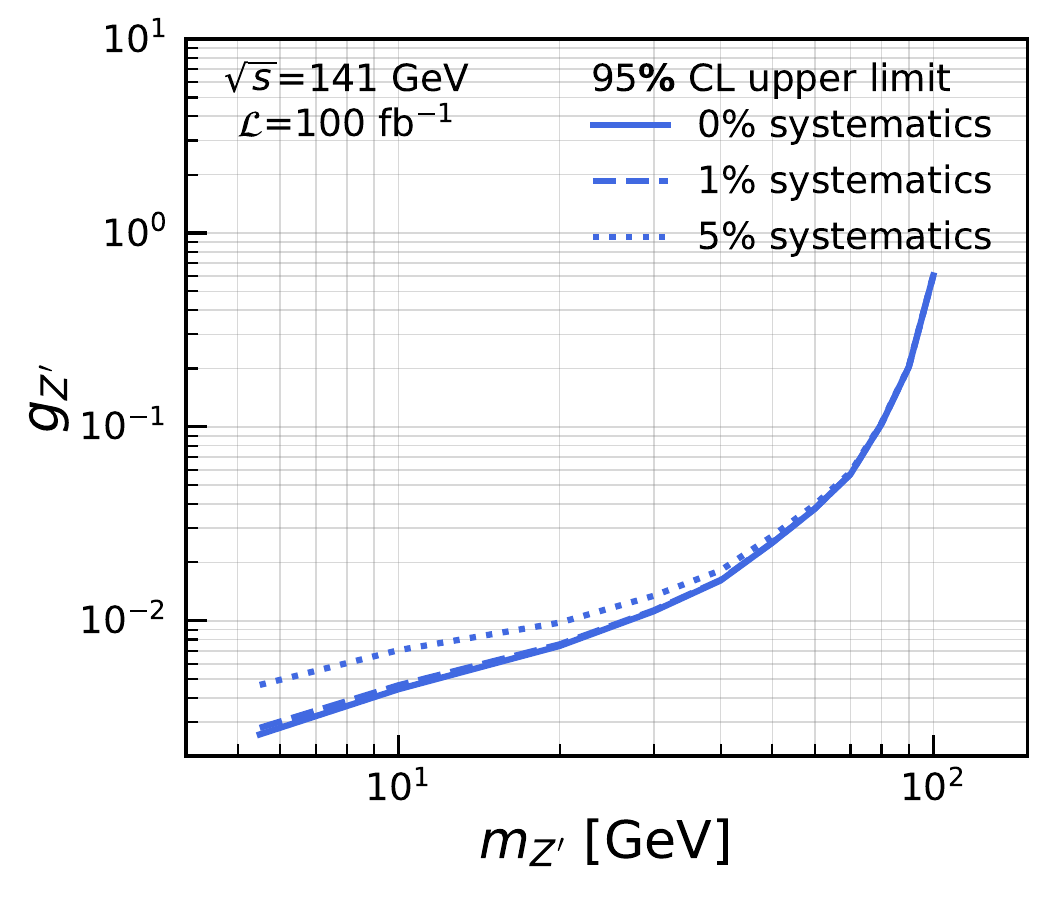}
}
\caption{Upper limit at $95\%$ CL on (a) the signal production cross-section, $\sigma(e^- p\to e^-e^+e^-j)$ and (b) the $Z'$-electron coupling, $g_{\rm Z'}$ as a function of $Z'$ mass, $m_{\rm Z'}$. The solid blue line corresponds to adding null systematic uncertainty, while dashed and dotted lines include $1\%$ and $5\%$ systematic uncertainties, respectively.}
\label{fig:ul_gz}
\end{figure}

Next, we evaluate the exclusion limits at 95$\%$ CL in the signal production cross-section and show it in Fig.~\ref{fig:ul_gz} (a), as a function of $m_{\rm Z'}$. The upper limit at 95$\%$ CL in $\sigma(e^- p\to e^- e^+ e^- j)$ is between 5.46 fb and 0.61 fb for $m_{\rm Z'}=5.5$ and 100 GeV, respectively. Upon adding systematic uncertainty $1\%$ ($5\%$), as shown by dashed (dotted) lines, the limits become weaker for the $Z'$ mass between 5.5 GeV and 70 GeV with 6.38 fb (17.70 fb) and 0.36 fb (0.37 fb), respectively. The effects of adding systematics are mild for heavier $Z'$ as the background yield, B, becomes comparatively smaller and the contribution from systematics becomes negligible in the signal significance formula. In Fig.~\ref{fig:ul_gz} (b), we illustrate the translated upper limits at 95$\%$ CL on the $Z'$-electron coupling, $g_{\rm Z'}$, with respect to $m_{\rm Z'}$. The upper limits are in the range $g_{\rm Z'}\lesssim$ (0.0026 - 0.60) for $m_{\rm Z'}\sim (5.5-100)$ GeV without systematics. The limits weaken between $m_{\rm Z'}\sim (5.5-70)$ GeV with $g_{\rm Z'}$ between $\lesssim$ 0.0028 (0.0047) and 0.057 (0.058) with $1\%$ ($5\%$) systematics.
So far we discussed about a $Z'$ that couples to $e$ only. However, in a realistic anomally free models, $Z'$ can couple to other fermions as well, apart from $e$. For completeness, we also discuss two of such leptophilic models, $L_e-L_\tau$ and $L_e-L_\mu $ in Appendix \ref{apx:A}.

\section{Interpretation of the results}
\label{sec:result}

In this section, we compare our derived exclusion limits with the existing constraints in the ALP parameter space defined by the ($m_a$, $g_{aee}$) plane and $Z'$ parameter space defined by the ($m_{Z'}$, $g_{\rm Z'}$) plane.

\begin{figure}[tb!]
    \centering 
    \includegraphics[scale=0.5]{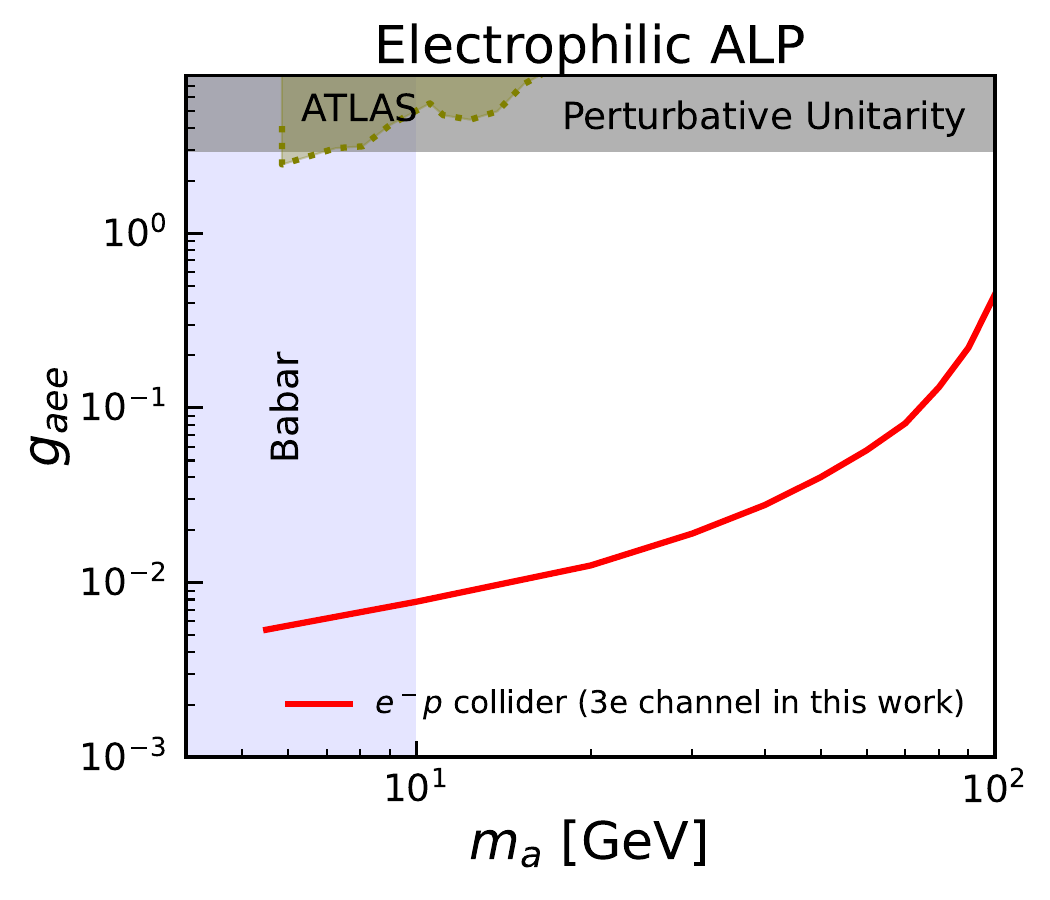}
    \caption{Projected constraint from $e-p$ collider at $95\%$ CL on the ALP-electron coupling, $g_{aee}$ as a function of ALP mass, $m_{\rm a}$ using the $e^- p\to e^-e^+e^-j$ channel shown by the region above the red solid line. Other existing constraints like perturbative limit \cite{Cornella:2019uxs}, BaBar \cite{BaBar:2014zli} are also shown. The constraint coming from ATLAS \cite{Biekotter:2025fll} through the loop induced $g_{a\gamma\gamma}$ coupling is also shown.}
    \label{fig:com_ge}
\end{figure}
In Fig.~\ref{fig:com_ge}, we show the constraints in the ALP parameter space. The solid red line corresponds to our derived $95\%$ CL upper limit on $g_{aee}$ from the three electron channel, $e^- p\to e^-e^+e^-j$, without any systematics. The translated upper limit in $g_{aee}$ is stringent in the low mass region where the $e^-e^+e^-j$ production cross-section is higher; conversely, the bounds become weaker at higher masses as the production rate diminishes, see Fig.~\ref{fig:ul_gaee_3e} (a). The blue shaded region is excluded by the \texttt{BaBar} experiment~\cite{BaBar:2014zli}. From LHC experiments, the ATLAS collaboration places a strict bound on the $g_{a\gamma\gamma}$ interaction in the considered ALP mass region~\cite{Biekotter:2025fll}. We translate this constraint into the $m_a-g_{aee}$ plane using Eq.~\ref{eq:gaggl} and show it with a dashed olive-shaded region in Fig.~\ref{fig:com_ge}.
Alongside experimental bounds, there are bounds from the perturbative unitarity of the ALP model. This is around $\sim \sqrt{8\pi/3}$~\cite{Cornella:2019uxs} and is shown with a shaded gray region.
Note that our derived limits are significantly stronger than the existing constraints for $m_a\gtrsim 10$ GeV. Previous work exploring ALP with electron coupling in the GeV region has considered coupling to tau as well \cite{Davoudiasl:2021mjy}, making purely electrophilic ALP elusive. Thus, our work presents a unique way to probe this unexplored region of purely electrophilic ALP.

\begin{figure}[tb!]
    \centering 
    \includegraphics[scale=0.5]{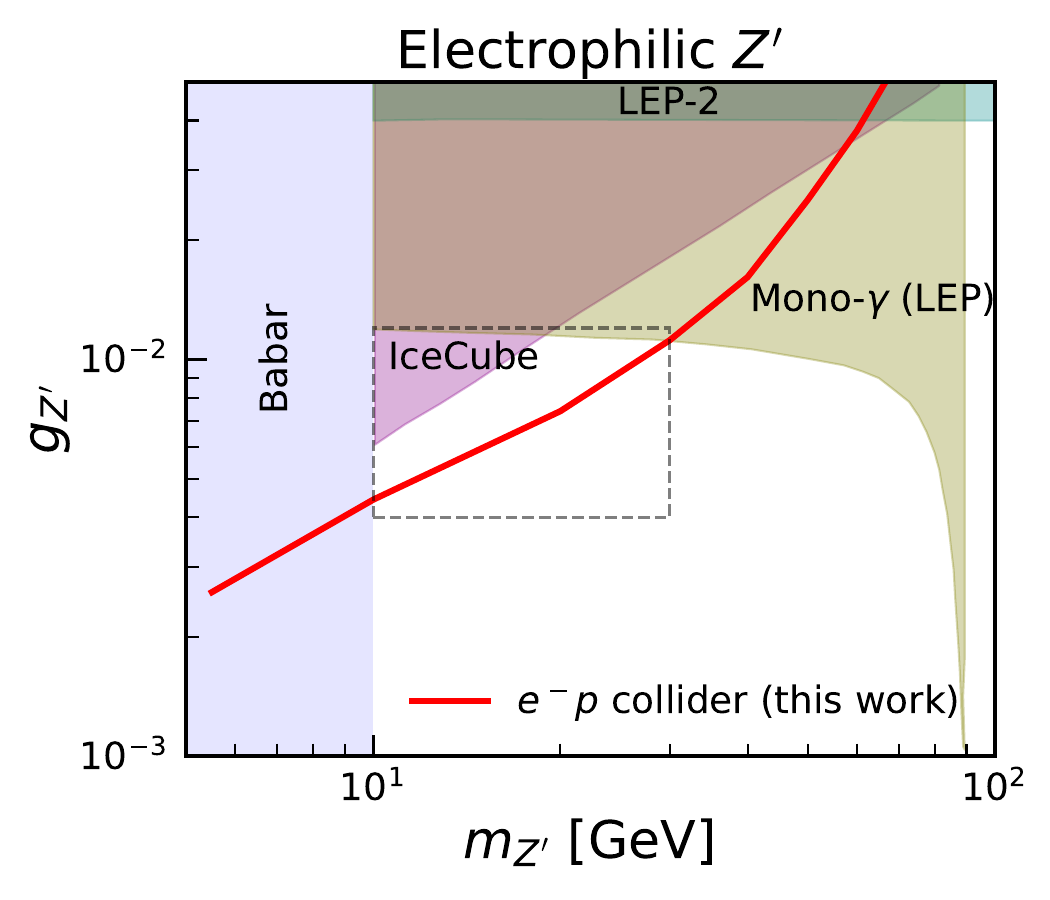}
    \caption{Projected constraint from $e-p$ collider at $95\%$ CL on the $Z'$-electron coupling, $g_{\rm Z'}$ as a function of $Z'$ mass, $m_{\rm Z'}$ using the $e^- p\to e^-e^+e^-j$ channel shown by the region above the red solid line. Other existing constraints like BaBar \cite{BaBar:2014zli}, LEP \cite{DELPHI:2003dlq,DELPHI:2008uka} are also displayed. Note that the IceCube constraint depends on the $\nu_e$ NSI \cite{IceCube:2022pbe}.}
    \label{fig:com_gz}
\end{figure}
Finally, in Fig.~\ref{fig:com_gz}, we compare our exclusion limits for the electrophilic $Z'$ scenario with the existing constraints from the LEP data~\cite{PhysRevD.84.014028,DELPHI:2003dlq,DELPHI:2008uka}, $e^+e^-$ collision experiment BaBar~\cite{BaBar:2014zli} and neutrino experiment IceCube~\cite{IceCube:2022pbe}. We refer to Refs.~\cite{GonzalezSuarez:2024dsp,Bauer:2018onh} for a detailed discussion on these constraints. In the low mass region below $\sim 10$ GeV, the constraints are coming from dark photon search in the $e^+e^-\to \gamma A'\to \gamma (e^+e^-/\mu^+\mu^-)$ final state with the \texttt{BaBar} detector. In the intermediate mass range with $m_{\rm Z'}\sim(10-80)$ GeV, the constraints include results from \texttt{IceCube} experiment, where the behavior of non-standard neutrino interactions (NSI) in matter propagation is probed. In the heavier $Z'$ mass range, the most stringent constraint comes from monophoton searches by the \texttt{DELPHI} collaboration 
at the \texttt{LEP} experiment and $e^+e^-\to e^+e^-$ searches at \texttt{LEP2}. Our derived exclusion limit at 95$\%$ CL on $g_{\rm Z'}$ is shown with the solid red line, considering null systematic uncertainty. The limit becomes less constrained with a higher $Z'$ mass as the production rate for the $e^-p\to e^-e^+e^-j$  process drops, as illustrated in Fig.~\ref{fig:ul_gz} (a).
Our limits provide a stronger bound than the existing constraints for $Z'$ mass between 10 GeV and 30 GeV, as highlighted with a dashed rectangle in Fig.~\ref{fig:com_gz}.
As mentioned earlier, in realistic UV complete models $Z'$ can couple to other SM fermions apart from $e$. 
In such a scenario, the corresponding branching ratio of $Z'$ changes, which may change the final projection limit on the coupling.
For completeness, we also discuss two such example models, $L_e-L_\tau$ and $L_e-L_\mu $, in Appendix \ref{apx:A} and present the future projected limit from EIC on such models.
There also exist more exotic leptophilic $U(1)_X$ models like $U(1)_{B_3-3 L_e}$ where the $Z'$ couples to $e$ and the third generation of quark.
Such a model can also be tested in EIC following our methodology.
Thus, our analysis serves as a pathway for probing various electrophilic $Z'$ models in EIC.

Before concluding, we briefly comment on the impact of polarized incoming beams on signal sensitivity at the EIC. The EIC is expected to provide polarized electron and proton beams.
In the present analysis, we assume unpolarized beams. For the benchmark
scenarios considered in this work, we do not expect longitudinal electron
beam polarization to lead to a substantial improvement in the inclusive
cut-and-count sensitivity. The dominant signal and background topologies in
the tri-electron channel proceed through $t$-channel $\gamma/Z$ exchange,
with the photon-mediated contribution being largely vector-like.
Since the photon couples equally to left- and right-handed electrons, a
choice of electron beam polarization does not significantly suppress the
dominant QED background. Moreover, the electrophilic $Z'$ benchmark
considered here has a purely vector coupling, $\bar e\gamma^\mu e Z'_\mu$,
so that the leading signal rate is also approximately insensitive to the
initial electron helicity after summing over final-state spins. Similarly,
for the pseudoscalar ALP coupling considered here, the inclusive rate is
not expected to receive a large polarization-induced enhancement in the
present counting analysis.
Beam polarization could, however, be useful in a more general extension of
the present framework with chiral or axial-vector $Z'$ couplings, where
left- and right-handed electron beams would probe different coupling
combinations. Such a polarization-dependent study lies beyond the scope of
the present work.

\section{Conclusion}
\label{sec:conclusion}
In this work, we studied the potential reach of the proposed $e-p$ collider in exploring the GeV scale electrophilic BSM sector. Although it has less center-of-mass energy compared to the LHC, the $e-p$ collider benefits from less QCD background contamination, leading to a cleaner signature of the signal events. As an example, we have considered two BSM scenarios: electrophilic ALP and electrophilic $Z'$. We briefly discuss the theoretical EFT framework in Section~\ref{sec:theory} to realize such BSM scenarios. Although the sub-GeV region of such BSM particles has been explored extensively \cite{BaBar:2014zli}, the GeV to sub-TeV region of electrophilic ALP and $Z'$ has not been investigated yet in the literature, which is the main goal of this work.

We performed the collider analysis for two scenarios: kinematic features, background discrimination strategy, and estimation of the exclusion limits are presented in Section~\ref{sec:collider} for the ALP and in Section~\ref{sec:analysis_zpr} for the $Z'$ boson. In case of the electrophilic ALP scenario, we studied the following three possible final states $e^-e^+e^-j$, $e^-\gamma\gamma j$, and $e^-\gamma j$, where the last two channels can arise via the interaction of the ALP-photon with an electron loop. We adopt the projected detector resolutions for the EIC collider in our analysis. For the collider analysis in the $e^-e^+e^-j$ channel, the main strategy is to reconstruct the invariant mass of the two hardest-$p_{T}$ electrons, giving rise to a resonance peak around the chosen ALP mass in the $m_{ee}$ distribution. However, the variance of the mass distribution worsens with a higher ALP mass as non-resonant ALP production dominates. We calculate the final signal efficiency and background yield at ${\cal L}=100~{\rm fb}^{-1}$ by fitting the $m_{ee}$ distribution with a Crystal Ball function and choosing a $1\sigma$ mass region around the fitted mean value for each ALP benchmark mass. The upper limits at $95\%$ CL are set on the $e^-e^+e^-j$ production cross-section, $\sigma(e^-p\to e^-e^+e^-j)$ that varies between 6.03 fb and 0.29 fb for ALP masses between 5.5 GeV and 100 GeV, without systematics. The limits weaken to 7.45 (23.19) fb and 0.35 (0.38) fb for an ALP mass between 5.5 GeV and 70 GeV, with $1\%$ ($5\%$) systematics. After translating these cross-section limits to ALP-electron coupling, we obtain the upper bound at $95\%$ CL on $g_{aee}$ between (0.005-0.64) for the ALP mass between 5.5 and 100 GeV. The limits on $g_{aee}$ weaken for low ALP masses when systematic uncertainties are added. 
Since forward object reconstruction could be viable at the EIC, we explicitly check the signal sensitivity by vetoing the forward jet configurations. We observe stronger exclusion limits with $\sim 10\%$ improvement on the signal production cross-section and $\sim 6\%$ on the $g_{aee}$, without any systematics, indicating that signal sensitivity might benefit from tagging/vetoing options on the forward objects at the EIC.
We also study the $e^-e^+e^-j$ channel with additional loop-induced ALP-photon coupling. However, our analysis shows that the changes in the final exclusion limits are negligible.

A similar strategy is adopted to perform the collider analysis in the $e^-p\to e^-\gamma\gamma j$ channel. The invariant mass of the two final state photons is reconstructed to obtain resonance peaks in the $m_{\gamma\gamma}$ distribution. After that, we fit the $m_{\gamma\gamma}$ distribution with a Crystal Ball function and get the signal efficiency and background yield by selecting the region $(\mu-\sigma,\mu+\sigma)$ from the fitted signal distribution. The exclusion limits at $95\%$ CL on the production cross-section of the $e^-\gamma\gamma j$ process vary in the range (19.33-0.47) fb for the ALP mass range between 5.5 GeV and 100 GeV, but, when translating to upper limits on $g_{aee}$, we obtain poor sensitivity to the ALP signal compared to the $e^-e^+e^-j$ channel. Finally, in the case of the $e^-\gamma j$ channel, in the ALP scenario, we performed an optimized cut-based analysis. The requirements for the three kinematic observables, $p_{T,e}$, $p_{T,\gamma}$, and $\Delta R_{e\gamma}$, are optimized to increase the signal significance and obtain the final signal efficiency and background yield for each ALP mass. This yielded an upper limit at $95\%$ CL on the $e^-\gamma j$ production cross-section with $\sigma(e^-p\to e^-\gamma j)<(45.66-8.45)$ fb for ALP masses between 5.5 and 100 GeV. Similar to the di-photon final state, we get weaker sensitivity to $g_{aee}$ upon translating the cross-section exclusion limits. Thus, in the ALP scenario, the most stringent constraint on the ALP-electron coupling comes from the three electron channel, $e^-p\to e^-e^+e^-j$.

In case of the electrophilic $Z'$ scenario, we performed a collider analysis on the $e^-e^+e^-j$ final state. We followed a similar strategy to that for the three electron channel in the ALP scenario. The invariant mass $m_{ee}$ is constructed from the hardest-$p_T$ electrons in the final state and fitted with a Crystal Ball function. The signal significance and background yields are obtained by selecting the $\pm 1\sigma$ region around the fitted mean value, $\mu$. The upper limit at $95\%$ CL on $\sigma(e^-p\to e^-e^+e^- j)$ varies between 5.46 fb and 0.36 fb with null systematics, 6.38 (17.70) fb, and 0.36 (0.37) fb with systematics $1\%$ ($5\%$), for $Z'$ masses between 5.5 GeV and 70 GeV. In the same mass range, translating these limits into $Z'$-electron coupling, we obtain $g_{Z'}<(0.26 - 5.6).10^{-2}$ with null systematics, (0.28 - 5.7).$10^{-2}$, and (0.47 - 5.8).$10^{-2}$ with systematics $1\%$ and $5\%$, respectively.

Finally, we present the possible projection from the $e-p$ collider in Section~\ref{sec:result}. For both electrophilic ALP and $Z'$, we show the future projection at $95\%$ CL. For comparison, we also list the other possible existing constraints in this scenario.
For ALP, we find that our projected limit is the strongest for $10~ {\rm GeV}\lesssim m_a\lesssim 100$ GeV.
For an electrophilic $Z'$ we find that our projected limit is competitive with existing limits and even the strongest in the mass regime $10~ {\rm GeV}\lesssim m_Z'\lesssim 30$ GeV.
Thus, we present a unique study of how the future $e-p$ collider can help the community explore the purely electrophilic ALP and $Z'$ gauge boson. Our analysis can be generalized to a wide class of BSM scenarios. We note that the experimental data from such facilities may provide opportunities to investigate these scenarios in future studies.

\section*{Acknowledgement}
The work of AA is funded by the Indian Association for the Cultivation of Science, Kolkata.
SJ is supported
by the National Natural Science Foundation of China (12425506, 12375101, 12090060, and
12090064) and the SJTU Double First Class start-up fund (WF220442604). SJ also acknowledges the financial support provided by the Indian Association for the Cultivation of Science, Kolkata, where part of the work was done.

\appendix
\section{Electrophilic $Z'$ in UV complete models}
\label{apx:A}

\begin{figure}[htb!]
\centering
\includegraphics[width=.45\textwidth]{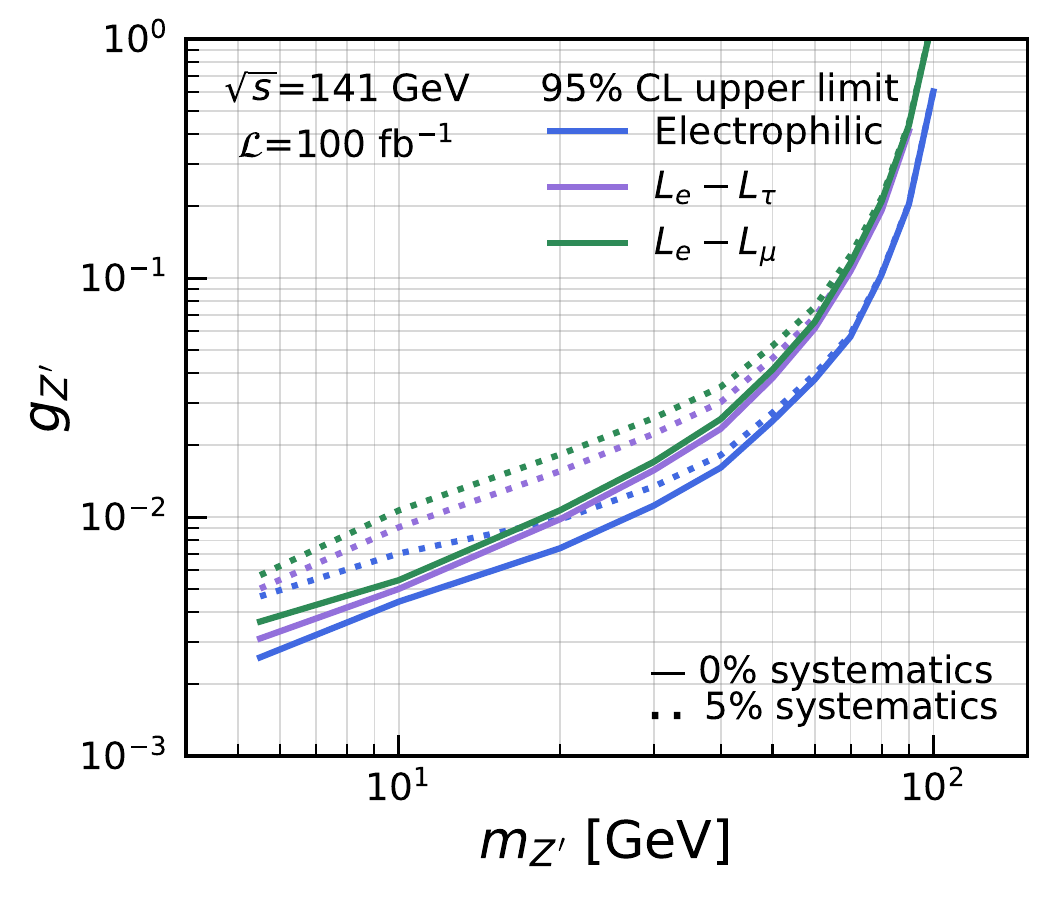}
\caption{The $95\%$ CL upper limits on $g_{z'}$ as a function of $Z'$ mass, $m_{Z'}$. Solid lines correspond to adding null systematic uncertainty, while dotted lines include $5\%$ systematic uncertainty.}
\label{fig:ul_zpr_model}
\end{figure}

Similar to the previous Section~\ref{sec:theory_zpr}  we consider a $Z'$ in $L_e-L_\tau$ and $L_e-L_\mu$ model with following interaction,
\begin{eqnarray}
   L_e-L_\tau:~~ &&\mathcal{L}_{\rm int}\supset - g_{\rm Z'} \left( \bar{e} \gamma^\mu e +\bar{\nu_e}\gamma^\mu \nu_e-\bar{\tau} \gamma^\mu \tau -\bar{\nu_\tau}\gamma^\mu \nu_\tau \right) Z'_\mu\\
     L_e-L_\mu:~~ &&\mathcal{L}_{\rm int}\supset - g_{\rm Z'} \left( \bar{e} \gamma^\mu e +\bar{\nu_e}\gamma^\mu \nu_e-\bar{\mu} \gamma^\mu \mu -\bar{\nu_\mu}\gamma^\mu \nu_\mu \right) Z'_\mu
\end{eqnarray}
Incorporating these interactions for respective models, we perform the same analysis as mentioned in Section~\ref{sec:analysis_zpr} and obtain the limit on the respective coupling. We present the limits for $L_e-L_\tau$ (magenta) and $L_e-L_\mu$ (green) in Fig.~\ref{fig:ul_zpr_model}.
The solid lines in the same plot correspond to adding null systematic uncertainty, while dotted lines include $5\%$ systematic uncertainty.
In the same figure, we also include previously obtained limits for purely electrophilic $Z'$.
Note that, in the presence of additional decay channels of $Z'$, the BR to electron decreases, leading to a slightly weaker upper limit on the cross-section for these models compared to the purely electrophilic case.
Indeed, the same effect has been reflected in the corresponding coupling limit (Fig.~\ref{fig:ul_zpr_model}) as the signal process scales as $\propto g_{Z'}^4$.

\begin{figure}[htb!]
\centering
\includegraphics[width=.45\textwidth]{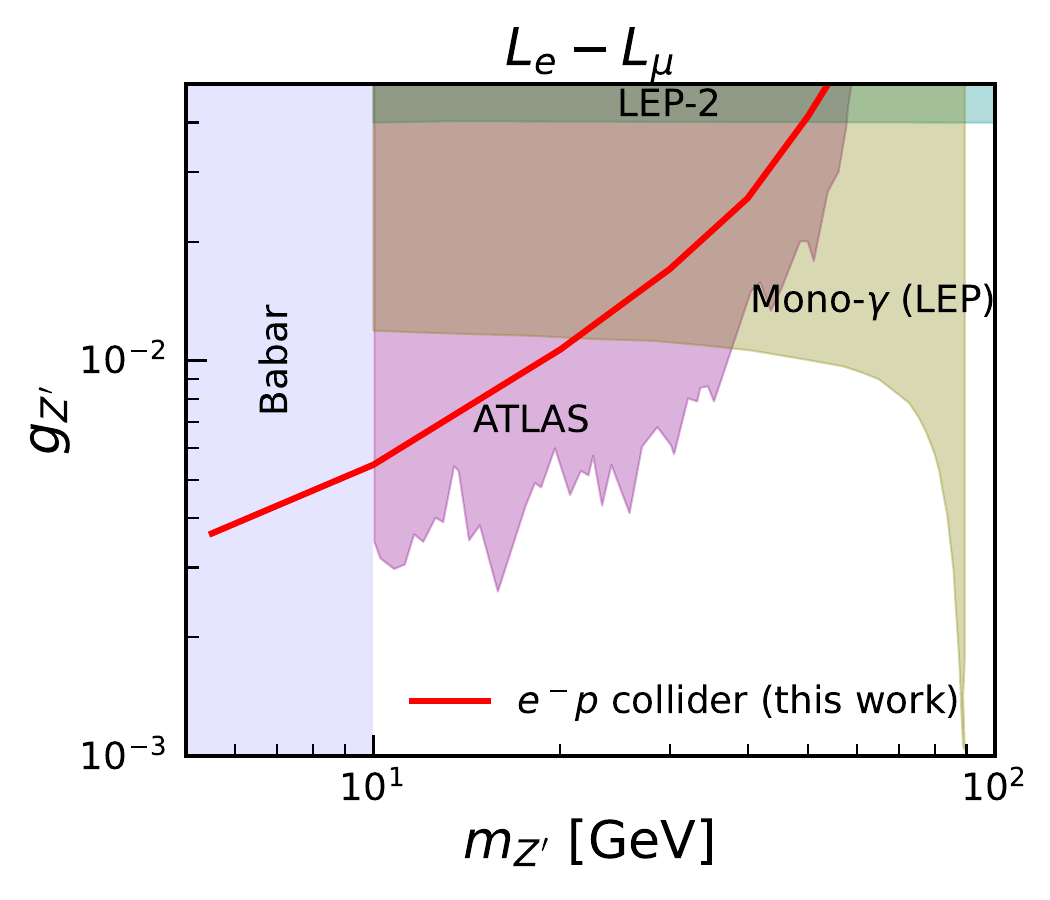}
\includegraphics[width=.45\textwidth]{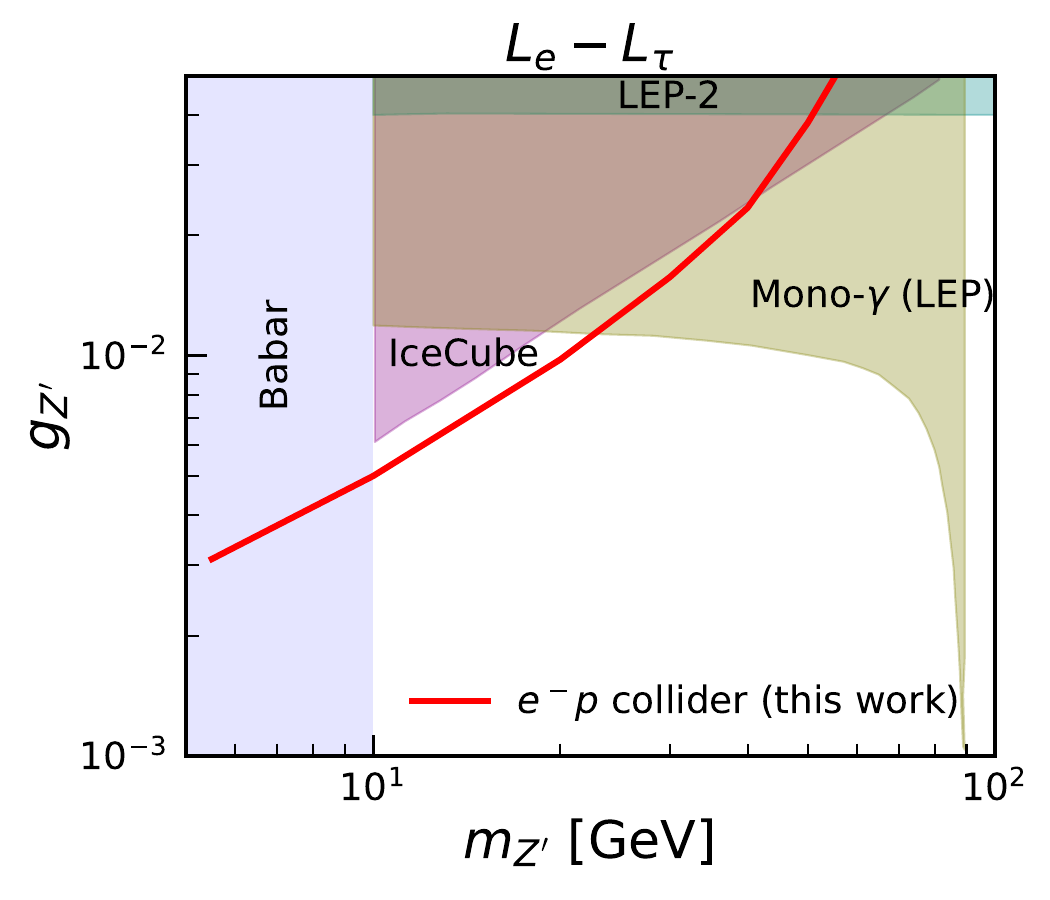}
\caption{Projected constraint from $e-p$ collider at $95\%$ CL on the $Z'$-electron coupling, $g_{\rm Z'}$ as a function of $Z'$ mass, $m_{\rm Z'}$ shown by the region above the red solid line in the $L_e-L_\mu$ model (left) and $L_e-L_\tau$ model (right). Other existing constraints like BaBar \cite{BaBar:2014zli}, LEP \cite{DELPHI:2003dlq,DELPHI:2008uka}, ATLAS \cite{ATLAS:2023vxg} are also displayed. Note that the IceCube constraint depends on the $\nu_e$ NSI \cite{IceCube:2022pbe}.}
\label{fig:UV_model}
\end{figure}

Finally, in Fig.~\ref{fig:UV_model}, we compare our exclusion limits for the $Z'$ realized in $L_e-L_\tau$ and $L_e-L_\mu$ models with the existing constraints from the LEP data~\cite{PhysRevD.84.014028,DELPHI:2003dlq,DELPHI:2008uka}, $e^+e^-$ collision experiment BaBar~\cite{BaBar:2014zli} and neutrino experiment IceCube~\cite{IceCube:2022pbe}.
Despite the stringent constraints, the projection obtained with our methodology can probe the region $10~{\rm GeV}\lesssim m_{Z'}\lesssim30$ GeV for the $L_e-L_\tau$ model.
A similar coupling strength can be probed for the $L_e-L_\mu$ model also. However, for muonphilic $Z'$, ATLAS \cite{ATLAS:2023vxg} in LHC provides the strongest constraint looking at the $4\mu$ final state, utilising the higher C.O.M energy of LHC.
Our projected limit becomes competitive with it in the mass range $m_{Z'}\sim15$ GeV despite the low C.O.M energy and can become stronger with higher luminosity.
Thus, EIC can shed light on the previously unexplored region of parameter space for these leptophilic models.
In the future, it will be interesting to look for other $Z'$ models e.g. $B_i-3L_e$, at EIC. 

\bibliographystyle{JHEP}
\bibliography{refs}

\providecommand{\href}[2]{#2}\begingroup\raggedright\begin{thebibliography}{100}

\bibitem{ATLAS:2012yve}
{\scshape ATLAS} collaboration, G.~Aad et~al., \emph{{Observation of a new
  particle in the search for the Standard Model Higgs boson with the ATLAS
  detector at the LHC}},
  \href{http://dx.doi.org/10.1016/j.physletb.2012.08.020}{\emph{Phys. Lett. B}
  {\bf 716} (2012) 1--29}, [\href{https://arxiv.org/abs/1207.7214}{{\tt
  1207.7214}}].

\bibitem{CMS:2012qbp}
{\scshape CMS} collaboration, S.~Chatrchyan et~al., \emph{{Observation of a New
  Boson at a Mass of 125 GeV with the CMS Experiment at the LHC}},
  \href{http://dx.doi.org/10.1016/j.physletb.2012.08.021}{\emph{Phys. Lett. B}
  {\bf 716} (2012) 30--61}, [\href{https://arxiv.org/abs/1207.7235}{{\tt
  1207.7235}}].

\bibitem{Zwicky:1933gu}
F.~Zwicky, \emph{{Die Rotverschiebung von extragalaktischen Nebeln}},
  \href{http://dx.doi.org/10.1007/s10714-008-0707-4}{\emph{Helv. Phys. Acta}
  {\bf 6} (1933) 110--127}.

\bibitem{Rubin:1970zza}
V.~C. Rubin and W.~K. Ford, Jr., \emph{{Rotation of the Andromeda Nebula from a
  Spectroscopic Survey of Emission Regions}},
  \href{http://dx.doi.org/10.1086/150317}{\emph{Astrophys. J.} {\bf 159} (1970)
  379--403}.

\bibitem{Clowe:2006eq}
D.~Clowe, M.~Bradac, A.~H. Gonzalez, M.~Markevitch, S.~W. Randall, C.~Jones
  et~al., \emph{{A direct empirical proof of the existence of dark matter}},
  \href{http://dx.doi.org/10.1086/508162}{\emph{Astrophys. J. Lett.} {\bf 648}
  (2006) L109--L113}, [\href{https://arxiv.org/abs/astro-ph/0608407}{{\tt
  astro-ph/0608407}}].

\bibitem{Planck:2018vyg}
{\scshape Planck} collaboration, N.~Aghanim et~al., \emph{{Planck 2018 results.
  VI. Cosmological parameters}},
  \href{http://dx.doi.org/10.1051/0004-6361/201833910}{\emph{Astron.
  Astrophys.} {\bf 641} (2020) A6},
  [\href{https://arxiv.org/abs/1807.06209}{{\tt 1807.06209}}].

\bibitem{DayaBay:2012fng}
{\scshape Daya Bay} collaboration, F.~P. An et~al., \emph{{Observation of
  electron-antineutrino disappearance at Daya Bay}},
  \href{http://dx.doi.org/10.1103/PhysRevLett.108.171803}{\emph{Phys. Rev.
  Lett.} {\bf 108} (2012) 171803}, [\href{https://arxiv.org/abs/1203.1669}{{\tt
  1203.1669}}].

\bibitem{RENO:2012mkc}
{\scshape RENO} collaboration, J.~K. Ahn et~al., \emph{{Observation of Reactor
  Electron Antineutrino Disappearance in the RENO Experiment}},
  \href{http://dx.doi.org/10.1103/PhysRevLett.108.191802}{\emph{Phys. Rev.
  Lett.} {\bf 108} (2012) 191802}, [\href{https://arxiv.org/abs/1204.0626}{{\tt
  1204.0626}}].

\bibitem{MINOS:2013xrl}
{\scshape MINOS} collaboration, P.~Adamson et~al., \emph{{Electron neutrino and
  antineutrino appearance in the full MINOS data sample}},
  \href{http://dx.doi.org/10.1103/PhysRevLett.110.171801}{\emph{Phys. Rev.
  Lett.} {\bf 110} (2013) 171801}, [\href{https://arxiv.org/abs/1301.4581}{{\tt
  1301.4581}}].

\bibitem{ParticleDataGroup:2024cfk}
{\scshape Particle Data Group} collaboration, S.~Navas et~al., \emph{{Review of
  particle physics}},
  \href{http://dx.doi.org/10.1103/PhysRevD.110.030001}{\emph{Phys. Rev. D} {\bf
  110} (2024) 030001}.

\bibitem{Sakharov:1967dj}
A.~D. Sakharov, \emph{{Violation of CP Invariance, C asymmetry, and baryon
  asymmetry of the universe}},
  \href{http://dx.doi.org/10.1070/PU1991v034n05ABEH002497}{\emph{Pisma Zh.
  Eksp. Teor. Fiz.} {\bf 5} (1967) 32--35}.

\bibitem{Peccei:1977hh}
R.~D. Peccei and H.~R. Quinn, \emph{{CP Conservation in the Presence of
  Instantons}},
  \href{http://dx.doi.org/10.1103/PhysRevLett.38.1440}{\emph{Phys. Rev. Lett.}
  {\bf 38} (1977) 1440--1443}.

\bibitem{Harris:2014hga}
P.~Harris, V.~V. Khoze, M.~Spannowsky and C.~Williams, \emph{{Constraining Dark
  Sectors at Colliders: Beyond the Effective Theory Approach}},
  \href{http://dx.doi.org/10.1103/PhysRevD.91.055009}{\emph{Phys. Rev. D} {\bf
  91} (2015) 055009}, [\href{https://arxiv.org/abs/1411.0535}{{\tt
  1411.0535}}].

\bibitem{Alekhin:2015byh}
S.~Alekhin et~al., \emph{{A facility to Search for Hidden Particles at the CERN
  SPS: the SHiP physics case}},
  \href{http://dx.doi.org/10.1088/0034-4885/79/12/124201}{\emph{Rept. Prog.
  Phys.} {\bf 79} (2016) 124201}, [\href{https://arxiv.org/abs/1504.04855}{{\tt
  1504.04855}}].

\bibitem{Essig:2013vha}
R.~Essig, J.~Mardon, M.~Papucci, T.~Volansky and Y.-M. Zhong,
  \emph{{Constraining Light Dark Matter with Low-Energy $e^+e^-$ Colliders}},
  \href{http://dx.doi.org/10.1007/JHEP11(2013)167}{\emph{JHEP} {\bf 11} (2013)
  167}, [\href{https://arxiv.org/abs/1309.5084}{{\tt 1309.5084}}].

\bibitem{DESI:2024mwx}
{\scshape DESI} collaboration, A.~G. Adame et~al., \emph{{DESI 2024 VI:
  cosmological constraints from the measurements of baryon acoustic
  oscillations}},
  \href{http://dx.doi.org/10.1088/1475-7516/2025/02/021}{\emph{JCAP} {\bf 02}
  (2025) 021}, [\href{https://arxiv.org/abs/2404.03002}{{\tt 2404.03002}}].

\bibitem{Raffelt:1996wa}
G.~G. Raffelt, \emph{{Stars as laboratories for fundamental physics}: {The
  astrophysics of neutrinos, axions, and other weakly interacting particles}}.
\newblock 5, 1996.

\bibitem{Accardi:2012qut}
A.~Accardi et~al., \emph{{Electron Ion Collider: The Next QCD Frontier}:
  {Understanding the glue that binds us all}},
  \href{http://dx.doi.org/10.1140/epja/i2016-16268-9}{\emph{Eur. Phys. J. A}
  {\bf 52} (2016) 268}, [\href{https://arxiv.org/abs/1212.1701}{{\tt
  1212.1701}}].

\bibitem{AbdulKhalek:2021gbh}
R.~Abdul~Khalek et~al., \emph{{Science Requirements and Detector Concepts for
  the Electron-Ion Collider}: {EIC Yellow Report}},
  \href{http://dx.doi.org/10.1016/j.nuclphysa.2022.122447}{\emph{Nucl. Phys. A}
  {\bf 1026} (2022) 122447}, [\href{https://arxiv.org/abs/2103.05419}{{\tt
  2103.05419}}].

\bibitem{Balkin:2023gya}
R.~Balkin, O.~Hen, W.~Li, H.~Liu, T.~Ma, Y.~Soreq et~al., \emph{{Probing
  axion-like particles at the Electron-Ion Collider}},
  \href{http://dx.doi.org/10.1007/JHEP02(2024)123}{\emph{JHEP} {\bf 02} (2024)
  123}, [\href{https://arxiv.org/abs/2310.08827}{{\tt 2310.08827}}].

\bibitem{Batell:2022ogj}
B.~Batell, T.~Ghosh, T.~Han and K.~Xie, \emph{{Heavy neutral leptons at the
  Electron-Ion Collider}},
  \href{http://dx.doi.org/10.1007/JHEP03(2023)020}{\emph{JHEP} {\bf 03} (2023)
  020}, [\href{https://arxiv.org/abs/2210.09287}{{\tt 2210.09287}}].

\bibitem{Jiang:2025frv}
X.-H. Jiang, Y.~Liu and B.~Yan, \emph{{Probing top-quark electroweak couplings
  indirectly at the Electron-Ion Collider}},
  \href{http://dx.doi.org/10.1103/w9wl-cjzq}{\emph{Phys. Rev. D} {\bf 112}
  (2025) L111303}, [\href{https://arxiv.org/abs/2507.21477}{{\tt 2507.21477}}].

\bibitem{Deng:2025hio}
Y.~Deng, X.-H. Jiang, T.~Liu and B.~Yan, \emph{{Testing lepton flavor
  universality at the Electron-Ion Collider}},
  \href{http://dx.doi.org/10.1007/JHEP06(2025)157}{\emph{JHEP} {\bf 06} (2025)
  157}, [\href{https://arxiv.org/abs/2503.02605}{{\tt 2503.02605}}].

\bibitem{Davoudiasl:2025rpn}
H.~Davoudiasl and H.~Liu, \emph{{Electron-ion collider as a discovery tool for
  invisible dark bosons}},
  \href{http://dx.doi.org/10.1103/gtsf-24x4}{\emph{Phys. Rev. D} {\bf 112}
  (2025) 075001}, [\href{https://arxiv.org/abs/2505.08871}{{\tt 2505.08871}}].

\bibitem{Bellafronte:2025ubi}
L.~Bellafronte, S.~Dawson, P.~P. Giardino and H.~Liu, \emph{{Probing
  Top-Quark{\textendash}Electron Interactions at Future Colliders}},
  \href{http://dx.doi.org/10.1103/7z6x-ygqq}{\emph{Phys. Rev. Lett.} {\bf 135}
  (2025) 251801}, [\href{https://arxiv.org/abs/2507.02039}{{\tt 2507.02039}}].

\bibitem{Balkin:2025rtc}
R.~Balkin, T.~Coren, A.~Jentsch, H.~Liu, M.~Ovchynnikov, Y.~Soreq et~al.,
  \emph{{Braking protons at the EIC: from invisible meson decay to new physics
  searches}},  \href{https://arxiv.org/abs/2601.00068}{{\tt 2601.00068}}.

\bibitem{Wen:2024cfu}
X.-K. Wen, B.~Yan, Z.~Yu and C.~P. Yuan, \emph{{Dihadron azimuthal asymmetry
  and light-quark dipole moments at the Electron-Ion Collider}},
  \href{https://arxiv.org/abs/2408.07255}{{\tt 2408.07255}}.

\bibitem{Wang:2024zns}
H.-L. Wang, X.-K. Wen, H.~Xing and B.~Yan, \emph{{Probing the four-fermion
  operators via the transverse double spin asymmetry at the Electron-Ion
  Collider}}, \href{http://dx.doi.org/10.1103/PhysRevD.109.095025}{\emph{Phys.
  Rev. D} {\bf 109} (2024) 095025},
  [\href{https://arxiv.org/abs/2401.08419}{{\tt 2401.08419}}].

\bibitem{Davoudiasl:2024vje}
H.~Davoudiasl, R.~Marcarelli and E.~T. Neil, \emph{{Flavor-violating ALPs,
  electron g-2, and the Electron-Ion Collider}},
  \href{http://dx.doi.org/10.1103/PhysRevD.109.115013}{\emph{Phys. Rev. D} {\bf
  109} (2024) 115013}, [\href{https://arxiv.org/abs/2402.17821}{{\tt
  2402.17821}}].

\bibitem{Davoudiasl:2023pkq}
H.~Davoudiasl, R.~Marcarelli and E.~T. Neil, \emph{{Displaced signals of hidden
  vectors at the Electron-Ion Collider}},
  \href{http://dx.doi.org/10.1103/PhysRevD.108.075017}{\emph{Phys. Rev. D} {\bf
  108} (2023) 075017}, [\href{https://arxiv.org/abs/2307.00102}{{\tt
  2307.00102}}].

\bibitem{Yan:2022npz}
B.~Yan, \emph{{Probing the dark photon via polarized DIS scattering at the HERA
  and EIC}},
  \href{http://dx.doi.org/10.1016/j.physletb.2022.137384}{\emph{Phys. Lett. B}
  {\bf 833} (2022) 137384}, [\href{https://arxiv.org/abs/2203.01510}{{\tt
  2203.01510}}].

\bibitem{Li:2021uww}
H.~T. Li, B.~Yan and C.~P. Yuan, \emph{{Jet charge: A new tool to probe the
  anomalous Zbb{\textasciimacron} couplings at the EIC}},
  \href{http://dx.doi.org/10.1016/j.physletb.2022.137300}{\emph{Phys. Lett. B}
  {\bf 833} (2022) 137300}, [\href{https://arxiv.org/abs/2112.07747}{{\tt
  2112.07747}}].

\bibitem{Yan:2021htf}
B.~Yan, Z.~Yu and C.~P. Yuan, \emph{{The anomalous Zbb{\textasciimacron}
  couplings at the HERA and EIC}},
  \href{http://dx.doi.org/10.1016/j.physletb.2021.136697}{\emph{Phys. Lett. B}
  {\bf 822} (2021) 136697}, [\href{https://arxiv.org/abs/2107.02134}{{\tt
  2107.02134}}].

\bibitem{Cirigliano:2021img}
V.~Cirigliano, K.~Fuyuto, C.~Lee, E.~Mereghetti and B.~Yan, \emph{{Charged
  Lepton Flavor Violation at the EIC}},
  \href{http://dx.doi.org/10.1007/JHEP03(2021)256}{\emph{JHEP} {\bf 03} (2021)
  256}, [\href{https://arxiv.org/abs/2102.06176}{{\tt 2102.06176}}].

\bibitem{Davoudiasl:2021mjy}
H.~Davoudiasl, R.~Marcarelli and E.~T. Neil, \emph{{Lepton-flavor-violating
  ALPs at the Electron-Ion Collider: a golden opportunity}},
  \href{http://dx.doi.org/10.1007/JHEP02(2023)071}{\emph{JHEP} {\bf 02} (2023)
  071}, [\href{https://arxiv.org/abs/2112.04513}{{\tt 2112.04513}}].

\bibitem{Liu:2021lan}
Y.~Liu and B.~Yan, \emph{{Searching for the axion-like particle at the EIC*}},
  \href{http://dx.doi.org/10.1088/1674-1137/acbbc0}{\emph{Chin. Phys. C} {\bf
  47} (2023) 043113}, [\href{https://arxiv.org/abs/2112.02477}{{\tt
  2112.02477}}].

\bibitem{Weinberg:1977ma}
S.~Weinberg, \emph{{A New Light Boson?}},
  \href{http://dx.doi.org/10.1103/PhysRevLett.40.223}{\emph{Phys. Rev. Lett.}
  {\bf 40} (1978) 223--226}.

\bibitem{Wilczek:1977pj}
F.~Wilczek, \emph{{Problem of Strong $P$ and $T$ Invariance in the Presence of
  Instantons}}, \href{http://dx.doi.org/10.1103/PhysRevLett.40.279}{\emph{Phys.
  Rev. Lett.} {\bf 40} (1978) 279--282}.

\bibitem{Kim:2008hd}
J.~E. Kim and G.~Carosi, \emph{{Axions and the Strong CP Problem}},
  \href{http://dx.doi.org/10.1103/RevModPhys.82.557}{\emph{Rev. Mod. Phys.}
  {\bf 82} (2010) 557--602}, [\href{https://arxiv.org/abs/0807.3125}{{\tt
  0807.3125}}].

\bibitem{Kim:1979if}
J.~E. Kim, \emph{{Weak Interaction Singlet and Strong CP Invariance}},
  \href{http://dx.doi.org/10.1103/PhysRevLett.43.103}{\emph{Phys. Rev. Lett.}
  {\bf 43} (1979) 103}.

\bibitem{Shifman:1979if}
M.~A. Shifman, A.~I. Vainshtein and V.~I. Zakharov, \emph{{Can Confinement
  Ensure Natural CP Invariance of Strong Interactions?}},
  \href{http://dx.doi.org/10.1016/0550-3213(80)90209-6}{\emph{Nucl. Phys. B}
  {\bf 166} (1980) 493--506}.

\bibitem{Zhitnitsky:1980tq}
A.~R. Zhitnitsky, \emph{{On Possible Suppression of the Axion Hadron
  Interactions. (In Russian)}}, {\emph{Sov. J. Nucl. Phys.} {\bf 31} (1980)
  260}.

\bibitem{Dine:1981rt}
M.~Dine, W.~Fischler and M.~Srednicki, \emph{{A Simple Solution to the Strong
  CP Problem with a Harmless Axion}},
  \href{http://dx.doi.org/10.1016/0370-2693(81)90590-6}{\emph{Phys. Lett. B}
  {\bf 104} (1981) 199--202}.

\bibitem{Biekotter:2025fll}
A.~Biek{\"o}tter and K.~Mimasu, \emph{{Axions and Axion-like particles:
  collider searches}}.
\newblock 8, 2025.
\newblock \href{https://arxiv.org/abs/2508.19358}{{\tt 2508.19358}}.

\bibitem{DiLuzio:2020wdo}
L.~Di~Luzio, M.~Giannotti, E.~Nardi and L.~Visinelli, \emph{{The landscape of
  QCD axion models}},
  \href{http://dx.doi.org/10.1016/j.physrep.2020.06.002}{\emph{Phys. Rept.}
  {\bf 870} (2020) 1--117}, [\href{https://arxiv.org/abs/2003.01100}{{\tt
  2003.01100}}].

\bibitem{Bharucha:2022lty}
A.~Bharucha, F.~Br\"ummer, N.~Desai and S.~Mutzel, \emph{{Axion-like particles
  as mediators for dark matter: beyond freeze-out}},
  \href{http://dx.doi.org/10.1007/JHEP02(2023)141}{\emph{JHEP} {\bf 02} (2023)
  141}, [\href{https://arxiv.org/abs/2209.03932}{{\tt 2209.03932}}].

\bibitem{Dine:1982ah}
M.~Dine and W.~Fischler, \emph{{The Not So Harmless Axion}},
  \href{http://dx.doi.org/10.1016/0370-2693(83)90639-1}{\emph{Phys. Lett. B}
  {\bf 120} (1983) 137--141}.

\bibitem{Preskill:1982cy}
J.~Preskill, M.~B. Wise and F.~Wilczek, \emph{{Cosmology of the Invisible
  Axion}}, \href{http://dx.doi.org/10.1016/0370-2693(83)90637-8}{\emph{Phys.
  Lett. B} {\bf 120} (1983) 127--132}.

\bibitem{Abbott:1982af}
L.~F. Abbott and P.~Sikivie, \emph{{A Cosmological Bound on the Invisible
  Axion}}, \href{http://dx.doi.org/10.1016/0370-2693(83)90638-X}{\emph{Phys.
  Lett. B} {\bf 120} (1983) 133--136}.

\bibitem{Co:2019jts}
R.~T. Co, L.~J. Hall and K.~Harigaya, \emph{{Axion Kinetic Misalignment
  Mechanism}},
  \href{http://dx.doi.org/10.1103/PhysRevLett.124.251802}{\emph{Phys. Rev.
  Lett.} {\bf 124} (2020) 251802},
  [\href{https://arxiv.org/abs/1910.14152}{{\tt 1910.14152}}].

\bibitem{Arias:2012az}
P.~Arias, D.~Cadamuro, M.~Goodsell, J.~Jaeckel, J.~Redondo and A.~Ringwald,
  \emph{{WISPy Cold Dark Matter}},
  \href{http://dx.doi.org/10.1088/1475-7516/2012/06/013}{\emph{JCAP} {\bf 06}
  (2012) 013}, [\href{https://arxiv.org/abs/1201.5902}{{\tt 1201.5902}}].

\bibitem{Jaeckel:2014qea}
J.~Jaeckel, J.~Redondo and A.~Ringwald, \emph{{3.55 keV hint for decaying
  axionlike particle dark matter}},
  \href{http://dx.doi.org/10.1103/PhysRevD.89.103511}{\emph{Phys. Rev. D} {\bf
  89} (2014) 103511}, [\href{https://arxiv.org/abs/1402.7335}{{\tt
  1402.7335}}].

\bibitem{Ghosh:2023tyz}
D.~K. Ghosh, A.~Ghoshal and S.~Jeesun, \emph{{Axion-like particle (ALP) portal
  freeze-in dark matter confronting ALP search experiments}},
  \href{http://dx.doi.org/10.1007/JHEP01(2024)026}{\emph{JHEP} {\bf 01} (2024)
  026}, [\href{https://arxiv.org/abs/2305.09188}{{\tt 2305.09188}}].

\bibitem{Marsh:2015xka}
D.~J.~E. Marsh, \emph{{Axion Cosmology}},
  \href{http://dx.doi.org/10.1016/j.physrep.2016.06.005}{\emph{Phys. Rept.}
  {\bf 643} (2016) 1--79}, [\href{https://arxiv.org/abs/1510.07633}{{\tt
  1510.07633}}].

\bibitem{Caloni:2022uya}
L.~Caloni, M.~Gerbino, M.~Lattanzi and L.~Visinelli, \emph{{Novel cosmological
  bounds on thermally-produced axion-like particles}},
  \href{http://dx.doi.org/10.1088/1475-7516/2022/09/021}{\emph{JCAP} {\bf 09}
  (2022) 021}, [\href{https://arxiv.org/abs/2205.01637}{{\tt 2205.01637}}].

\bibitem{Cadamuro:2010cz}
D.~Cadamuro, S.~Hannestad, G.~Raffelt and J.~Redondo, \emph{{Cosmological
  bounds on sub-MeV mass axions}},
  \href{http://dx.doi.org/10.1088/1475-7516/2011/02/003}{\emph{JCAP} {\bf 02}
  (2011) 003}, [\href{https://arxiv.org/abs/1011.3694}{{\tt 1011.3694}}].

\bibitem{Turner:1987by}
M.~S. Turner, \emph{{Axions from SN 1987a}},
  \href{http://dx.doi.org/10.1103/PhysRevLett.60.1797}{\emph{Phys. Rev. Lett.}
  {\bf 60} (1988) 1797}.

\bibitem{Caputo:2024oqc}
A.~Caputo and G.~Raffelt, \emph{{Astrophysical Axion Bounds: The 2024
  Edition}}, \href{http://dx.doi.org/10.22323/1.454.0041}{\emph{PoS} {\bf
  COSMICWISPers} (2024) 041}, [\href{https://arxiv.org/abs/2401.13728}{{\tt
  2401.13728}}].

\bibitem{Fiorillo:2025gnd}
D.~F.~G. Fiorillo, {\'A}.~Gil~Muyor, H.-T. Janka, G.~G. Raffelt and
  E.~Vitagliano, \emph{{Axion-photon conversion in transient compact stars:
  Systematics, constraints, and opportunities}},
  \href{http://dx.doi.org/10.1088/1475-7516/2026/03/053}{\emph{JCAP} {\bf 03}
  (2026) 053}, [\href{https://arxiv.org/abs/2509.13322}{{\tt 2509.13322}}].

\bibitem{Candon:2025sdm}
F.~R. Cand{\'o}n, D.~F.~G. Fiorillo, {\'A}.~Gil~Muyor, H.-T. Janka, G.~G.
  Raffelt and E.~Vitagliano, \emph{{Stripped-Envelope Supernovae for QCD Axion
  Detection}},  \href{https://arxiv.org/abs/2511.13815}{{\tt 2511.13815}}.

\bibitem{Bauer:2017ris}
M.~Bauer, M.~Neubert and A.~Thamm, \emph{{Collider Probes of Axion-Like
  Particles}}, \href{http://dx.doi.org/10.1007/JHEP12(2017)044}{\emph{JHEP}
  {\bf 12} (2017) 044}, [\href{https://arxiv.org/abs/1708.00443}{{\tt
  1708.00443}}].

\bibitem{OPAL:2002vhf}
{\scshape OPAL} collaboration, G.~Abbiendi et~al., \emph{{Multiphoton
  production in e+ e- collisions at s**(1/2) = 181-GeV to 209-GeV}},
  \href{http://dx.doi.org/10.1140/epjc/s2002-01074-5}{\emph{Eur. Phys. J. C}
  {\bf 26} (2003) 331--344}, [\href{https://arxiv.org/abs/hep-ex/0210016}{{\tt
  hep-ex/0210016}}].

\bibitem{Jaeckel:2015jla}
J.~Jaeckel and M.~Spannowsky, \emph{{Probing MeV to 90 GeV axion-like particles
  with LEP and LHC}},
  \href{http://dx.doi.org/10.1016/j.physletb.2015.12.037}{\emph{Phys. Lett. B}
  {\bf 753} (2016) 482--487}, [\href{https://arxiv.org/abs/1509.00476}{{\tt
  1509.00476}}].

\bibitem{Yue:2021iiu}
C.-X. Yue, H.-Y. Zhang and H.~Wang, \emph{{Production of axion-like particles
  via vector boson fusion at future electron-positron colliders}},
  \href{http://dx.doi.org/10.1140/epjc/s10052-022-10007-7}{\emph{Eur. Phys. J.
  C} {\bf 82} (2022) 88}, [\href{https://arxiv.org/abs/2112.11604}{{\tt
  2112.11604}}].

\bibitem{Tian:2022rsi}
M.~Tian, Z.~S. Wang and K.~Wang, \emph{{Search for long-lived axions with far
  detectors at future lepton colliders}},
  \href{https://arxiv.org/abs/2201.08960}{{\tt 2201.08960}}.

\bibitem{BESIII:2022rzz}
{\scshape BESIII} collaboration, M.~Ablikim et~al., \emph{{Search for an
  axion-like particle in radiative J/\ensuremath{\psi} decays}},
  \href{http://dx.doi.org/10.1016/j.physletb.2023.137698}{\emph{Phys. Lett. B}
  {\bf 838} (2023) 137698}, [\href{https://arxiv.org/abs/2211.12699}{{\tt
  2211.12699}}].

\bibitem{Adhikary:2024mzi}
A.~Adhikary, A.~Bharucha, L.~Feligioni and M.~Frigerio, \emph{{Prospects for
  sub-EW-scale ALP searches via {\ensuremath{\gamma}}+bb{\textasciimacron}
  signatures at the LHC using jet substructure techniques}},
  \href{http://dx.doi.org/10.1103/bf6h-573f}{\emph{Phys. Rev. D} {\bf 112}
  (2025) 055042}, [\href{https://arxiv.org/abs/2410.09033}{{\tt 2410.09033}}].

\bibitem{CMS:2012cve}
{\scshape CMS} collaboration, S.~Chatrchyan et~al., \emph{{Search for Exclusive
  or Semi-Exclusive Photon Pair Production and Observation of Exclusive and
  Semi-Exclusive Electron Pair Production in $pp$ Collisions at $\sqrt{s}=7$
  TeV}}, \href{http://dx.doi.org/10.1007/JHEP11(2012)080}{\emph{JHEP} {\bf 11}
  (2012) 080}, [\href{https://arxiv.org/abs/1209.1666}{{\tt 1209.1666}}].

\bibitem{ATLAS:2014jdv}
{\scshape ATLAS} collaboration, G.~Aad et~al., \emph{{Search for Scalar
  Diphoton Resonances in the Mass Range $65-600$ GeV with the ATLAS Detector in
  $pp$ Collision Data at $\sqrt{s}$ = 8 $TeV$}},
  \href{http://dx.doi.org/10.1103/PhysRevLett.113.171801}{\emph{Phys. Rev.
  Lett.} {\bf 113} (2014) 171801}, [\href{https://arxiv.org/abs/1407.6583}{{\tt
  1407.6583}}].

\bibitem{Mimasu:2014nea}
K.~Mimasu and V.~Sanz, \emph{{ALPs at Colliders}},
  \href{http://dx.doi.org/10.1007/JHEP06(2015)173}{\emph{JHEP} {\bf 06} (2015)
  173}, [\href{https://arxiv.org/abs/1409.4792}{{\tt 1409.4792}}].

\bibitem{Brivio:2017ije}
I.~Brivio, M.~B. Gavela, L.~Merlo, K.~Mimasu, J.~M. No, R.~del Rey et~al.,
  \emph{{ALPs Effective Field Theory and Collider Signatures}},
  \href{http://dx.doi.org/10.1140/epjc/s10052-017-5111-3}{\emph{Eur. Phys. J.
  C} {\bf 77} (2017) 572}, [\href{https://arxiv.org/abs/1701.05379}{{\tt
  1701.05379}}].

\bibitem{Ebadi:2019gij}
J.~Ebadi, S.~Khatibi and M.~Mohammadi~Najafabadi, \emph{{New probes for
  axionlike particles at hadron colliders}},
  \href{http://dx.doi.org/10.1103/PhysRevD.100.015016}{\emph{Phys. Rev. D} {\bf
  100} (2019) 015016}, [\href{https://arxiv.org/abs/1901.03061}{{\tt
  1901.03061}}].

\bibitem{Bonilla:2022pxu}
J.~Bonilla, I.~Brivio, J.~Machado-Rodr\'\i{}guez and J.~F. de~Troc\'oniz,
  \emph{{Nonresonant searches for axion-like particles in vector boson
  scattering processes at the LHC}},
  \href{http://dx.doi.org/10.1007/JHEP06(2022)113}{\emph{JHEP} {\bf 06} (2022)
  113}, [\href{https://arxiv.org/abs/2202.03450}{{\tt 2202.03450}}].

\bibitem{Mitridate:2023tbj}
A.~Mitridate, M.~Papucci, C.~Wang, C.~Pe\~na and S.~Xie, \emph{{Energetic
  long-lived particles in the CMS muon chambers}},
  \href{https://arxiv.org/abs/2304.06109}{{\tt 2304.06109}}.

\bibitem{Dutta:2023abe}
B.~Dutta, D.~Kim and H.~Kim, \emph{{A Novel Beam-Dump Measurement with the LHC
  General-Purpose Detectors}},  \href{https://arxiv.org/abs/2305.16383}{{\tt
  2305.16383}}.

\bibitem{Bao:2022onq}
Y.~Bao, J.~Fan and L.~Li, \emph{{Electroweak ALP Searches at a Muon Collider}},
   \href{https://arxiv.org/abs/2203.04328}{{\tt 2203.04328}}.

\bibitem{CHARM:1985anb}
{\scshape CHARM} collaboration, F.~Bergsma et~al., \emph{{Search for Axion Like
  Particle Production in 400-{GeV} Proton - Copper Interactions}},
  \href{http://dx.doi.org/10.1016/0370-2693(85)90400-9}{\emph{Phys. Lett. B}
  {\bf 157} (1985) 458--462}.

\bibitem{Bjorken:1988as}
J.~D. Bjorken, S.~Ecklund, W.~R. Nelson, A.~Abashian, C.~Church, B.~Lu et~al.,
  \emph{{Search for Neutral Metastable Penetrating Particles Produced in the
  SLAC Beam Dump}},
  \href{http://dx.doi.org/10.1103/PhysRevD.38.3375}{\emph{Phys. Rev. D} {\bf
  38} (1988) 3375}.

\bibitem{Blumlein:1990ay}
J.~Blumlein et~al., \emph{{Limits on neutral light scalar and pseudoscalar
  particles in a proton beam dump experiment}},
  \href{http://dx.doi.org/10.1007/BF01548556}{\emph{Z. Phys. C} {\bf 51} (1991)
  341--350}.

\bibitem{Dobrich:2017gcm}
B.~D\"obrich, \emph{{Axion-like Particles from Primakov production in
  beam-dumps}},
  \href{http://dx.doi.org/10.23727/CERN-Proceedings-2018-001.253}{\emph{CERN
  Proc.} {\bf 1} (2018) 253}, [\href{https://arxiv.org/abs/1708.05776}{{\tt
  1708.05776}}].

\bibitem{NA64:2020qwq}
{\scshape NA64} collaboration, D.~Banerjee et~al., \emph{{Search for Axionlike
  and Scalar Particles with the NA64 Experiment}},
  \href{http://dx.doi.org/10.1103/PhysRevLett.125.081801}{\emph{Phys. Rev.
  Lett.} {\bf 125} (2020) 081801},
  [\href{https://arxiv.org/abs/2005.02710}{{\tt 2005.02710}}].

\bibitem{Afik:2023mhj}
Y.~Afik, B.~D\"obrich, J.~Jerhot, Y.~Soreq and K.~Tobioka, \emph{{Probing
  Long-lived Axions at the KOTO Experiment}},
  \href{https://arxiv.org/abs/2303.01521}{{\tt 2303.01521}}.

\bibitem{Ema:2023tjg}
Y.~Ema, Z.~Liu and R.~Plestid, \emph{{Searching for axions with kaon decay at
  rest}},  \href{https://arxiv.org/abs/2308.08589}{{\tt 2308.08589}}.

\bibitem{BaBar:2014zli}
{\scshape BaBar} collaboration, J.~P. Lees et~al., \emph{{Search for a Dark
  Photon in $e^+e^-$ Collisions at BaBar}},
  \href{http://dx.doi.org/10.1103/PhysRevLett.113.201801}{\emph{Phys. Rev.
  Lett.} {\bf 113} (2014) 201801}, [\href{https://arxiv.org/abs/1406.2980}{{\tt
  1406.2980}}].

\bibitem{Inguglia:2016acz}
G.~Inguglia, \emph{{Belle II studies of missing energy decays and searches for
  dark photon production}},
  \href{http://dx.doi.org/10.22323/1.265.0263}{\emph{PoS} {\bf DIS2016} (2016)
  263}, [\href{https://arxiv.org/abs/1607.02089}{{\tt 1607.02089}}].

\bibitem{Okada:2016tci}
N.~Okada and S.~Okada, \emph{{$Z^\prime$-portal right-handed neutrino dark
  matter in the minimal U(1)$_X$ extended Standard Model}},
  \href{http://dx.doi.org/10.1103/PhysRevD.95.035025}{\emph{Phys. Rev. D} {\bf
  95} (2017) 035025}, [\href{https://arxiv.org/abs/1611.02672}{{\tt
  1611.02672}}].

\bibitem{Mohapatra:2020bze}
R.~N. Mohapatra and N.~Okada, \emph{{Freeze-in Dark Matter from a Minimal B-L
  Model and Possible Grand Unification}},
  \href{http://dx.doi.org/10.1103/PhysRevD.101.115022}{\emph{Phys. Rev. D} {\bf
  101} (2020) 115022}, [\href{https://arxiv.org/abs/2005.00365}{{\tt
  2005.00365}}].

\bibitem{Okada:2021nwo}
H.~Okada, Y.~Orikasa and Y.~Shoji, \emph{{Radiative dark matter and neutrino
  masses from an alternative U(1) B-L gauge symmetry}},
  \href{http://dx.doi.org/10.1088/1475-7516/2021/07/006}{\emph{JCAP} {\bf 07}
  (2021) 006}, [\href{https://arxiv.org/abs/2102.10944}{{\tt 2102.10944}}].

\bibitem{Ma:2015raa}
E.~Ma and R.~Srivastava, \emph{{Dirac or inverse seesaw neutrino masses from
  gauged $B–L$ symmetry}},
  \href{http://dx.doi.org/10.1142/S0217732315300207}{\emph{Mod. Phys. Lett. A}
  {\bf 30} (2015) 1530020}, [\href{https://arxiv.org/abs/1504.00111}{{\tt
  1504.00111}}].

\bibitem{Qi:2024pqe}
X.~Qi and H.~Sun, \emph{{Dark matter, leptogenesis and $Z^\prime $ in the $B-L$
  model}}, \href{http://dx.doi.org/10.1140/epjc/s10052-025-14972-7}{\emph{Eur.
  Phys. J. C} {\bf 85} (2025) 1224},
  [\href{https://arxiv.org/abs/2407.21292}{{\tt 2407.21292}}].

\bibitem{Bonilla:2017lsq}
C.~Bonilla, T.~Modak, R.~Srivastava and J.~W.~F. Valle,
  \emph{{$U(1)_{B_3-3L_\mu}$ gauge symmetry as a simple description of $b\to s$
  anomalies}}, \href{http://dx.doi.org/10.1103/PhysRevD.98.095002}{\emph{Phys.
  Rev. D} {\bf 98} (2018) 095002},
  [\href{https://arxiv.org/abs/1705.00915}{{\tt 1705.00915}}].

\bibitem{Bauer:2018onh}
M.~Bauer, P.~Foldenauer and J.~Jaeckel, \emph{{Hunting All the Hidden
  Photons}}, \href{http://dx.doi.org/10.1007/JHEP07(2018)094}{\emph{JHEP} {\bf
  07} (2018) 094}, [\href{https://arxiv.org/abs/1803.05466}{{\tt 1803.05466}}].

\bibitem{Coloma:2020gfv}
P.~Coloma, M.~C. Gonzalez-Garcia and M.~Maltoni, \emph{{Neutrino oscillation
  constraints on U(1)' models: from non-standard interactions to long-range
  forces}}, \href{http://dx.doi.org/10.1007/JHEP01(2021)114}{\emph{JHEP} {\bf
  01} (2021) 114}, [\href{https://arxiv.org/abs/2009.14220}{{\tt 2009.14220}}].

\bibitem{Coloma:2022umy}
P.~Coloma, P.~Coloma, M.~C. Gonzalez-Garcia, M.~C. Gonzalez-Garcia, M.~Maltoni,
  M.~Maltoni et~al., \emph{{Constraining new physics with Borexino Phase-II
  spectral data}}, \href{http://dx.doi.org/10.1007/JHEP07(2022)138}{\emph{JHEP}
  {\bf 07} (2022) 138}, [\href{https://arxiv.org/abs/2204.03011}{{\tt
  2204.03011}}].

\bibitem{AtzoriCorona:2022moj}
M.~Atzori~Corona, M.~Cadeddu, N.~Cargioli, F.~Dordei, C.~Giunti, Y.~F. Li
  et~al., \emph{{Probing light mediators and $(g-2)_{\mu}$ through detection of
  coherent elastic neutrino nucleus scattering at COHERENT}},
  \href{http://dx.doi.org/10.1007/JHEP05(2022)109}{\emph{JHEP} {\bf 05} (2022)
  109}, [\href{https://arxiv.org/abs/2202.11002}{{\tt 2202.11002}}].

\bibitem{Majumdar:2021vdw}
A.~Majumdar, D.~K. Papoulias and R.~Srivastava, \emph{{Dark matter detectors as
  a novel probe for light new physics}},
  \href{http://dx.doi.org/10.1103/PhysRevD.106.013001}{\emph{Phys. Rev. D} {\bf
  106} (2022) 013001}, [\href{https://arxiv.org/abs/2112.03309}{{\tt
  2112.03309}}].

\bibitem{Chakraborty:2021apc}
K.~Chakraborty, A.~Das, S.~Goswami and S.~Roy, \emph{{Constraining general U(1)
  interactions from neutrino-electron scattering measurements at DUNE near
  detector}}, \href{http://dx.doi.org/10.1007/JHEP04(2022)008}{\emph{JHEP} {\bf
  04} (2022) 008}, [\href{https://arxiv.org/abs/2111.08767}{{\tt 2111.08767}}].

\bibitem{Demirci:2023tui}
M.~Demirci and M.~F. Mustamin, \emph{{Solar neutrino constraints on light
  mediators through coherent elastic neutrino-nucleus scattering}},
  \href{http://dx.doi.org/10.1103/PhysRevD.109.015021}{\emph{Phys. Rev. D} {\bf
  109} (2024) 015021}, [\href{https://arxiv.org/abs/2312.17502}{{\tt
  2312.17502}}].

\bibitem{CMS:2016cfx}
{\scshape CMS} collaboration, V.~Khachatryan et~al., \emph{{Search for narrow
  resonances in dilepton mass spectra in proton-proton collisions at $\sqrt{s}$
  = 13 TeV and combination with 8 TeV data}},
  \href{http://dx.doi.org/10.1016/j.physletb.2017.02.010}{\emph{Phys. Lett. B}
  {\bf 768} (2017) 57--80}, [\href{https://arxiv.org/abs/1609.05391}{{\tt
  1609.05391}}].

\bibitem{ATLAS:2019erb}
{\scshape ATLAS} collaboration, G.~Aad et~al., \emph{{Search for high-mass
  dilepton resonances using 139 fb$^{-1}$ of $pp$ collision data collected at
  $\sqrt{s}=$13 TeV with the ATLAS detector}},
  \href{http://dx.doi.org/10.1016/j.physletb.2019.07.016}{\emph{Phys. Lett. B}
  {\bf 796} (2019) 68--87}, [\href{https://arxiv.org/abs/1903.06248}{{\tt
  1903.06248}}].

\bibitem{Das:2016zue}
A.~Das, S.~Oda, N.~Okada and D.-s. Takahashi, \emph{{Classically conformal
  U(1)' extended standard model, electroweak vacuum stability, and LHC Run-2
  bounds}}, \href{http://dx.doi.org/10.1103/PhysRevD.93.115038}{\emph{Phys.
  Rev. D} {\bf 93} (2016) 115038},
  [\href{https://arxiv.org/abs/1605.01157}{{\tt 1605.01157}}].

\bibitem{Accomando:2017qcs}
E.~Accomando, L.~Delle~Rose, S.~Moretti, E.~Olaiya and C.~H.
  Shepherd-Themistocleous, \emph{{Extra Higgs boson and Z$^{\prime}$ as portals
  to signatures of heavy neutrinos at the LHC}},
  \href{http://dx.doi.org/10.1007/JHEP02(2018)109}{\emph{JHEP} {\bf 02} (2018)
  109}, [\href{https://arxiv.org/abs/1708.03650}{{\tt 1708.03650}}].

\bibitem{Essig:2009nc}
R.~Essig, P.~Schuster and N.~Toro, \emph{{Probing Dark Forces and Light Hidden
  Sectors at Low-Energy e+e- Colliders}},
  \href{http://dx.doi.org/10.1103/PhysRevD.80.015003}{\emph{Phys. Rev. D} {\bf
  80} (2009) 015003}, [\href{https://arxiv.org/abs/0903.3941}{{\tt
  0903.3941}}].

\bibitem{LHCb:2017trq}
{\scshape LHCb} collaboration, R.~Aaij et~al., \emph{{Search for Dark Photons
  Produced in 13 TeV $pp$ Collisions}},
  \href{http://dx.doi.org/10.1103/PhysRevLett.120.061801}{\emph{Phys. Rev.
  Lett.} {\bf 120} (2018) 061801},
  [\href{https://arxiv.org/abs/1710.02867}{{\tt 1710.02867}}].

\bibitem{Bross:1989mp}
A.~Bross, M.~Crisler, S.~H. Pordes, J.~Volk, S.~Errede and J.~Wrbanek, \emph{{A
  Search for Shortlived Particles Produced in an Electron Beam Dump}},
  \href{http://dx.doi.org/10.1103/PhysRevLett.67.2942}{\emph{Phys. Rev. Lett.}
  {\bf 67} (1991) 2942--2945}.

\bibitem{COHERENT:2021xmm}
{\scshape COHERENT} collaboration, D.~Akimov et~al., \emph{{Measurement of the
  Coherent Elastic Neutrino-Nucleus Scattering Cross Section on CsI by
  COHERENT}},
  \href{http://dx.doi.org/10.1103/PhysRevLett.129.081801}{\emph{Phys. Rev.
  Lett.} {\bf 129} (2022) 081801},
  [\href{https://arxiv.org/abs/2110.07730}{{\tt 2110.07730}}].

\bibitem{Escudero:2019gzq}
M.~Escudero, D.~Hooper, G.~Krnjaic and M.~Pierre, \emph{{Cosmology with A Very
  Light L$_{\mu}$ \ensuremath{-} L$_{\tau}$ Gauge Boson}},
  \href{http://dx.doi.org/10.1007/JHEP03(2019)071}{\emph{JHEP} {\bf 03} (2019)
  071}, [\href{https://arxiv.org/abs/1901.02010}{{\tt 1901.02010}}].

\bibitem{Ghosh:2024cxi}
D.~K. Ghosh, P.~Ghosh, S.~Jeesun and R.~Srivastava, \emph{{Neff at CMB
  challenges U(1)X light gauge boson scenarios}},
  \href{http://dx.doi.org/10.1103/PhysRevD.110.075032}{\emph{Phys. Rev. D} {\bf
  110} (2024) 075032}, [\href{https://arxiv.org/abs/2404.10077}{{\tt
  2404.10077}}].

\bibitem{Kolb:1987qy}
E.~W. Kolb and M.~S. Turner, \emph{{Supernova SN 1987a and the Secret
  Interactions of Neutrinos}},
  \href{http://dx.doi.org/10.1103/PhysRevD.36.2895}{\emph{Phys. Rev. D} {\bf
  36} (1987) 2895}.

\bibitem{Croon:2020lrf}
D.~Croon, G.~Elor, R.~K. Leane and S.~D. McDermott, \emph{{Supernova Muons: New
  Constraints on $Z$' Bosons, Axions and ALPs}},
  \href{http://dx.doi.org/10.1007/JHEP01(2021)107}{\emph{JHEP} {\bf 01} (2021)
  107}, [\href{https://arxiv.org/abs/2006.13942}{{\tt 2006.13942}}].

\bibitem{Akita:2023iwq}
K.~Akita, S.~H. Im, M.~Masud and S.~Yun, \emph{{Limits on heavy neutral
  leptons, $Z'$ bosons and majorons from high-energy supernova neutrinos}},
  \href{https://arxiv.org/abs/2312.13627}{{\tt 2312.13627}}.

\bibitem{Li:2023vpv}
S.-P. Li and X.-J. Xu, \emph{{Production rates of dark photons and Z' in the
  Sun and stellar cooling bounds}},
  \href{http://dx.doi.org/10.1088/1475-7516/2023/09/009}{\emph{JCAP} {\bf 09}
  (2023) 009}, [\href{https://arxiv.org/abs/2304.12907}{{\tt 2304.12907}}].

\bibitem{Eberhart:2025lyu}
A.~Eberhart, M.~Fedele, F.~Kahlhoefer, E.~Ravensburg and R.~Ziegler,
  \emph{{Leptophilic ALPs in Laboratory Experiments}},
  \href{https://arxiv.org/abs/2504.05873}{{\tt 2504.05873}}.

\bibitem{Alloul:2013bka}
A.~Alloul, N.~D. Christensen, C.~Degrande, C.~Duhr and B.~Fuks,
  \emph{{FeynRules 2.0 - A complete toolbox for tree-level phenomenology}},
  \href{http://dx.doi.org/10.1016/j.cpc.2014.04.012}{\emph{Comput. Phys.
  Commun.} {\bf 185} (2014) 2250--2300},
  [\href{https://arxiv.org/abs/1310.1921}{{\tt 1310.1921}}].

\bibitem{Alwall:2014hca}
J.~Alwall, R.~Frederix, S.~Frixione, V.~Hirschi, F.~Maltoni, O.~Mattelaer
  et~al., \emph{{The automated computation of tree-level and next-to-leading
  order differential cross sections, and their matching to parton shower
  simulations}}, \href{http://dx.doi.org/10.1007/JHEP07(2014)079}{\emph{JHEP}
  {\bf 07} (2014) 079}, [\href{https://arxiv.org/abs/1405.0301}{{\tt
  1405.0301}}].

\bibitem{NNPDF:2014otw}
{\scshape NNPDF} collaboration, R.~D. Ball et~al., \emph{{Parton distributions
  for the LHC Run II}},
  \href{http://dx.doi.org/10.1007/JHEP04(2015)040}{\emph{JHEP} {\bf 04} (2015)
  040}, [\href{https://arxiv.org/abs/1410.8849}{{\tt 1410.8849}}].

\bibitem{Adkins:2022jfp}
J.~K. Adkins et~al., \emph{{Design of the ECCE detector for the Electron Ion
  Collider}}, \href{http://dx.doi.org/10.1016/j.nima.2025.170240}{\emph{Nucl.
  Instrum. Meth. A} {\bf 1073} (2025) 170240},
  [\href{https://arxiv.org/abs/2209.02580}{{\tt 2209.02580}}].

\bibitem{Cowan:2010js}
G.~Cowan, K.~Cranmer, E.~Gross and O.~Vitells, \emph{{Asymptotic formulae for
  likelihood-based tests of new physics}},
  \href{http://dx.doi.org/10.1140/epjc/s10052-011-1554-0}{\emph{Eur. Phys. J.
  C} {\bf 71} (2011) 1554}, [\href{https://arxiv.org/abs/1007.1727}{{\tt
  1007.1727}}].

\bibitem{Cowan:2012}
G.~Cowan, \emph{{Discovery sensitivity for a counting experiment with
  background uncertainty}},  tech. rep., Royal Holloway, London, U.K. (2012).

\bibitem{Pitt:2024utg}
M.~Pitt, \emph{{Physics Perspectives with the ePIC Far-Forward and Far-Backward
  detectors}}, \href{http://dx.doi.org/10.22323/1.469.0259}{\emph{PoS} {\bf
  DIS2024} (2025) 259}, [\href{https://arxiv.org/abs/2409.02811}{{\tt
  2409.02811}}].

\bibitem{Oreglia:1980cs}
M.~Oreglia, \emph{{A Study of the Reactions $\psi^\prime \to \gamma \gamma
  \psi$}},  other thesis, 12, 1980.

\bibitem{ATLAS-CONF-2014-031}
\emph{{Search for scalar diphoton resonances in the mass range 65-600 GeV with
  the ATLAS detector in pp collision data at $\sqrt{s}$ = 8 TeV}},  tech. rep.,
  CERN, Geneva, 2014.

\bibitem{Cornella:2019uxs}
C.~Cornella, P.~Paradisi and O.~Sumensari, \emph{{Hunting for ALPs with Lepton
  Flavor Violation}},
  \href{http://dx.doi.org/10.1007/JHEP01(2020)158}{\emph{JHEP} {\bf 01} (2020)
  158}, [\href{https://arxiv.org/abs/1911.06279}{{\tt 1911.06279}}].

\bibitem{DELPHI:2003dlq}
{\scshape DELPHI} collaboration, J.~Abdallah et~al., \emph{{Photon events with
  missing energy in e+ e- collisions at s**(1/2) = 130-GeV to 209-GeV}},
  \href{http://dx.doi.org/10.1140/epjc/s2004-02051-8}{\emph{Eur. Phys. J. C}
  {\bf 38} (2005) 395--411}, [\href{https://arxiv.org/abs/hep-ex/0406019}{{\tt
  hep-ex/0406019}}].

\bibitem{DELPHI:2008uka}
{\scshape DELPHI} collaboration, J.~Abdallah et~al., \emph{{Search for one
  large extra dimension with the DELPHI detector at LEP}},
  \href{http://dx.doi.org/10.1140/epjc/s10052-009-0874-9}{\emph{Eur. Phys. J.
  C} {\bf 60} (2009) 17--23}, [\href{https://arxiv.org/abs/0901.4486}{{\tt
  0901.4486}}].

\bibitem{IceCube:2022pbe}
{\scshape IceCube} collaboration, R.~Abbasi et~al., \emph{{Non-standard
  neutrino interactions in IceCube}},
  \href{http://dx.doi.org/10.22323/1.398.0245}{\emph{PoS} {\bf EPS-HEP2021}
  (2022) 245}.

\bibitem{PhysRevD.84.014028}
P.~J. Fox, R.~Harnik, J.~Kopp and Y.~Tsai, \emph{Lep shines light on dark
  matter}, \href{http://dx.doi.org/10.1103/PhysRevD.84.014028}{\emph{Phys. Rev.
  D} {\bf 84} (Jul, 2011) 014028}.

\bibitem{GonzalezSuarez:2024dsp}
R.~Gonzalez~Suarez, B.~Pattnaik and J.~Zurita, \emph{{Leptophilic Z' bosons at
  the FCC-ee: Discovery opportunities}},
  \href{http://dx.doi.org/10.1103/PhysRevD.111.035029}{\emph{Phys. Rev. D} {\bf
  111} (2025) 035029}, [\href{https://arxiv.org/abs/2410.12903}{{\tt
  2410.12903}}].

\bibitem{ATLAS:2023vxg}
{\scshape ATLAS} collaboration, G.~Aad et~al., \emph{{Search for a new Z' gauge
  boson in $4\mu$ events with the ATLAS experiment}},
  \href{http://dx.doi.org/10.1007/JHEP07(2023)090}{\emph{JHEP} {\bf 07} (2023)
  090}, [\href{https://arxiv.org/abs/2301.09342}{{\tt 2301.09342}}].

\end{thebibliography}\endgroup

\end{document}